\documentclass[10pt,aps,prd,showpacs,superscriptaddress,eqsecnum,longbibliography,nofootinbib]{revtex4-2}
\usepackage{mathrsfs,amsmath,amsthm,latexsym,amssymb,amsfonts,epsfig,cancel,enumerate,graphicx,subcaption,txfonts,overpic}
\usepackage{multirow}
\usepackage{xcolor}
\definecolor{navy}{RGB}{0,0,150}
\usepackage[colorlinks, linkcolor=blue, citecolor=blue, urlcolor=blue, plainpages=false, pdfstartview=FitH]{hyperref}
\usepackage{appendix}
\allowdisplaybreaks
\newcommand{\GZU}{College of Physics, Guizhou University, Guiyang 550025, China}
\newcommand{\BJC}{Beijing Capital Air Environmental Science and Technology Co., Ltd., Beijing 100176, China}
\newcommand{\SU}{School of Physics and Electronic Engineering, Shanxi Normal University, Taiyuan 030031, China}

\begin{document}
	
	\title{Thin accretion disk around Schwarzschild-like black hole in bumblebee gravity}
	
	\author{Ziqiang Cai}
	\email{gs.zqcai24@gzu.edu.cn}
	\affiliation{\GZU}
	
	\author{Zhenglong Ban}
	\email{gs.zlban22@gzu.edu.cn}
	\affiliation{\GZU}
	
	\author{Lu Wang}
	\email{beautifulmind@163.com}
	\affiliation{\BJC}
	
	\author{Haiyuan Feng}
	\email{fenghaiyuanphysics@gmail.com}
	\affiliation{\SU}
	
	\author{Zheng-Wen Long}
	\thanks{Corresponding author}
	\email{zwlong@gzu.edu.cn}
	\affiliation{\GZU}

\begin{abstract}
The physical properties and optical appearance of a thin accretion disk surrounding a Schwarzschild-like black hole (BH) are investigated within the framework of bumblebee gravity. To understand how the Lorentz symmetry breaking (LSB) parameter $l$ affects the disk's behavior, we analyze main characteristics such as energy flux, temperature distribution, and emission spectrum. In addition, direct and secondary images of the accretion disk are generated and examined to explore how both the observational inclination angle and the LSB parameter $l$ shape the visual profile. Furthermore, we compute the redshift and observed flux distributions of the disk from the perspective of distant observers at various inclination angles. Our results indicate that the redshift factor grows as $l$ decreases. When the parameter $l$ assumes negative values, the BH exhibits enhanced luminosity with decreasing $l$. These findings highlight the crucial influence of the LSB parameter $l$ on the observable features of BHs.	
\end{abstract}
	
\maketitle
\section{Introduction}
General Relativity (GR) is widely recognized as the standard theory of gravity, offering a remarkably successful explanation for gravitational phenomena at the classical level \cite{ Will:2014kxa,LIGOScientific:2016aoc}. It stands out as one of the exemplary field theories in physics, providing precise descriptions of nature that are applicable even within the realm of particle physics. GR, while preserving Lorentz invariance in the absence of quantum effects, is considered inadequate for describing gravitational interactions and spacetime geometry at the quantum level, particularly when spacetime quantization is taken into account in high-energy regimes where LSB may occur. This limitation suggests that GR might require modifications to be unified with quantum mechanics into a comprehensive theory of gravity. Although direct tests of such quantum gravity (QG) theories are currently out of reach, being feasible only at the Planck scale ($10^{19}$ GeV), low-energy observations have hinted at potential signals of QG models, notably those associated with LSB \cite{Kostelecky:1989jw,Casana:2017jkc}. The bumblebee gravity model represents the simplest framework in which Lorentz symmetry is spontaneously broken, attributed to the non-zero vacuum expectation value of a singular vector field, referred to as the bumblebee field \cite{Kostelecky:2000mm,Kostelecky:2003fs}. The bumblebee model is a known gravity model that extends the standard formalism of GR. A static and spherically symmetric Schwarzschild-like BH solution was obtained in bumblebee gravity \cite{Casana:2017jkc}.
	
Furthermore, beyond purely theoretical explorations, the astronomical detection of real BH images has garnered considerable interest and excitement within the scientific community. This pursuit not only advances our understanding of these enigmatic objects but also provides crucial tests for various theories of gravity under extreme conditions. In 2019, the Event Horizon Telescope (EHT) Collaboration achieved a milestone by capturing the first direct visual evidence of a BH \cite{EventHorizonTelescope:2019dse,EventHorizonTelescope:2019uob,EventHorizonTelescope:2019jan,EventHorizonTelescope:2019ths,EventHorizonTelescope:2019pgp,EventHorizonTelescope:2019ggy}, specifically the supermassive object at the center of the M87$^{*}$ galaxy. The extraordinary image of the BH M87$^{*}$ captured by the EHT provides valuable insights for testing gravitational theories. These observations open new avenues for exploring BH properties in alternative theories of gravity. Building on this success, the EHT Collaboration has since released additional polarized images of the M87$^{*}$ BH \cite{EventHorizonTelescope:2021btj}, further enhancing our understanding of these enigmatic objects and their surrounding environments. These observations not only mark significant advancements in observational astronomy but also offer critical tests for various theories of gravity under extreme conditions.

Images of accretion disks surrounding BHs have fascinated observational astronomers since the 1970s. Early research in this field led to the proposal of a standard model for geometrically thin and optically thick accretion disks by Shakura et al. \cite{Shakura:1972te}, which has since become a cornerstone in understanding these phenomena. This foundational work was further developed into the relativistic framework by Novikov and Thorne \cite{Page:1974he}, resulting in the well-known Novikov-Thorne model. In real astrophysical observations, BHs are typically surrounded by accretion disks, making these models crucial for interpreting observed data and advancing our knowledge of BH environments. Techniques for imaging thin accretion disks primarily consist of two methods: the semi-analytic approach and ray-tracing combined with radiative transfer. Luminet utilized the semi-analytic method to generate direct and secondary images of a thin accretion disk around a Schwarzschild BH, and he calculated the disk's brightness using an analytical formula for radiation flux derived in \cite{Luminet:1979nyg}. To simulate the accretion structures surrounding BHs, researchers have utilized a range of numerical ray-tracing codes (e.g., \cite{17,18,19,Broderick:2005my,Dexter:2016cdk,Vincent:2011wz,24,Cunha:2016bjh}). Additionally, the optical characteristics and physical properties of thin accretion disks in diverse background spacetimes have been thoroughly examined \cite{Hou:2022eev,Zhang:2024lsf,Gyulchev:2019tvk,Shaikh:2019hbm,Bambi:2019tjh,Johannsen:2016uoh,Gates:2020sdh,Okyay:2021nnh}. In parallel, studies such as \cite{Huang:2023ilm,Guo:2023grt,Liu:2021lvk,Guo:2022rql} have delved into the imaging of BHs and naked singularities within various modified gravity frameworks.
	
The aim of this paper is to investigate the physical properties and optical appearance of a thin accretion disk surrounding a Schwarzschild-like BH within the bumblebee gravity framework \cite{Casana:2017jkc}, using the Novikov-Thorne model \cite{Page:1974he}. BHs are typically surrounded by thin accretion disks. Due to the strong gravitational field, these thin accretion disks emit high-energy fluxes \cite{Collodel:2021gxu,Feng:2024iqj,He:2022lrc,Wu:2024sng,Liu:2024brf,Abbas:2023rzk,Feng:2023iha,Feng:2022bst}. Adopting the methodology from \cite{You:2024uql}, we will generate both direct and secondary images of the accretion disk as seen by a distant observer. Additionally, we will illustrate the distribution of redshift and observed flux on the photographic plate to analyze potential deviations from the Schwarzschild BH scenario.
	
The structure of this paper is as follows. In Sec. \ref{section2}, we briefly review the bumblebee gravity and its Schwarzschild-like solution. We also analyze photon geodesics in the equatorial plane of the Schwarzschild-like BH and plot the photon orbits around the BH. In Sec. \ref{section3}, based on the properties of accretion disks surrounding Schwarzschild-like BH, we investigate the radiant energy flux, radiation temperature, and observed luminosity. In Sec. \ref{section4}, we use the $(\varphi(b))$ diagram to analyze the imaging of BH accretion disks and plot their direct and secondary images of the accretion disk, comparing them with those of Schwarzschild BHs. Finally, we study the distribution of radiation flux and redshift of the thin accretion disk as perceived by distant observers at various inclination angles. In Sec. \ref{section5}, we summarize our results. 
	
\section{GEODESICS OF THE SCHWARZSCHILD-LIKE BH}
\label{section2}
\subsection{Schwarzschild-like BH in bumblebee gravity}	
The bumblebee gravity models are the simplest examples of field theories with spontaneous Lorentz and diffeomorphism violations. In these scenarios, the spontaneous LSB is induced by a potential whose functional form possesses a minimum which breaks the $U(1)$ symmetry. For a single bumblebee field $B_{\mu}$ coupled to gravity and matter, the action can be written as \cite{Casana:2017jkc}
\begin{equation}
	S_{B}=\int d^{4}x\mathcal{L}_{B},\label{SB}
\end{equation}
the Lagrangian density $\mathcal{L}_{B}$ is given by
\begin{equation}
	\mathcal{L}_{B}=\frac{e}{2\kappa}R+\frac{e}{2\kappa}\xi B^{\mu}B^{\nu}R_{\mu\nu}-\frac{1}{4}eB_{\mu\nu}B^{\mu\nu}-eV(B^{\mu})+\mathcal{L}_{M},\label{LB}
\end{equation}
where $\kappa=8\pi G_{N}$ is the gravitational coupling, $e\equiv\sqrt{-g}$ is the determinant of the vierbein and $\xi$ is the real coupling constant which controls the nonminimal gravity bumblebee interaction, and $\mathcal{L}_{M}$ defines the matter and other field contents and their couplings to the bumblebee field. The bumblebee field strength is defined as
\begin{equation}
		B_{\mu\nu}=\partial_{\mu}B_{\nu}-\partial_{\nu}B_{\mu}.\label{Bmu}
\end{equation}
From the Lagrangian density (\ref{LB}), we obtain the modified Einstein equations
\begin{equation}
		R_{\mu\nu}-\frac{1}{2}Rg_{\mu\nu}=\kappa T_{\mu\nu},\label{Eeq}
\end{equation}
where $T_{\mu\nu}$ is the total energy-momentum tensor, which results from the contributions of the matter sector $(T_{\mu\nu}^{M})$ and the bumblebee field $(T_{\mu\nu}^{B})$; thus, we write
\begin{equation}
		T_{\mu\nu}=T_{\mu\nu}^{M}+T_{\mu\nu}^{B},\label{Tmu}
\end{equation}
with
\begin{equation}
		\begin{split}
			T_{\mu \nu}^{B} & = -B_{\mu \alpha} B_{\nu}^{\alpha}-\frac{1}{4} B_{\alpha \beta} B^{\alpha \beta} g_{\mu \nu}-V g_{\mu \nu}+2 V^{\prime} B_{\mu} B_{\nu} \\
			& +\frac{\xi}{\kappa}\left[\frac{1}{2} B^{\alpha} B^{\beta} R_{\alpha \beta} g_{\mu \nu}-B_{\mu} B^{\alpha} R_{\alpha \nu}-B_{\nu} B^{\alpha} R_{\alpha \mu}\right. \\
			& +\frac{1}{2} \nabla_{\alpha} \nabla_{\mu}\left(B^{\alpha} B_{\nu}\right)+\frac{1}{2} \nabla_{\alpha} \nabla_{\nu}\left(B^{\alpha} B_{\mu}\right) \\
			& \left.-\frac{1}{2} \nabla^{2}\left(B_{\mu} B_{\nu}\right)-\frac{1}{2} g_{\mu \nu} \nabla_{\alpha} \nabla_{\beta}\left(B^{\alpha} B^{\beta}\right)\right],
		\end{split}
	\label{TBmu}
\end{equation}
where the prime means differentiation with respect to the argument. The equation of motion for the bumblebee field from (\ref{LB}) is given by
\begin{equation}
		\nabla^{\mu}B_{\mu\nu}=J_{\nu}^{M}+J_{\nu}^{B},\label{dataBmu}
\end{equation}
where $J_{\nu}^{M}$ is the matter current and $J_{\nu}^{B}$ is the bumblebee field current which takes the form
\begin{equation}
		J_{\nu}^{B}=2V^{\prime}B_{\nu}-\frac{\xi}{\kappa}B^{\mu}R_{\mu\nu}.\label{JBmu}
\end{equation}
A static and spherically symmetric solution in bumblebee gravity, called the Schwarzschild-like BH, was obtained in \cite{Casana:2017jkc}. Its geometry is given by the following line element
\begin{equation}
		ds^{2}=-f(r)dt^{2}+(1+l)\frac{1}{f(r)}dr^{2}+r^{2}\left(d\theta^{2}+\sin^{2}\theta d \phi^{2}\right),\label{dugui}
\end{equation}
where $f(r)=1-2M/r$, with $M$ is the mass of the BH and $l$ is a constant characterizing the LSB. When $l=0$, the metric (\ref{dugui}) reduces to the Schwarzschild solution. Since $g_{11}>0$, we have theoretical constraints on the parameter $l\colon l>-1$. In \cite{Casana:2017jkc}, an upper-bound for $l$ was found: $l \leq 10^{-13}$. The horizon of BH is at $r_{h}=2M$.

\subsection{Null and Time-like Geodesics in Schwarzschild-like BH Spacetimes}		
The Lagrangian $\mathcal{L}$ for a point particle in the spacetime (\ref{dugui}) is given by 
\begin{equation}
		\mathcal{L}=\frac{1}{2}g_{\mu\nu}\frac{dx^{\mu}}{d\lambda}\frac{dx^{\nu}}{d\lambda}=\frac{1}{2}\varepsilon,\label{La}
\end{equation}
where $\lambda$ is the affine parameter. The Lagrangian (\ref{La}) corresponds to massive particles when $\varepsilon=-1$, while it corresponds to photons when $\varepsilon$ takes the value of $0$. We only consider the orbits in the equatorial plane $\theta=\pi/2$. The conserved energy $E$ and the conserved angular momentum $L$ can be calculated, respectively, as 
\begin{equation}
		E=-g_{tt}\frac{dt}{d\lambda}=\left(1-\frac{2M}{r}\right)\frac{dt}{d\lambda},\label{energy}
\end{equation}
\begin{equation}
		L=g_{\phi\phi}\frac{d\phi}{d\lambda}=r^{2}\frac{d\phi}{d\lambda}.\label{momentum}
\end{equation}
Using Eqs. (\ref{La}), (\ref{energy}), and (\ref{momentum}) we can get the light propagation equations,
\begin{equation}
		\frac{dt}{d\lambda}=\frac{E}{f(r)},\label{tdian}
\end{equation}
\begin{equation}
		\left(\frac{dr}{d\lambda}\right)^{2}=\frac{1}{1+l}\left[E^{2}-f(r)\frac{L^{2}}{r^{2}}\right],\label{rdian2}
\end{equation}
\begin{equation}
		\frac{d\phi}{d\lambda}=\frac{L}{r^{2}}.\label{faidian}
\end{equation}
Let $\lambda^{\prime}=L\lambda$, the light propagation equation can be rewritten as
\begin{equation}
		\frac{dt}{d\lambda^{\prime}}=\frac{1}{bf(r)},\label{2tdian}
\end{equation}
\begin{equation}
		\frac{d\phi}{d\lambda^{\prime}}=\frac{1}{r^{2}},\label{2faidian}
\end{equation}
\begin{equation}
		\left(\frac{dr}{d\lambda^{\prime}}\right)^{2}=\frac{1}{1+l}\left[\frac{1}{b^{2}}-\frac{f(r)}{r^{2}}\right],\label{2rdian2}
\end{equation}
where $b=L/E$ is the impact parameter associated with the light ray. From the light propagation equation, we can derive an ordinary differential equation of the radius $r$ in terms of the azimuthal angle $\phi$ on the orbital plane as
\begin{equation}
		\left(\frac{dr}{d\phi}\right)^{2}=\frac{r^{4}}{1+l}\left(\frac{1}{b^{2}}-\frac{f(r)}{r^{2}}\right)\equiv V_{eff},\label{Veff}
\end{equation}
where $V_{eff}$ is the effective potential of the photon. The radius $r_{ph}$ of the photon sphere formed by a bounded orbit of light is determined by
\begin{equation}
		V_{eff}|_{r=r_{ph}}=0,\label{Veff0}
\end{equation}
\begin{equation}
		\frac{dV_{eff}}{dr}|_{r=r_{ph}}=0.\label{dVeff0}
\end{equation}
By solving Eqs. (\ref{Veff0}), (\ref{dVeff0}) we can get  the critical compact parameter $b_{c}$ corresponding to the photon sphere as
\begin{equation}
		b_{c}=\frac{r_{ph}}{\sqrt{f(r_{ph})}}.\label{bc}
\end{equation}
	
Now let's introduce a crucial concept, the innermost stable circular orbit (ISCO) of a time-like particle. The orbit equation of the time-like geodesic reads
\begin{equation}
		\left(\frac{dr}{d\phi}\right)^{2}=\frac{r^{4}}{1+l}\left(\frac{1}{b^{2}}-\frac{f(r)}{r^{2}}-\frac{f(r)}{L^{2}}\right)=\tilde{V}_{eff}.\label{Veff2}
\end{equation}
The radius $r_{isco}$ of the ISCO is given by 
\begin{equation}
		\tilde{V}_{eff}|_{r=r_{isco}}=\frac{d\tilde{V}_{eff}}{dr}|_{r=r_{isco}}=\frac{d^{2}\tilde{V}_{eff}}{dr^{2}}|_{r=r_{isco}}=0.\label{Veff20}
\end{equation}
In the analysis of the image of thin accretion disks, the ISCO  plays a crucial role as it defines the inner boundary of the disk. For radii $r<r_{isco}$, equatorial circular orbits become unstable, meaning that any perturbation will cause particles to either fall into the BH or escape to greater distances. Thus, the ISCO determines the inner edge of the accretion disk by marking the threshold at which orbits begin to destabilize. Specifically, the ISCO signifies the critical boundary where internal orbits lose stability, thereby establishing the disk's innermost extent.
	
\begin{figure*}[htbp]
	\centering
	\begin{subfigure}{0.38\textwidth}
		\includegraphics[width=\linewidth]{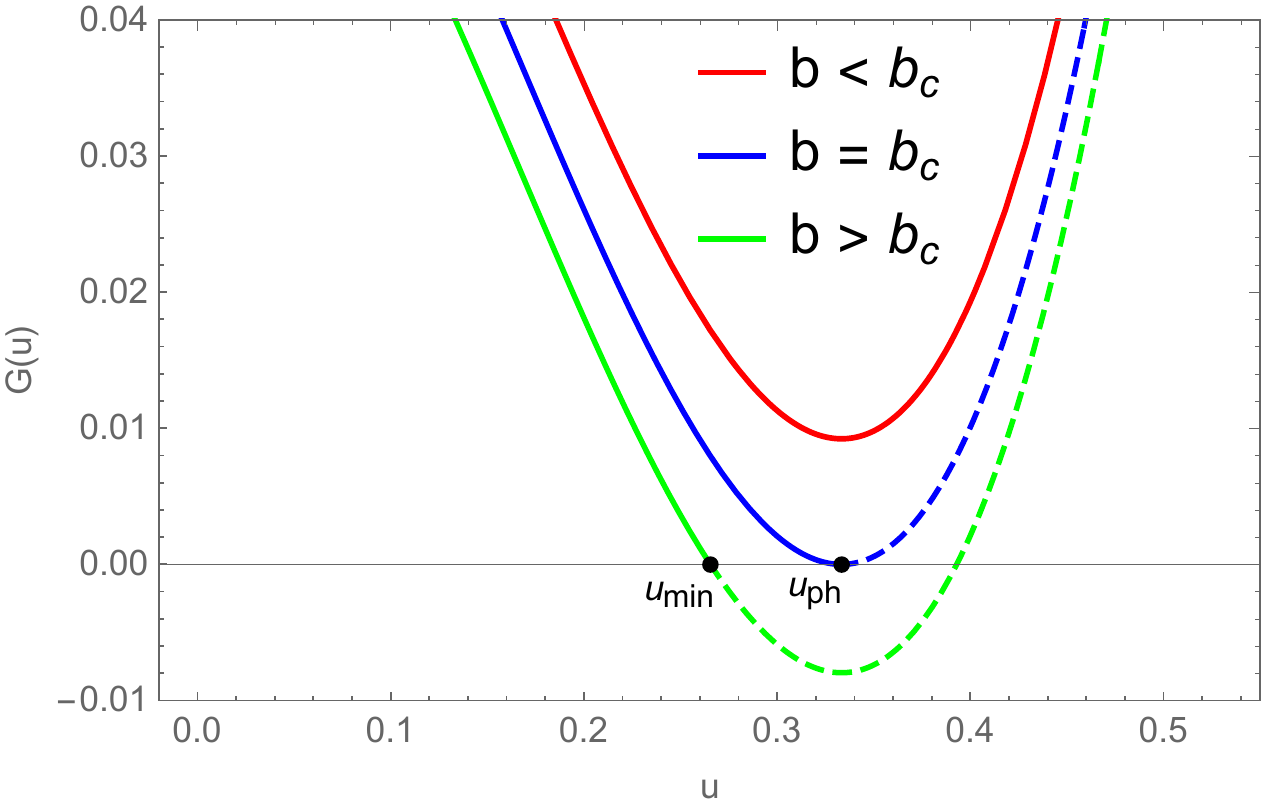}
	\end{subfigure}%
	\hspace{0.13\textwidth}
	\begin{subfigure}{0.38\textwidth}
		\includegraphics[width=\linewidth]{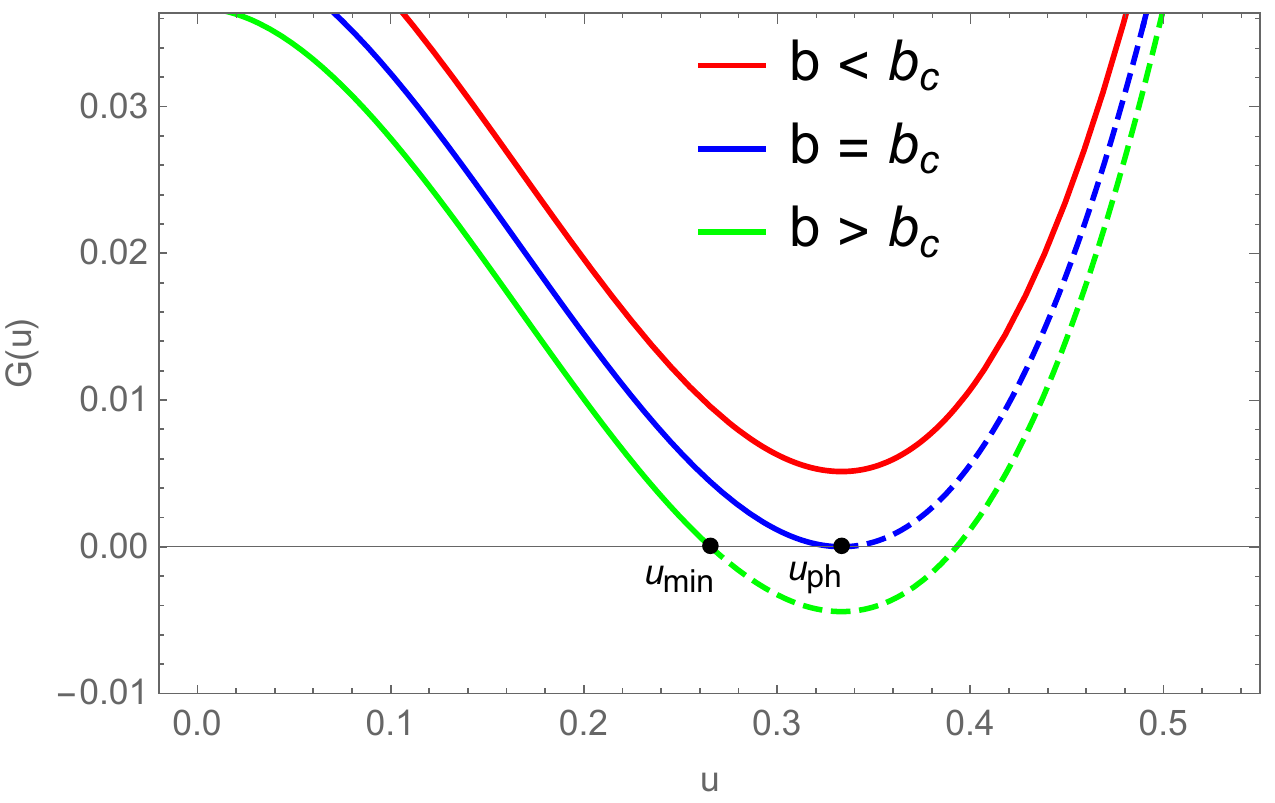}
	\end{subfigure}
	
	\caption{The functions $G(u)$ for different values of $l$, as a function of $u$. Left panel: $l=-0.5$. Right panel: $l=-0.1$. $u_{ph}=1/r_{ph}$, $u_{min}$ is the  minimum positive root of $G(u)$ when $b>b_{c}$.}
	\label{Gu}
\end{figure*}

In order to calculate the photon orbit, we introduce a new variable $u=1/r$, and the equation of motion for photon can be rewritten as
\begin{equation}
		\left(\frac{du}{d\phi}\right)^{2}=\frac{1}{1+l}\left[\frac{1}{b^{2}}-u^{2}f(\frac{1}{u})\right]\equiv G(u).\label{gu}
\end{equation}
We have plotted the $G(u)$ images with different parameters $l$, as shown in Fig.~\ref{Gu}. When $b>b_{c}$, $G(u)$ has two roots; when $b=b_{c}$, $G(u)$ has one root; and when $b<b_{c}$, $G(u)$ has no roots. In Fig.~\ref{Gu}, the values of $u_{min}$ and $u_{ph}$ are identical for the two different values of $l$. According to Eq. (\ref{gu}), the roots of $G(u)=0$ are independent of the parameter $l$.
	
In scenarios where $b<b_{c}$, our consideration is limited to the trajectory outside the event horizon, from which we derive the total change in the azimuthal angle $\varphi$ (We employ $\phi$ and $\varphi$ to represent the space-time coordinate and the change in $\phi$ associated with the photon's motion, respectively):
\begin{equation}
		\varphi = \int_{0}^{u_{h}} \frac{1}{\sqrt{G(u)}} \, du, \quad b < b_{c}.\label{fai1}
\end{equation}
Here, $u_{h}=1/r_{h}$, where $r_{h}$ is the radius of the outermost horizon. In the case where $b>b_{c}$, the overall variation in the azimuthal angle $\varphi$ for a trajectory characterized by an impact parameter $b$ can be determined by
\begin{equation}
		\varphi=2\int_{0}^{u_{min}}\frac{1}{\sqrt{G(u)}} \, du, \quad b > b_{c},\label{fai2}
\end{equation}
where $u_{min}$ is the minimum positive root of $G(u)=0$ when $b>b_{c}$.
	
\begin{figure}[htbp]
	\centering
	\begin{subfigure}{0.4\textwidth}
		\includegraphics[width=2.6in, height=3.5in, keepaspectratio]{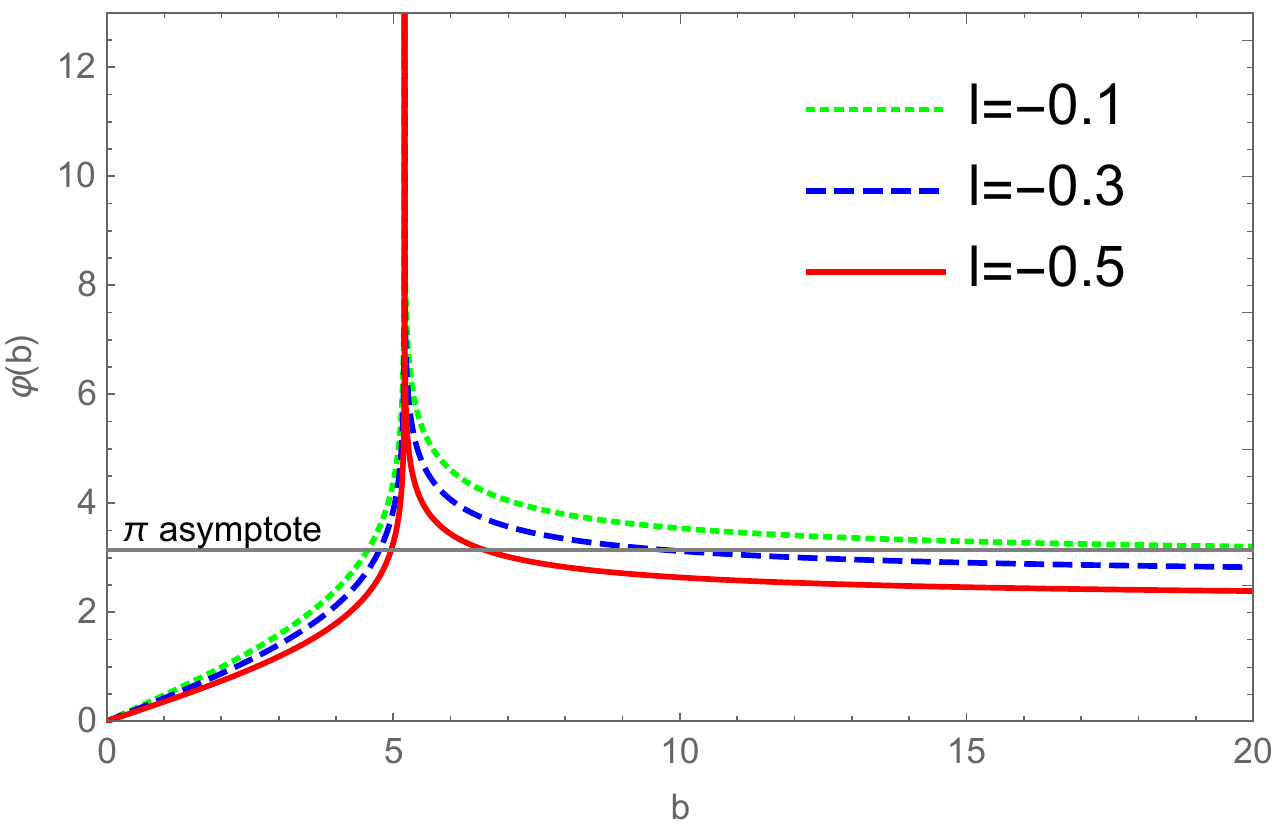}
	\end{subfigure}
	\caption{The azimuthal angle $\varphi(b)$ for different values of $l$, as a function of $b$.}
	\label{faib}
\end{figure}

\begin{figure*}[htbp]
	\centering
	
	\begin{subfigure}{0.3\textwidth}
		\centering
		\begin{overpic}[width=\textwidth]{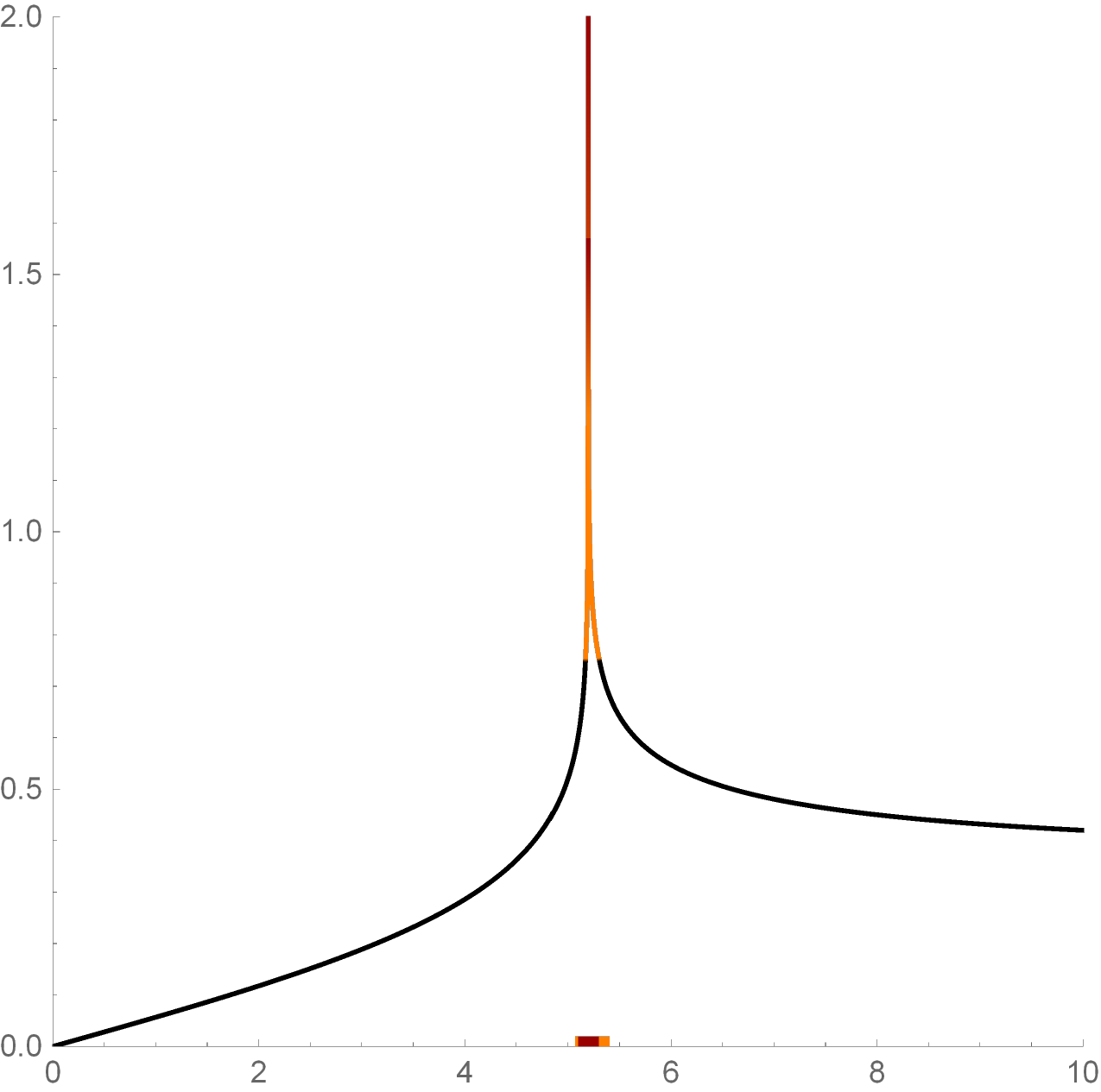}
			\put(103,2){\color{black} $b$}
			\put(0,100){\color{black} $n=\phi/2\pi$}
		\end{overpic}
		\includegraphics[width=\textwidth]{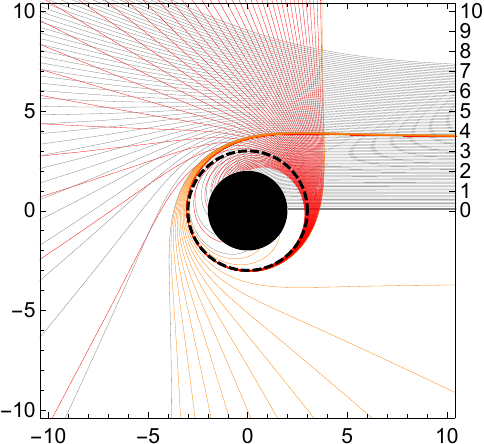}
	\end{subfigure}
	\hfill
	\begin{subfigure}{0.3\textwidth}
		\centering
		\begin{overpic}[width=\textwidth]{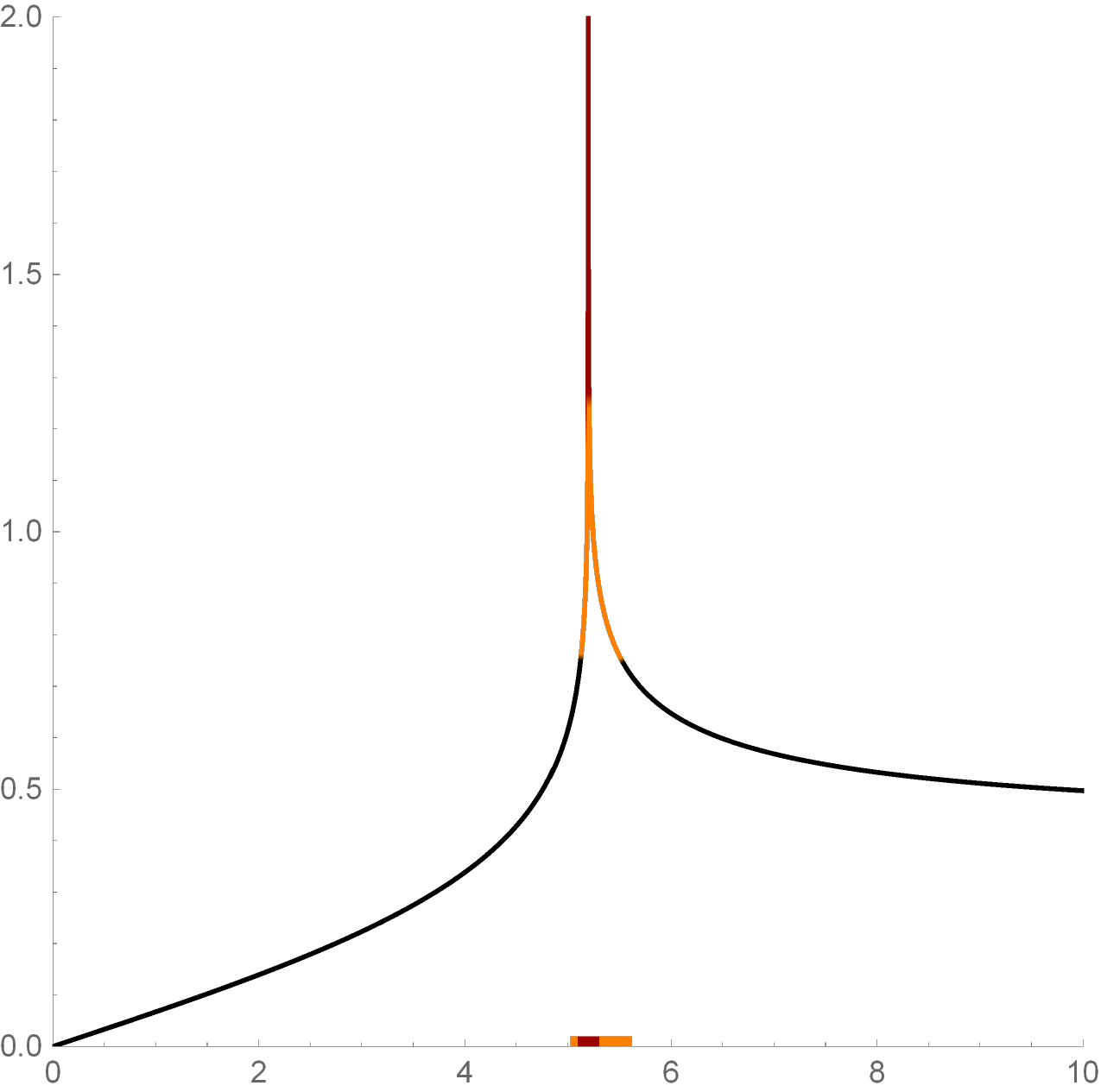}
			\put(103,2){\color{black} $b$}
			\put(0,100){\color{black} $n=\phi/2\pi$}
		\end{overpic}
		\includegraphics[width=\textwidth]{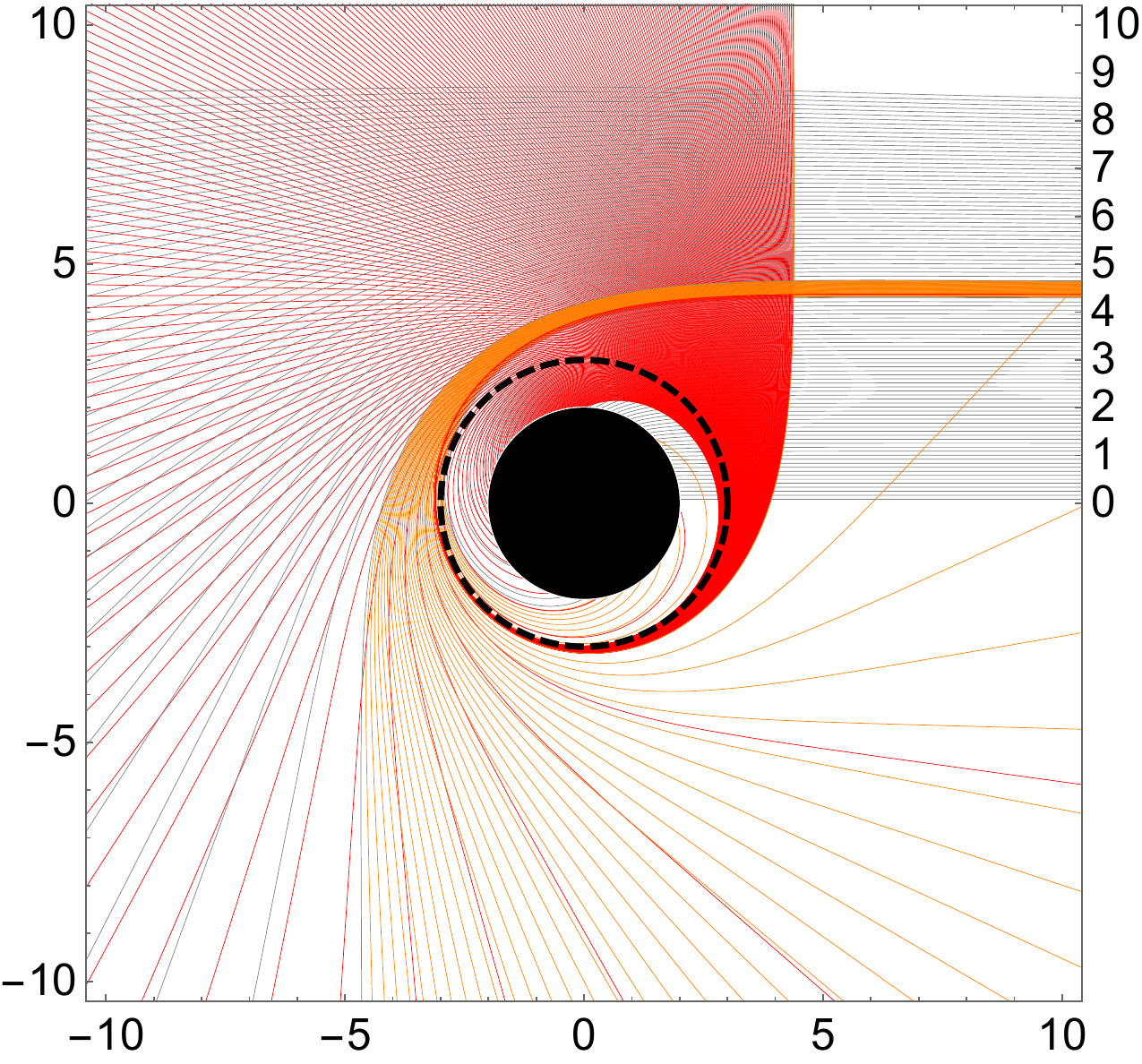}
	\end{subfigure}
	\hfill
	\begin{subfigure}{0.3\textwidth}
		\centering
		\begin{overpic}[width=\textwidth]{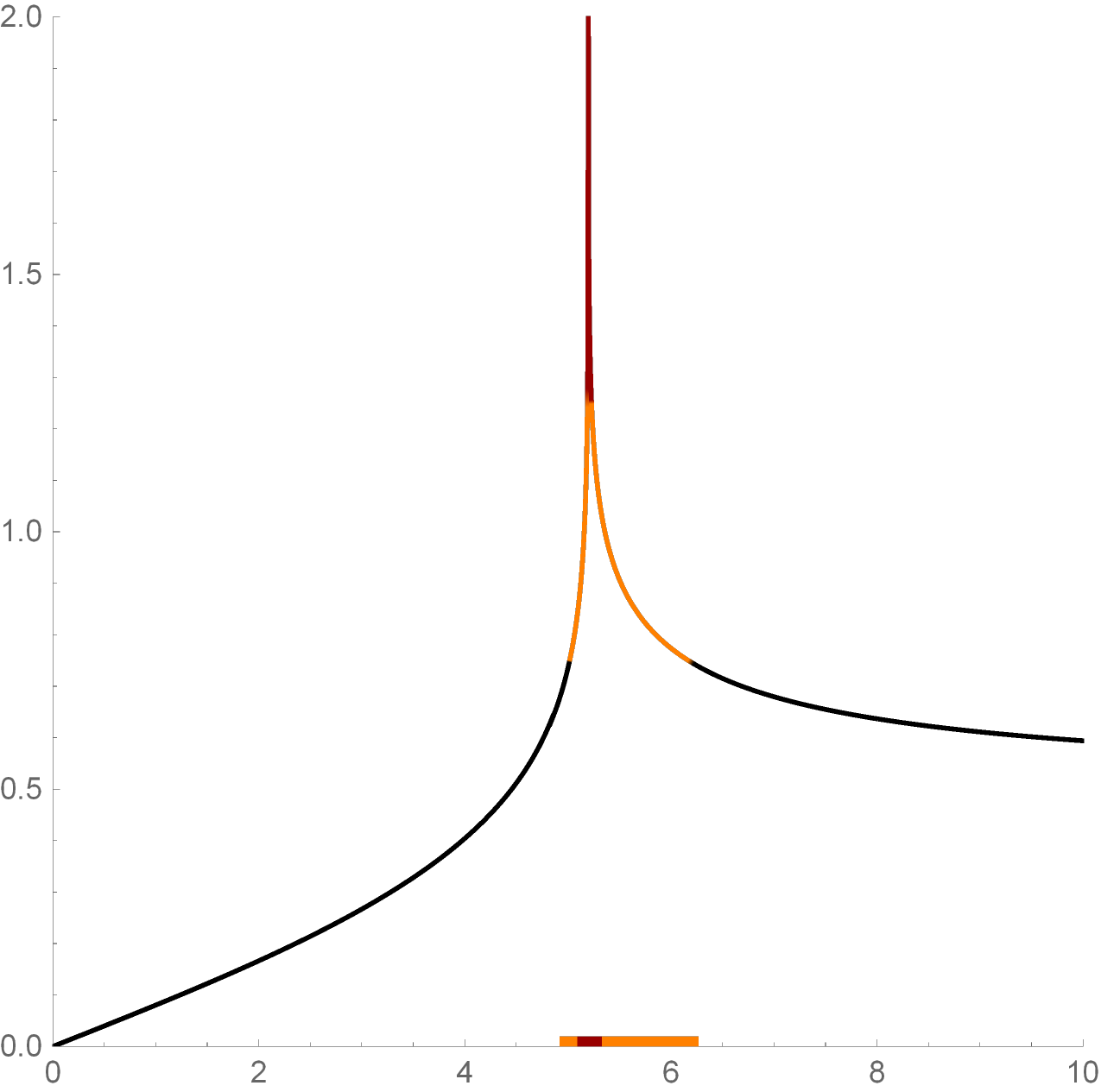}
			\put(103,2){\color{black} $b$}
			\put(0,100){\color{black} $n=\phi/2\pi$}
		\end{overpic}
		\includegraphics[width=\textwidth]{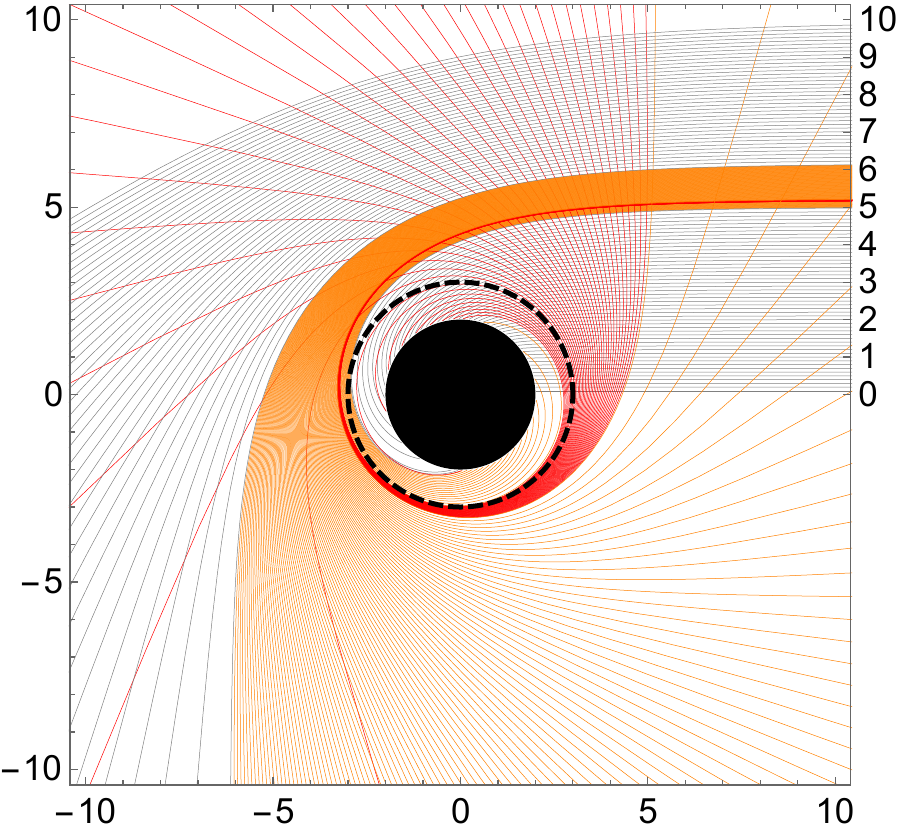}
	\end{subfigure}
	
	\caption{The behavior of photon trajectories around a Schwarzschild-like BH in bumblebee gravity as a function of the impact parameter $b$ for different values of the parameter $l$. The upper panel displays the total number of orbits $(n = \phi/2\pi)$, classifying trajectories into three categories based on $n$: direct emission $(n<3/4)$ depicted in black, lensed trajectories $(3/4 < n < 5/4)$ shown in orange, and photon ring trajectories $(n > 5/4)$ colored in red. In the lower panel, selected photon paths are visualized using Euclidean polar coordinates $(r, \phi)$. The impact parameter spacing is adjusted to $1/10$, $1/100$, and $1/1000$ for direct emissions, lensed paths, and photon rings, respectively. Three scenarios are analyzed: setting $l=-0.5$ in the first column; $l=-0.3$ in the second column; and $l=0$ in the third column.}
	\label{wanqu}
\end{figure*}

For a photon with an impact parameter $b=b_{c}$, it will reach $u_{ph}$ and subsequently engage in perpetual circular motion. Fig.~\ref{faib} shows the azimuthal angle $\varphi$ as a function of the impact parameter $b$. It can be observed from the figure that $\varphi(b)=\pi$ is an asymptote of the graph. As $l$ increases, the entire graph of $\varphi(b)$ will shift upwards.
In order to interpret the observed features of radiation emitted near a BH, Ref. \cite{Gralla:2019xty} introduced a classification of photon trajectories into direct, lensed, and photon ring contributions. Let $n=\frac{\phi}{2\pi}$ denote the total number of orbits completed by a photon trajectory. This quantity is generally a function of the impact parameter $b$, and we denote its solution as
\begin{equation}
	\label{nb}
	n(b) = \frac{2m - 1}{4}, \quad m = 1, 2, 3, \dotsc
\end{equation}
by $b_{m}^{\pm}$. It is important to note that $b_{m}^{-}<b_{c}$ and $b_{m}^{+}>b_{c}$. Based on this distinction, we can classify all photon trajectories as follows:
\begin{itemize}
	\item direct: $\dfrac{1}{4} < n < \dfrac{3}{4} \Rightarrow b \in (b_{1}^{-}, b_{2}^{-}) \cup (b_{2}^{+}, \infty)$
	\item lensed: $\dfrac{3}{4} < n < \dfrac{5}{4} \Rightarrow b \in (b_{2}^{-}, b_{3}^{-}) \cup (b_{3}^{+}, b_{2}^{+})$
	\item photon ring: $n > \dfrac{5}{4} \Rightarrow b \in (b_{3}^{-}, b_{3}^{+})$
\end{itemize}
The physical significance of this classification can be understood from the trajectory plots in Fig.~\ref{wanqu}. Assuming emission from the north pole (on the far right), trajectories with $1/4 < n < 3/4$ intersect the equatorial plane once, those with $3/4 < n < 5/4$ intersect it twice, and trajectories with $n > 5/4$ intersect it at least three times.
\section{Physical properties of thin accretion disks}
\label{section3}
\begin{figure}[htbp]
	\centering
	\begin{subfigure}{0.4\textwidth}
		\includegraphics[width=2.6in, height=3.5in, keepaspectratio]{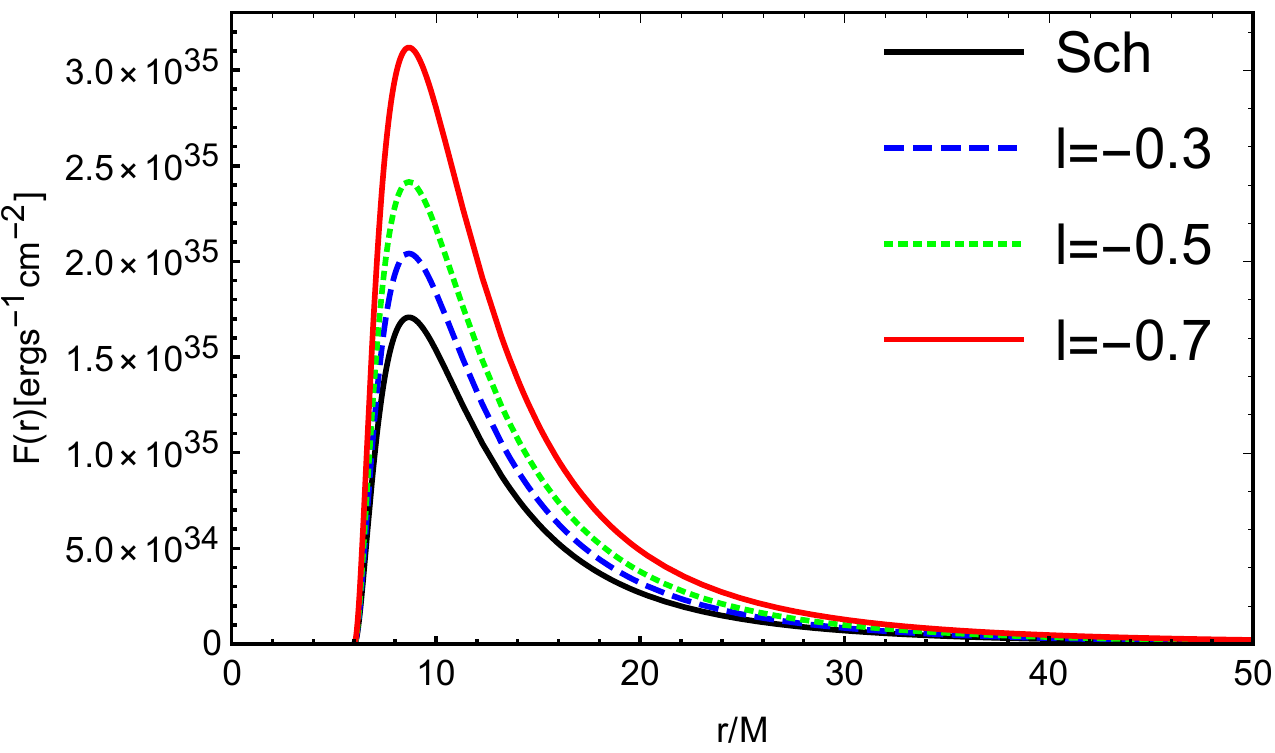}
	\end{subfigure}
	\caption{The energy flux $F(r)$ from a disk around a Schwarzschild-like BH in bumblebee gravity for different values of $l$.}
	\label{Fr}
\end{figure}

\begin{figure}[htbp]
	\centering
	\begin{subfigure}{0.4\textwidth}
		\includegraphics[width=2.6in, height=3.5in, keepaspectratio]{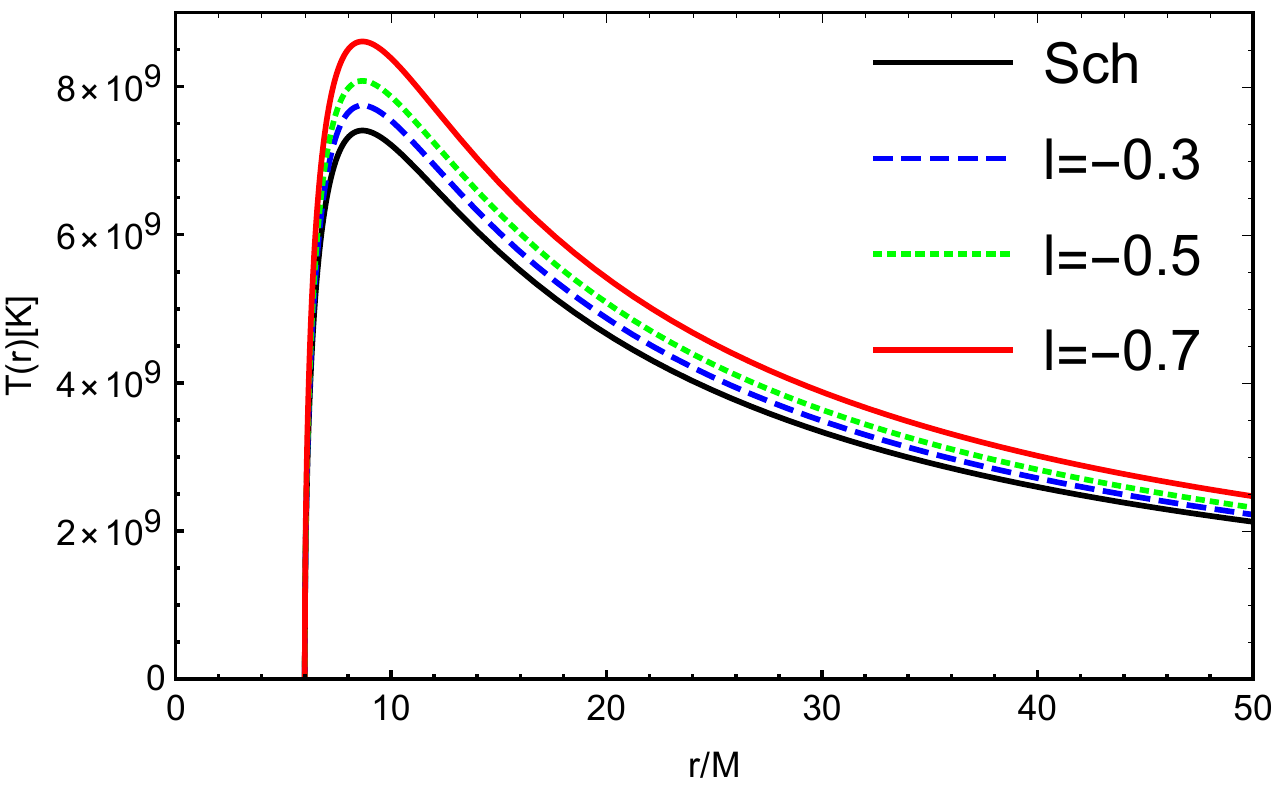}
	\end{subfigure}
	\caption{Variety of the disk temperature $T(r)$ with different parameters $l$ for the thin disk around a Schwarzschild-like BH in bumblebee gravity.}
	\label{Tr}
\end{figure}

In this section, we will explore the accretion processes occurring in thin disks surrounding Schwarzschild-like BH. A detailed discussion will be provided on how the parameter $l$ affects the radiant energy flux, radiation temperature, and observable luminosity. The following values are adopted for the physical constants and the characteristics of the thin accretion disk in our analysis: the speed of light $c=2.997\times10^{10}\text{cms}^{-1}$, the accretion rate $\dot{M}_{0}=2\times10^{-6}M_{\odot}\text{yr}^{-1}$, the unit of year $1\text{yr}=3.156\times10^{7}\text{s}$, the Stefan-Boltzmann constant $\sigma_\text{SB}=5.67\times10^{-5}\text{ergs}^{-1}\text{cm}^{-2}\text{K}^{-4}$, the Planck constant $h=6.625\times10^{-27}\text{ergs}$, the Boltzmann constant $k_{B}=1.38\times10^{-16}\text{erg}\text{K}^{-1}$, the solar mass $M_{\odot}=1.989\times10^{33}\text{g}$, and the mass of BH $M=2\times10^{6}M_{\odot}$, are used in \cite{He:2022lrc}. The surface radiant energy flux is determined based on the methods outlined in \cite{Page:1974he}
\begin{equation}
		F(r)=-\frac{\dot{M}_{0}\Omega_{,r}}{4\pi\sqrt{-g}(E-\Omega L)^{2}}\int^{r}_{r_{isco}}(E-\Omega L)L_{,r}dr.\label{fr}
\end{equation}
The equation is commonly used in the literature for cylindrical coordinates. However, for applications in spherical coordinates, it must be reformulated as detailed in \cite{Collodel:2021gxu}
\begin{equation}
		F(r)=-\frac{-c^{2}\dot{M}_{0}\Omega_{,r}}{4\pi\sqrt{-g/g_{\theta\theta}}(E-\Omega L)^{2}}\int^{r}_{r_{isco}}(E-\Omega L)L_{,r}dr,\label{fr2}
\end{equation}
where $\dot{M}_{0}$ stands for the mass accretion rate and $g$ refers to the metric determinant. $E$, $L$, and $\Omega$ denote the energy, angular momentum, and angular velocity of the particle in the circular orbit, respectively. As shown in Fig.~\ref{Fr}, the energy flux $F(r)$ of a disk around a Schwarzschild-like BH varies with different values of $l$. The black line represents the energy flux of a Schwarzschild (Sch) BH. As $l$ increases, the energy flux decreases. For a general static spherically symmetric metric $ds^{2}=g_{tt}dt^{2}+g_{rr}dr^{2}+g_{\theta\theta}d\theta^{2}+g_{\varphi\varphi}d\varphi^{2}$, $E$, $L$, and $\Omega$ are denoted as
\begin{equation}
		E=-\frac{g_{tt}}{\sqrt{-g_{tt}-g_{\phi\phi}\Omega^{2}}},\label{nengliang}
\end{equation}
\begin{equation}
		L=\frac{g_{\phi\phi}\Omega}{\sqrt{-g_{tt}-g_{\phi\phi}\Omega^{2}}},\label{dongliang}
\end{equation}
\begin{equation}
		\Omega=\frac{d\phi}{dt}=\sqrt{-\frac{g_{tt,r}}{g_{\phi\phi,r}}}.\label{jdongliang}
\end{equation}
Within the Novikov-Thorne model framework, the accreted material reaches a state of thermodynamic equilibrium, which implies that the radiation emitted by the disk closely resembles that of a perfect black body. The relationship between the disk's radiation temperature $T(r)$ and its energy flux $F(r)$ is given by the Stefan-Boltzmann law: $F(r)=\sigma_\text{SB}T^{4}(r)$, where $\sigma_\text{SB}$ represents the Stefan-Boltzmann constant. In Fig.~\ref{Tr}, we present the radiation temperature $T(r)$ for the region surrounding a Schwarzschild-like BH. As the parameter $l$ increases, the radiation temperature $T(r)$ decreases.

\begin{figure}[htbp]
	\centering
	\begin{subfigure}{0.4\textwidth}
		\includegraphics[width=2.6in, height=3.5in, keepaspectratio]{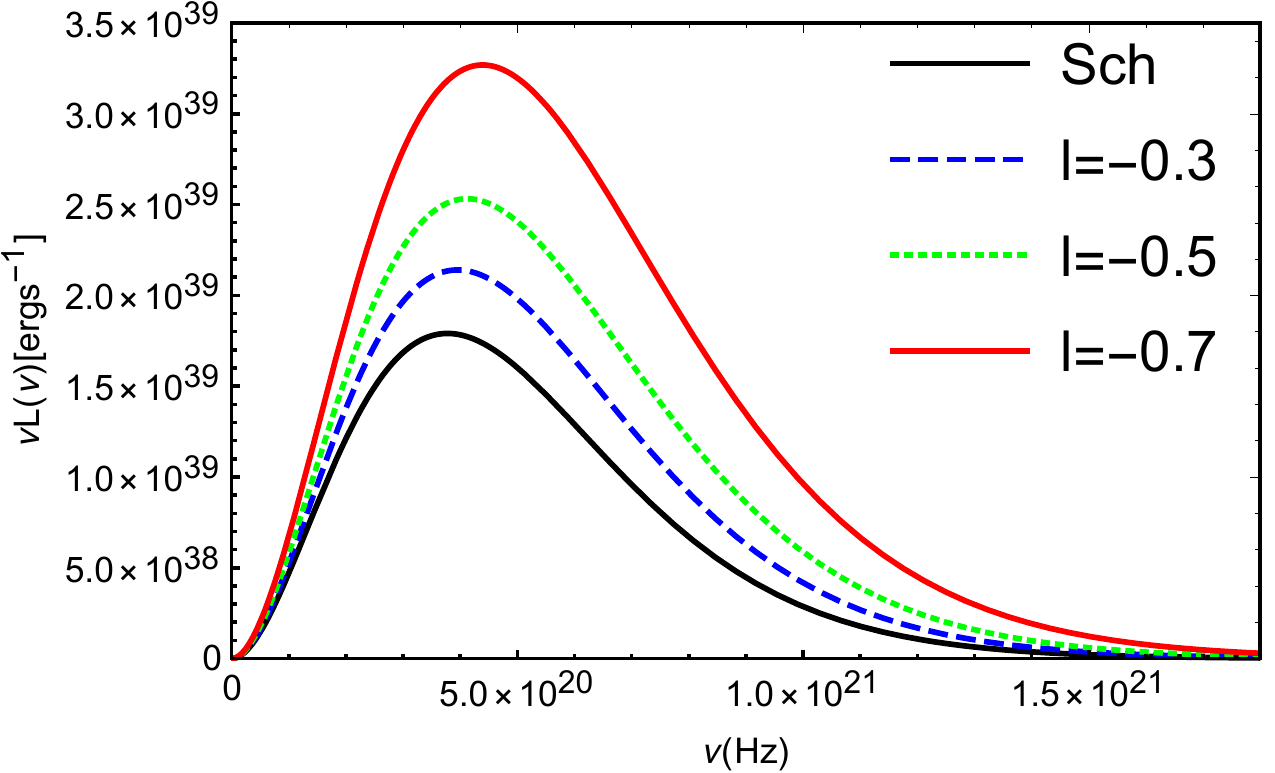}
	\end{subfigure}
	\caption{The emission spectrum $\nu L(\nu)$ of the accretion disk around a Schwarzschild-like BH in bumblebee gravity for different values of $l$, as a function of frequency $\nu$.}
	\label{vLv}
\end{figure}

The red-shifted black body spectrum of the observed luminosity $L(\nu)$ for a thin accretion disk around a BH is provided in \cite{Torres:2002td}
\begin{equation}
		L(\nu)=4\pi d^{2}I(\nu)=\frac{8\pi h\cos\gamma}{c^{2}}\int_{r_{i}}^{r_{f}}\int_{0}^{2\pi}\frac{\nu_{e}^{3}r}{e^{\frac{h\nu_{e}}{k_{B}T}}-1}drd\phi,\label{lv}
\end{equation}
where, $d$ represents the distance to the disk center, $I(\nu)$ denotes the thermal energy flux radiated by the disk, $h$ is the Planck constant, $k_{B}$ is the Boltzmann constant, and $\gamma$ is the disk inclination angle, which we will set to zero. The quantities $r_{f}$ and $r_{i}$ correspond to the outer and inner radii of the disk's edges, respectively. Assuming that the flux over the disk surface approaches zero, we select $r_{i}=r_{isco}$ and $r_{f}\rightarrow\infty$ for calculating the luminosity $L(\nu)$ of the disk. The emitted frequency is given by $\nu_{e}=\nu(1+z)$, where the redshift factor $z$ can be expressed as follows \cite{Luminet:1979nyg}
\begin{equation}
		1+z=\frac{1+\Omega b\sin\theta\cos\alpha}{\sqrt{-g_{tt}-g_{\phi\phi}\Omega^{2}}}.\label{hongyi}
\end{equation}
In this case, we designate $\alpha$ from Eq. (17) in Ref. \cite{Luminet:1979nyg} as $\alpha^{\prime}$. The relationship between $\alpha$ and $\alpha^{\prime}$ is defined by $\alpha=\frac{\pi}{2}-\alpha^{\prime}$. For the sake of illustration and simplicity in calculations, we neglect light bending \cite{Bhattacharyya:2000kt}. We thus write $b\cos\alpha = r\sin\phi$ and the redshift factor $z$ can be rewritten as:
\begin{equation}
		1+z=\frac{1+\Omega r\sin\theta\sin\phi}{\sqrt{-g_{tt}-g_{\phi\phi}\Omega^{2}}}.\label{hongyi2}
\end{equation}
Fig.~\ref{vLv} presents the variations observed in the spectral energy distribution. Following the same pattern as seen with the energy flux and disk temperature, we observe that for negative $l$, the disk around a Schwarzschild-like BH is more luminous than one around a Schwarzschild BH in GR. 
\section{Image of the thin accretion disk around Schwarzschild-like BHs}
\label{section4}
\subsection{Observation coordinate system}
\begin{figure}[htbp]
	\centering
	\begin{subfigure}{0.42\textwidth}
		\includegraphics[width=3.5in, height=3.5in, keepaspectratio]{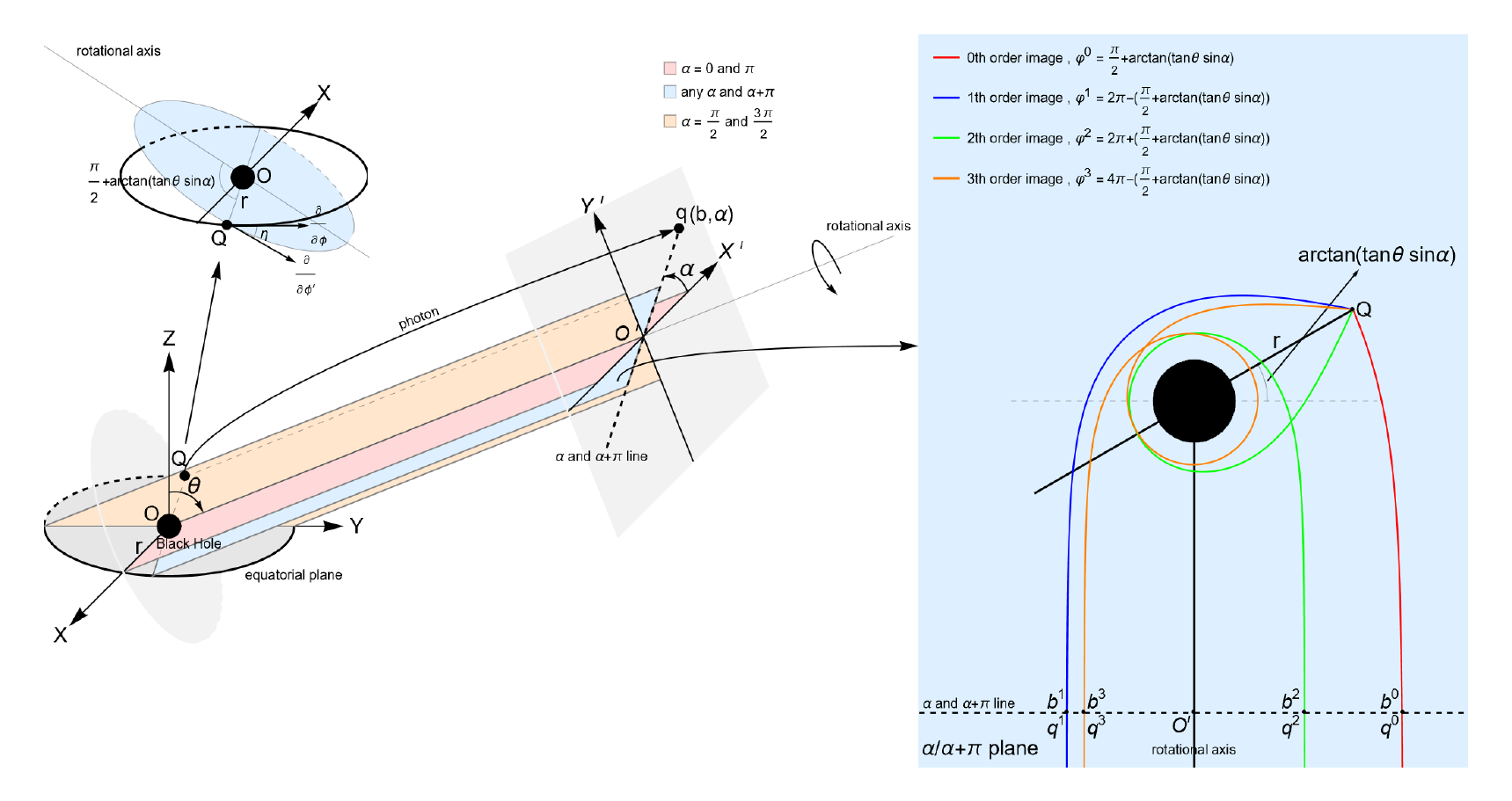}
	\end{subfigure}
	\caption{Coordinate system is indicated in Ref. \cite{You:2024uql}.}
	\label{coordinate}
\end{figure}

To analyze the imaging of a thin accretion disk, we employ an observational coordinate system as illustrated in Fig.~\ref{coordinate}. The observer is positioned at $(\infty,\theta,0)$ within the BH's spherical coordinate system $(r,\theta,\phi)$, with the origin at the BH's center $(r=0)$.
	
In this observer-based $O^{\prime}X^{\prime}Y^{\prime}$ coordinate system, consider a photon that originates from point $q(b,\alpha)$ and propagates perpendicularly to the $O^{\prime}X^{\prime}Y^{\prime}$ plane. Here, $b$ represents the photon's impact parameter. This photon intersects the accretion disk at point $Q(r,\frac{\pi}{2},\phi)$. By applying the principle of optical path reversibility, a photon emitted from $Q(r,\frac{\pi}{2},\phi)$ within the accretion disk will follow a trajectory that ultimately reaches the image point $q(b,\alpha)$ in the observer's field of view.
	
When the radial distance $r$ is held constant, the resulting image corresponds to an orbit of constant radius. As depicted on the left side of Fig.~\ref{coordinate}, every $\alpha/\alpha+\pi$ plane intersects the constant-$r$ orbit within the equatorial plane at two distinct points, with their azimuthal angles $\phi$ differing by $\pi$. For our coordinate system, the $X^{\prime}$-axis is defined by setting $\alpha=0$, while the $X$-axis is aligned with $\phi=0$. Geometrically, this setup allows us to determine the angle $\varphi$ formed between the rotational axis and the line segment $OQ$
\begin{equation}
		\varphi=\frac{\pi}{2}+\arctan(\tan\theta\sin\alpha).\label{faijiao}
\end{equation}
	
As the impact parameter  $b$ approaches $b_{c}$, the degree of light bending increases, potentially causing a single source point $Q$ to produce multiple image points $q$. These image points are labeled according to their increasing azimuthal angles $\varphi$ as $q^{n}$ $(n\in\mathbb{N})$, where $n$ denotes the order of the image.
	
As illustrated in Fig.~\ref{coordinate} (right side), all even-order images of $Q$ appear on the same side $(\alpha)$ as the source point $Q$, while all odd-order images are located on the opposite side $(\alpha+\pi)$. The angular deflections responsible for generating the $n^{th}$-order image are denoted by $\varphi^{n}$
\begin{equation}
		\varphi^{n}= 
		\begin{cases} 
			\frac{n}{2}2\pi+(-1)^{n}[\frac{\pi}{2}+\arctan(\tan\theta\sin\alpha)], & \text{when $n$ is even,} \\
			\frac{n+1}{2}2\pi+(-1)^{n}[\frac{\pi}{2}+\arctan(\tan\theta\sin\alpha)], & \text{when $n$ is odd,}
		\end{cases}
	\label{faijiaon}
\end{equation}
the parameter $n$ denotes the order of the image observed. Specifically, when $n=0$, it corresponds to the primary, or direct, visualization of the accretion disk as seen by an observer. Higher values of $n$, such as $1$, $2$, $3$, and so forth, indicate successive orders of images, capturing secondary, tertiary, and further iterative representations, respectively.
\subsection{Direct and secondary images of BH}
\begin{figure*}[htbp]
	\centering
	\begin{subfigure}{0.38\textwidth}
		\includegraphics[width=\linewidth]{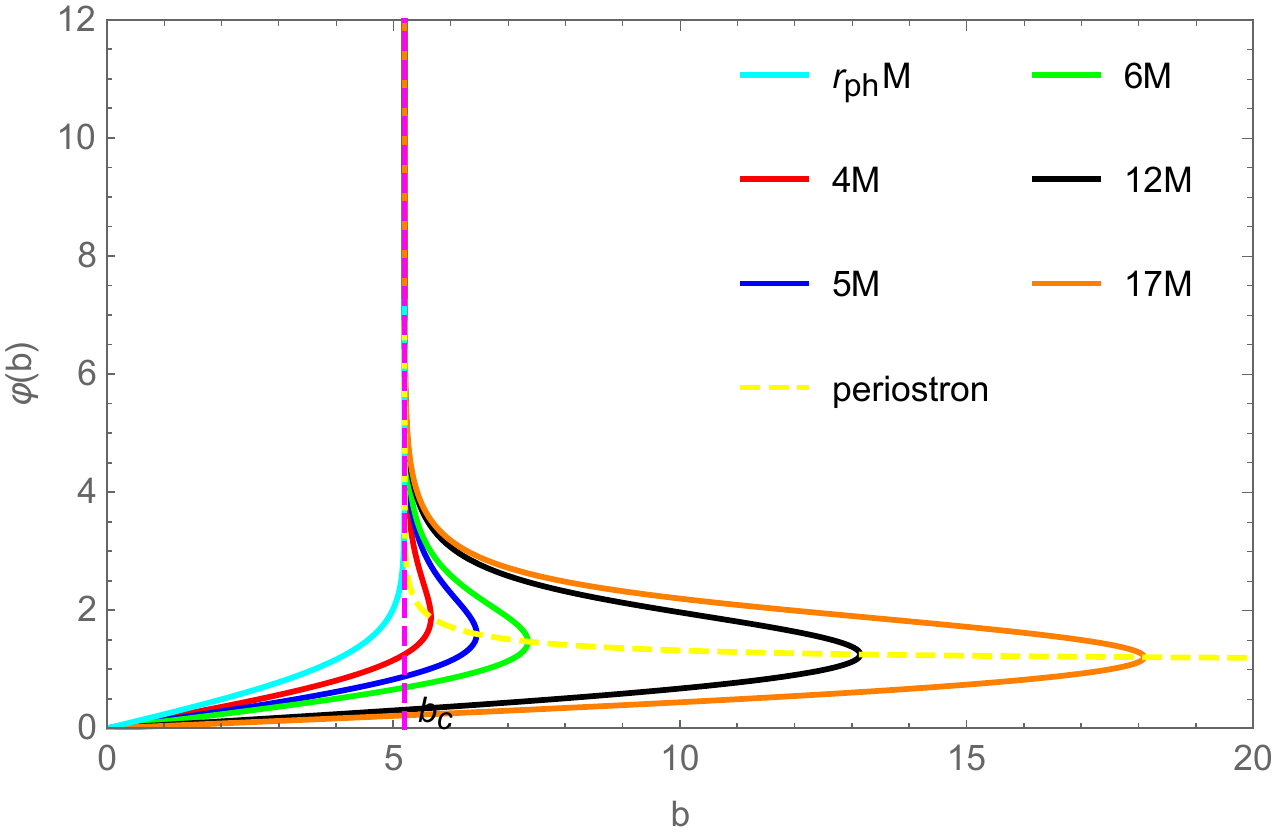}
	\end{subfigure}%
	\hspace{0.13\textwidth}
	\begin{subfigure}{0.38\textwidth}
		\includegraphics[width=\linewidth]{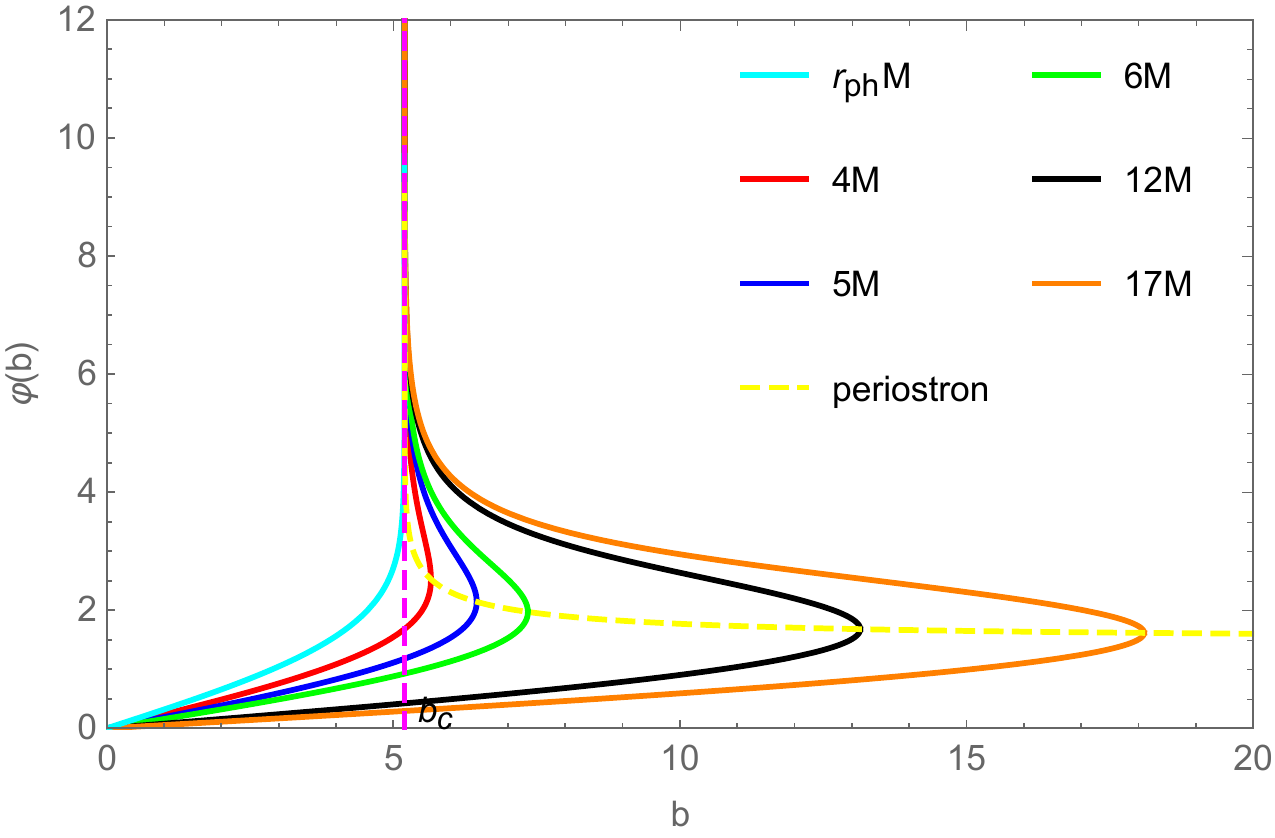}
	\end{subfigure}
	
	\caption{Deflection angle $\varphi(b)$ corresponding to intersections as a function of $b$ for different $r$. Left panel: $l=-0.5$. Right panel: $l=-0.1$.}
	\label{Faib}
\end{figure*}

Photons arriving from infinity with varying impact parameters $b$ intersect the equal-$r$ orbit at different points. Fig.~\ref{Faib} depicts the function $\varphi(b)$. As can be seen from the figure, as $l$ increases, the graph of $\varphi(b)$ shifts upward. The yellow dashed line in the figure is designated as $\varphi_{1}(b)$. $\varphi_{1}(b)$ represents the deflection angle of a photon with impact parameter $b$ when it reaches the perihelion of its trajectory. When a photon is incident from infinity, it deflects under the gravitational pull of the black hole. The cumulative deflection angle at the moment when the photon arrives at the closest position to the black hole (i.e., the perihelion $r_{pe}$) is the physical quantity described by $\varphi_{1}(b)$. Taking this line as a dividing boundary, the colored curves lying below it are denoted as $\varphi_{2}(b)$, while those above it are labeled as $\varphi_{3}(b)$.
\begin{figure*}[htbp]
	\centering
	
	\begin{subfigure}{0.28\textwidth}
		\includegraphics[width=\linewidth]{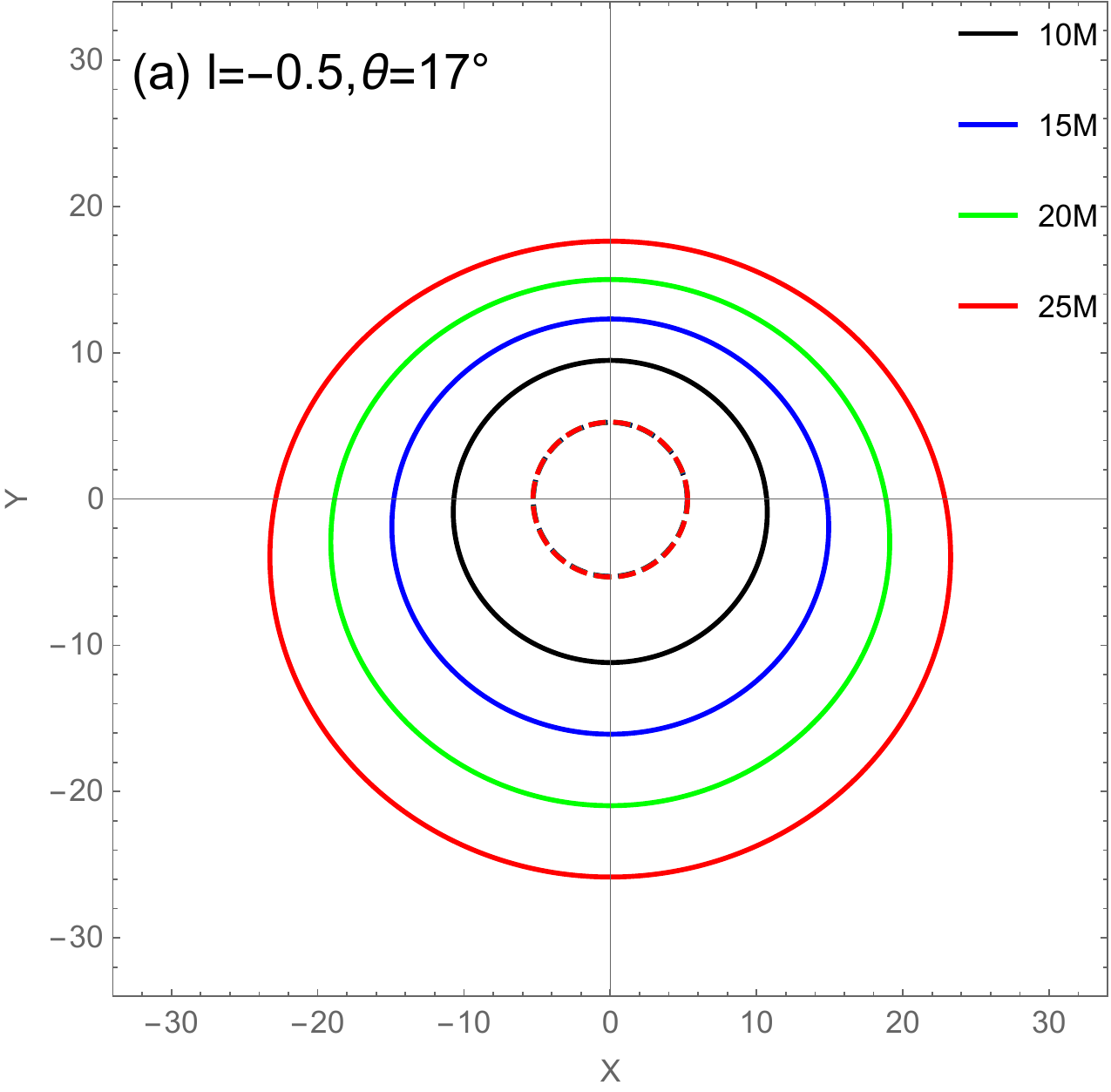}
	\end{subfigure}
	\hspace{0.02\textwidth}
	\begin{subfigure}{0.28\textwidth}
		\includegraphics[width=\linewidth]{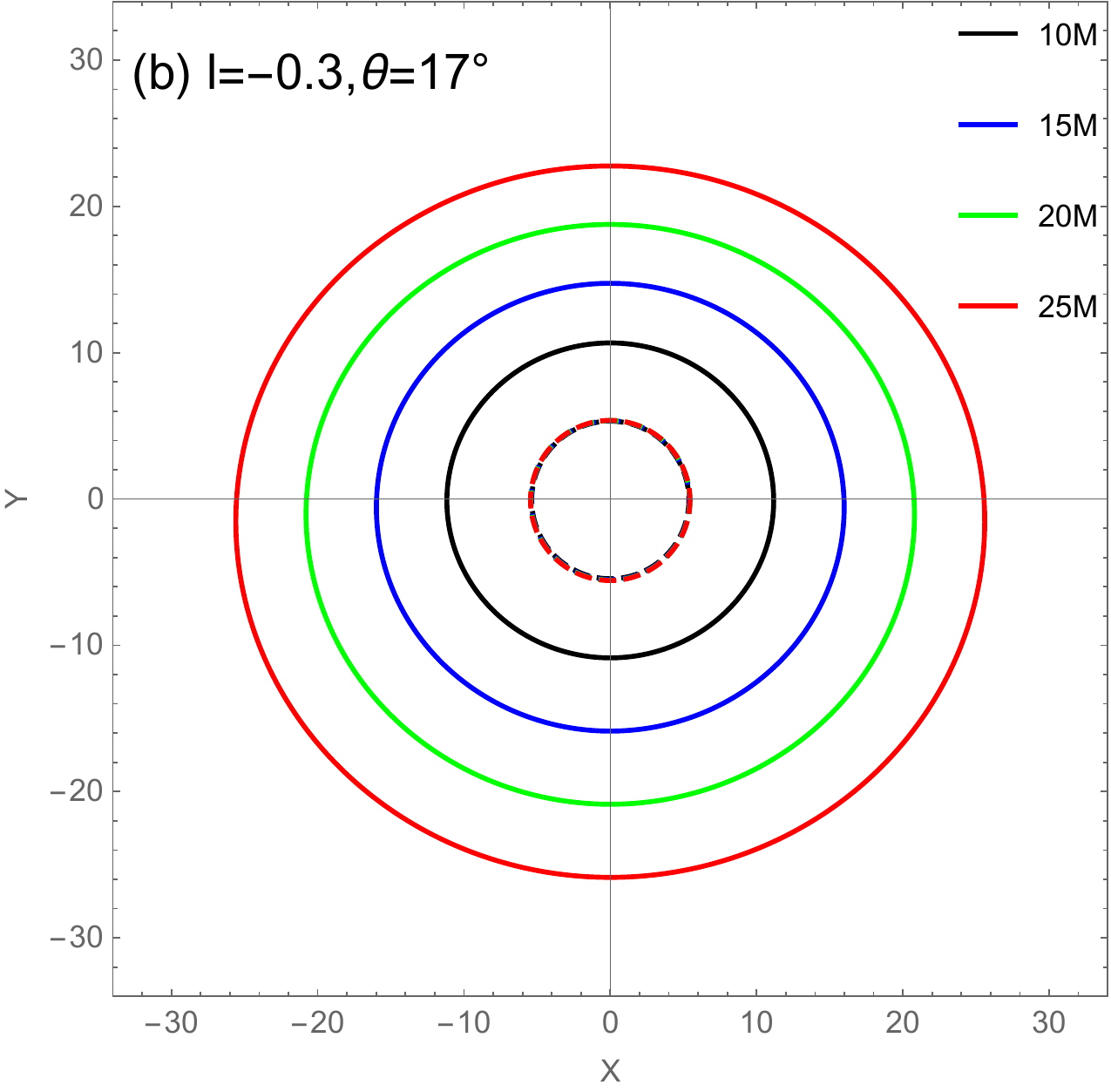}
	\end{subfigure}
	\hspace{0.02\textwidth}
	\begin{subfigure}{0.28\textwidth}
		\includegraphics[width=\linewidth]{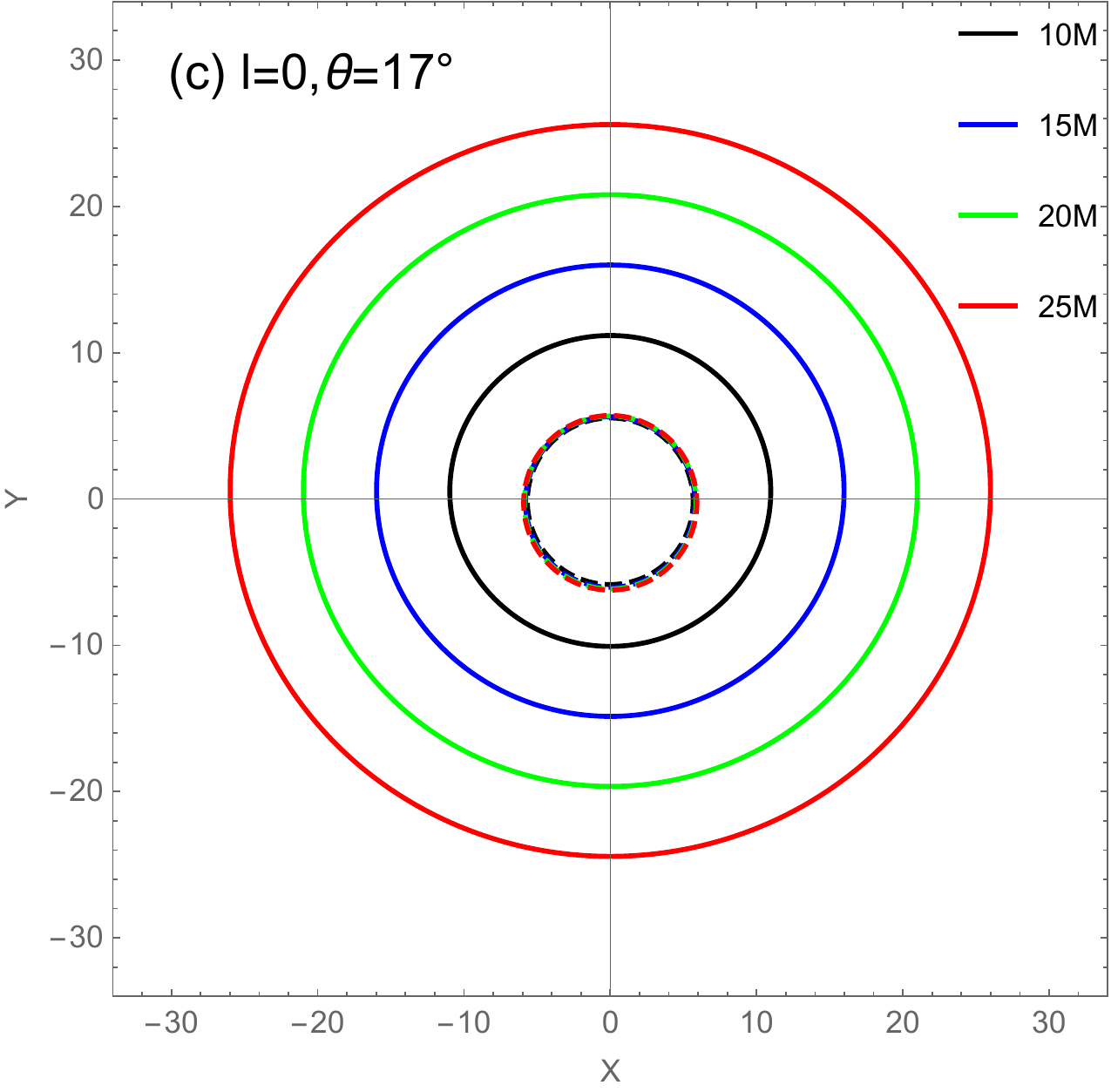}
	\end{subfigure}
	
	\vspace{0.3cm}
	
	\begin{subfigure}{0.28\textwidth}
		\includegraphics[width=\linewidth]{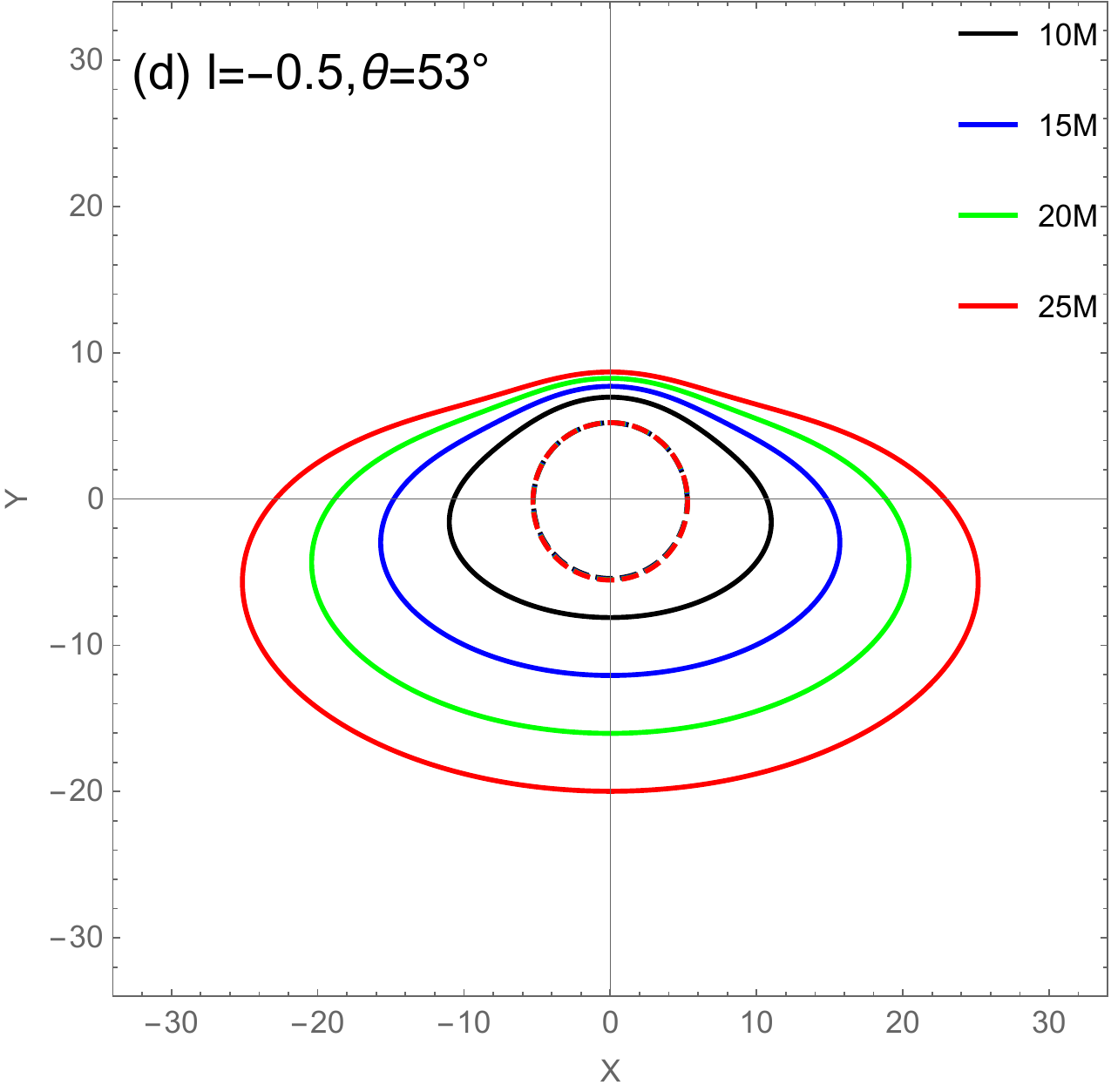}
	\end{subfigure}
	\hspace{0.02\textwidth}
	\begin{subfigure}{0.28\textwidth}
		\includegraphics[width=\linewidth]{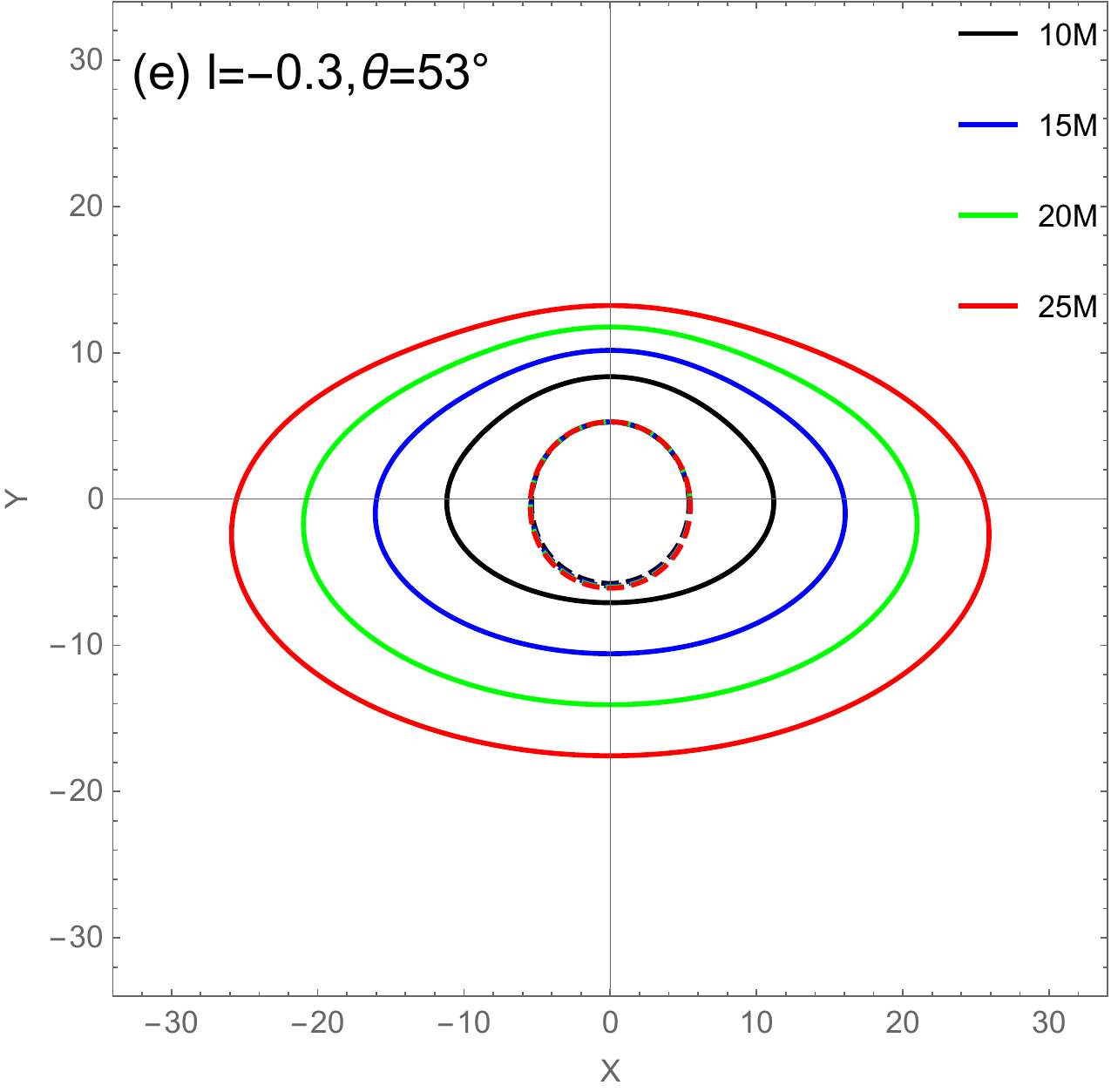}
	\end{subfigure}
	\hspace{0.02\textwidth}
	\begin{subfigure}{0.28\textwidth}
		\includegraphics[width=\linewidth]{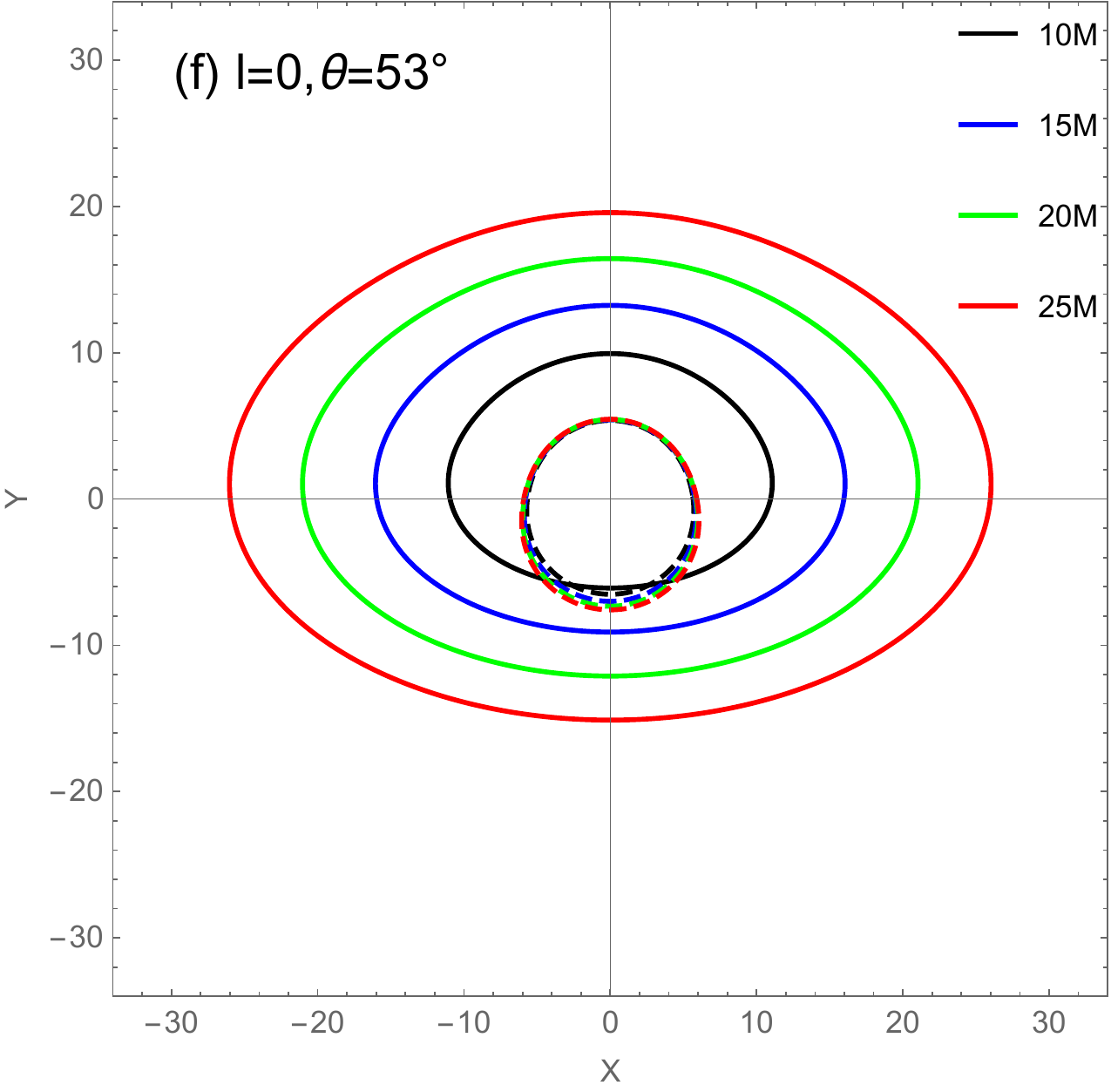}
	\end{subfigure}
	
	\vspace{0.3cm}
	
	\begin{subfigure}{0.28\textwidth}
		\includegraphics[width=\linewidth]{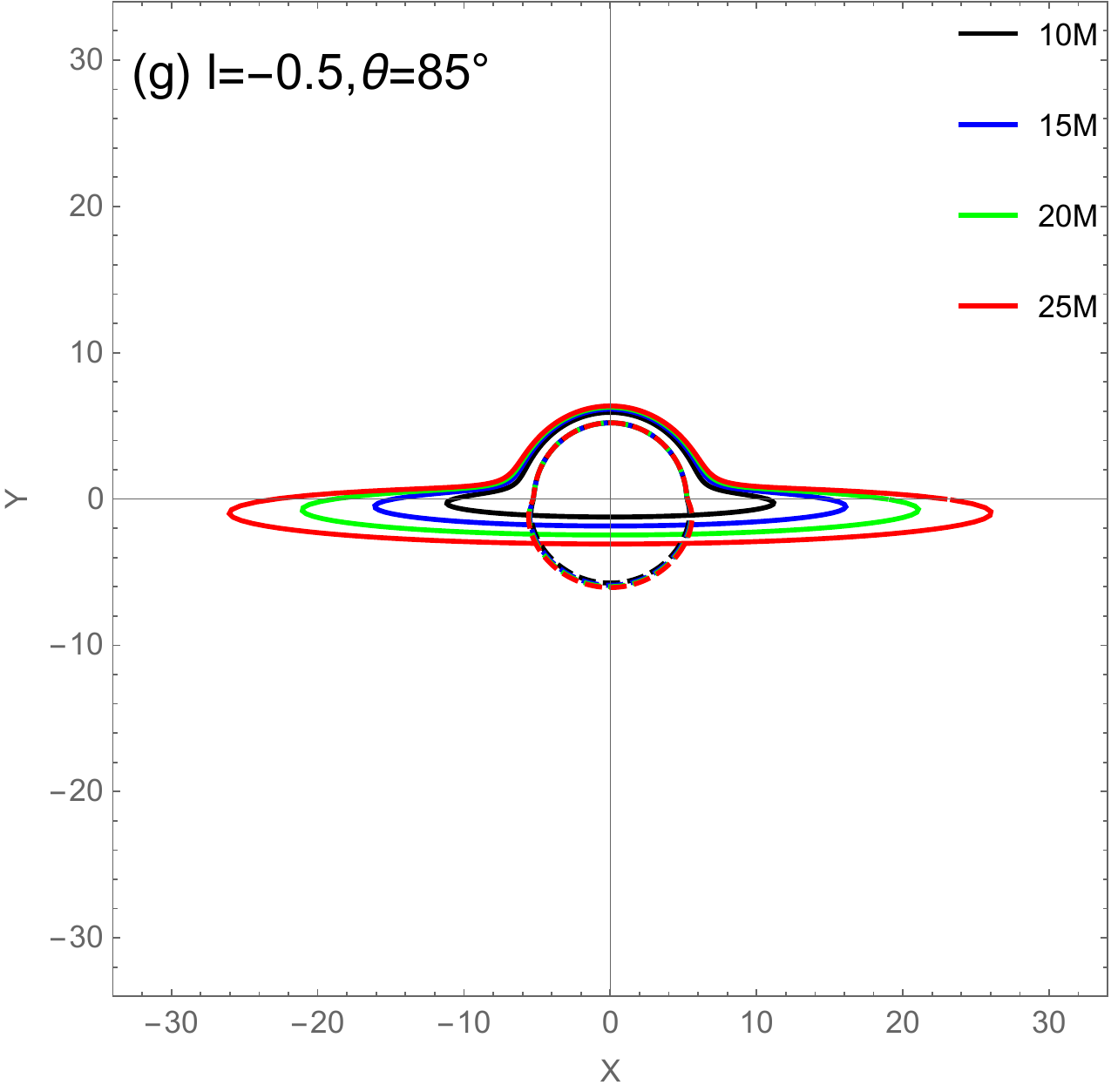}
	\end{subfigure}
	\hspace{0.02\textwidth}
	\begin{subfigure}{0.28\textwidth}
		\includegraphics[width=\linewidth]{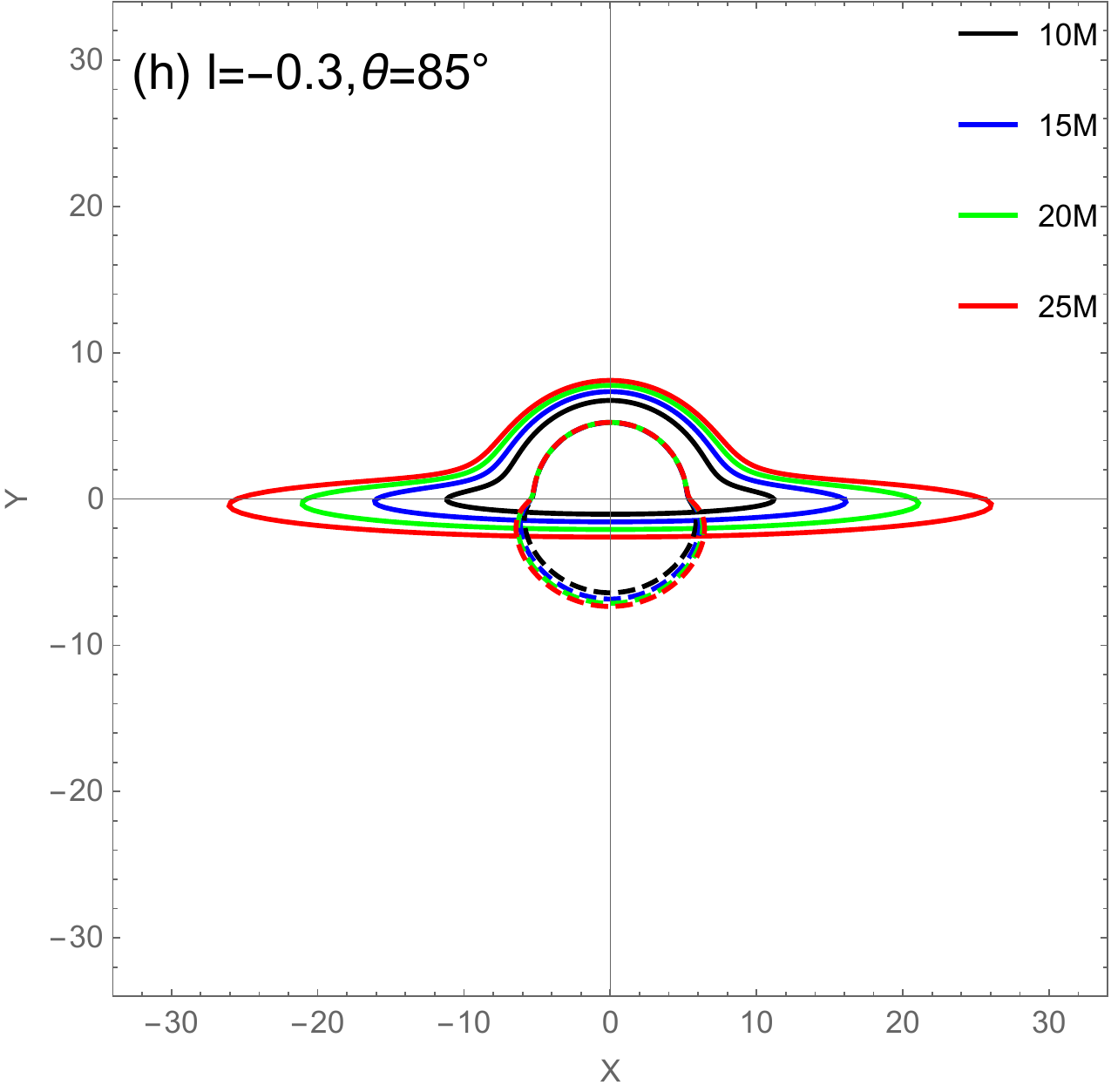}
	\end{subfigure}
	\hspace{0.02\textwidth}
	\begin{subfigure}{0.28\textwidth}
		\includegraphics[width=\linewidth]{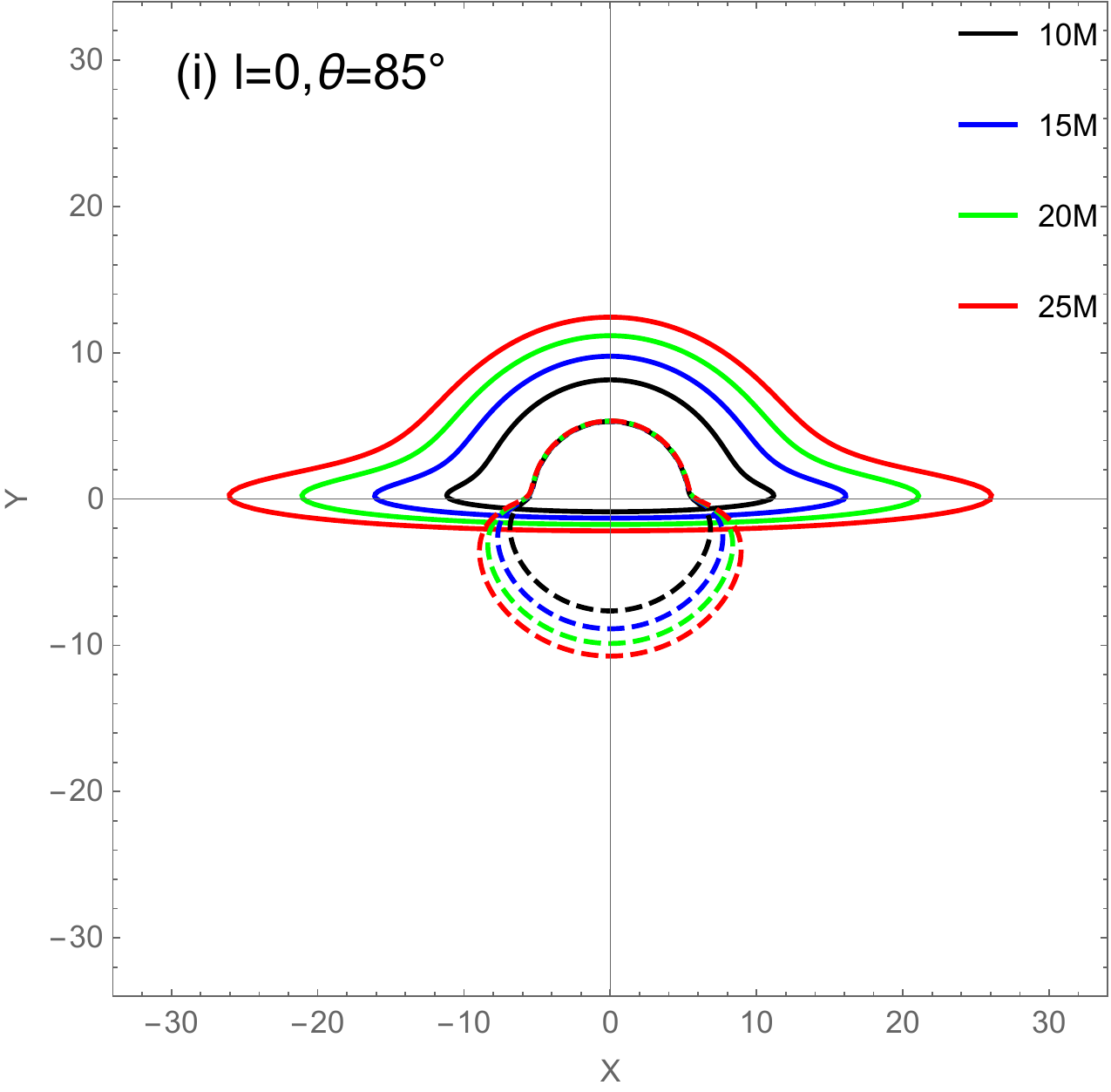}
	\end{subfigure}
	
	\caption{Direct and secondary image of the thin accretion disk.}
	\label{diskimage}
\end{figure*}
Therefore, we can define:
\begin{equation}
		\varphi_{1}(b)=\int_{0}^{u_{min}}\frac{1}{\sqrt{G(u)}}du,\label{faione}
\end{equation}
\begin{equation}
		\varphi_{2}(b)=2\int_{0}^{u_{r}}\frac{1}{\sqrt{G(u)}}du,\label{faitwo}
\end{equation}
\begin{equation}
		\varphi_{3}(b)=2\int_{0}^{u_{min}}\frac{1}{\sqrt{G(u)}}du-\int_{0}^{u_{r}}\frac{1}{\sqrt{G(u)}}du.\label{faithree}
\end{equation}
Each colored curve in the figures denotes an equal-$r$ orbit, with points $(b,\varphi)$ indicating the deflection angle $\varphi$ for photons arriving at the equal-$r$ orbit with impact parameter $b$. The purple dashed line intersects these curves at their peak points, representing the deflection angle when photons reach their closest approach perihelion $r_{pe}$.
	
By solving the system of Eqs. (\ref{faijiao}), (\ref{faione}), and (\ref{faitwo}) simultaneously and employing numerical integration methods to find all $(b,\alpha)$ pairs, one can obtain the projection of the accretion disk in the observer's plane.

Fig.~\ref{diskimage} displays the direct and secondary images of representative stable circular orbits around Schwarzschild-like BHs, observed by a remote observer at various inclination angles. Each column, from top to bottom, corresponds to inclination angles of $17^{\circ}$, $53^{\circ}$, and $85^{\circ}$, while each row, from left to right, represents $l$ values of $-0.5$, $-0.3$, and $0$, respectively. These images correspond to stable circular orbits with radii of $r=10$, $15$, $20$, $25$, moving from the innermost to the outermost. The rightmost column corresponds to the Schwarzschild BH. As $l$ increases, the cap shape contracts inward, whereas with decreasing $l$, the cap shape stretches outward.
	
\subsection{Observed flux and Redshift factor}
\begin{figure*}[htbp]
		\centering
		\begin{tabular}{ccc}
			\begin{minipage}[t]{0.28\textwidth}
				\centering
				\begin{overpic}[width=0.75\textwidth]{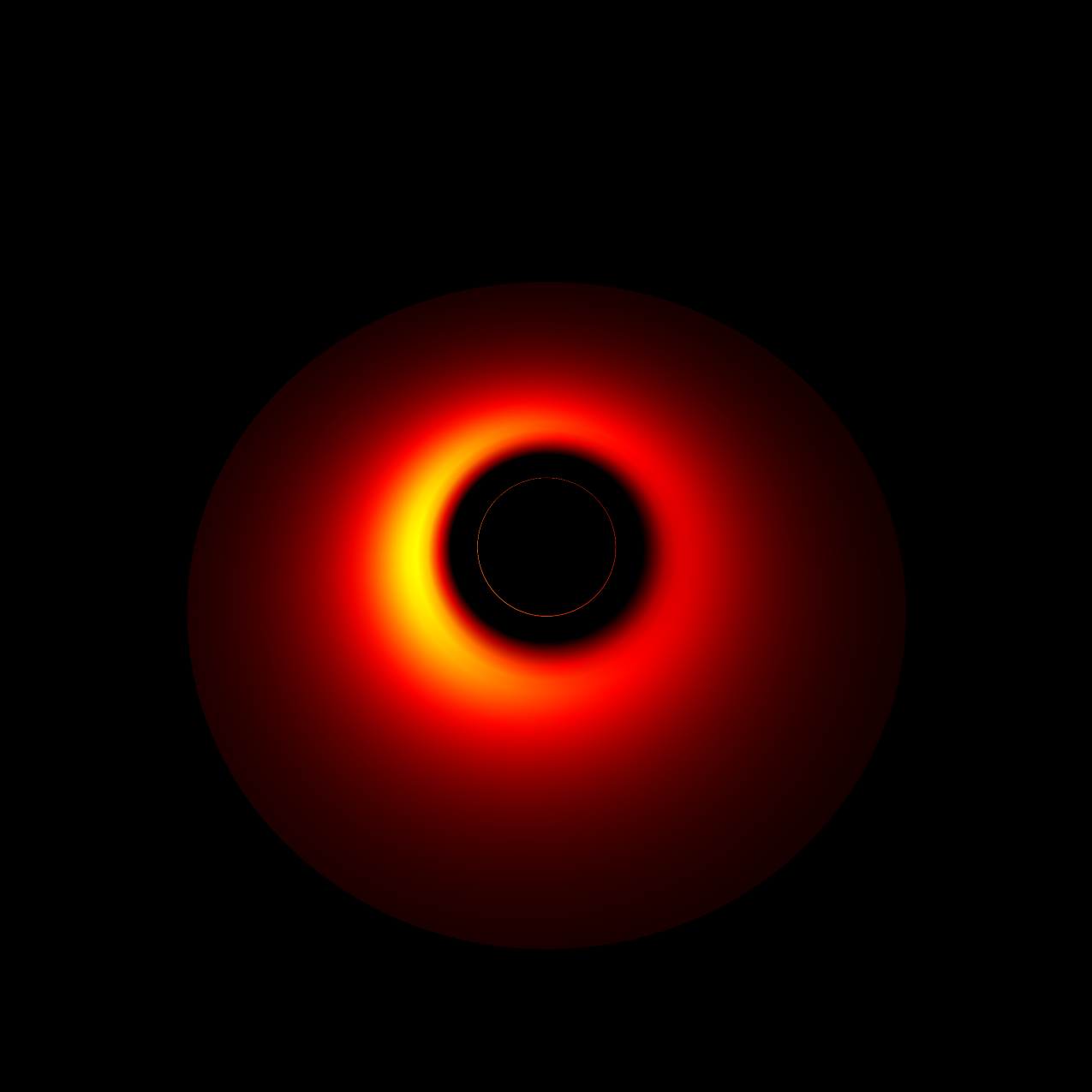} 
					\put(17,103){\color{black}\large $l=-0.5,\theta=17^{\circ}$} 
					\put(-10,48){\color{black} Y}
					\put(48,-10){\color{black} X}
				\end{overpic}
			\end{minipage}
			&
			\begin{minipage}[t]{0.28\textwidth}
				\centering
				\begin{overpic}[width=0.75\textwidth]{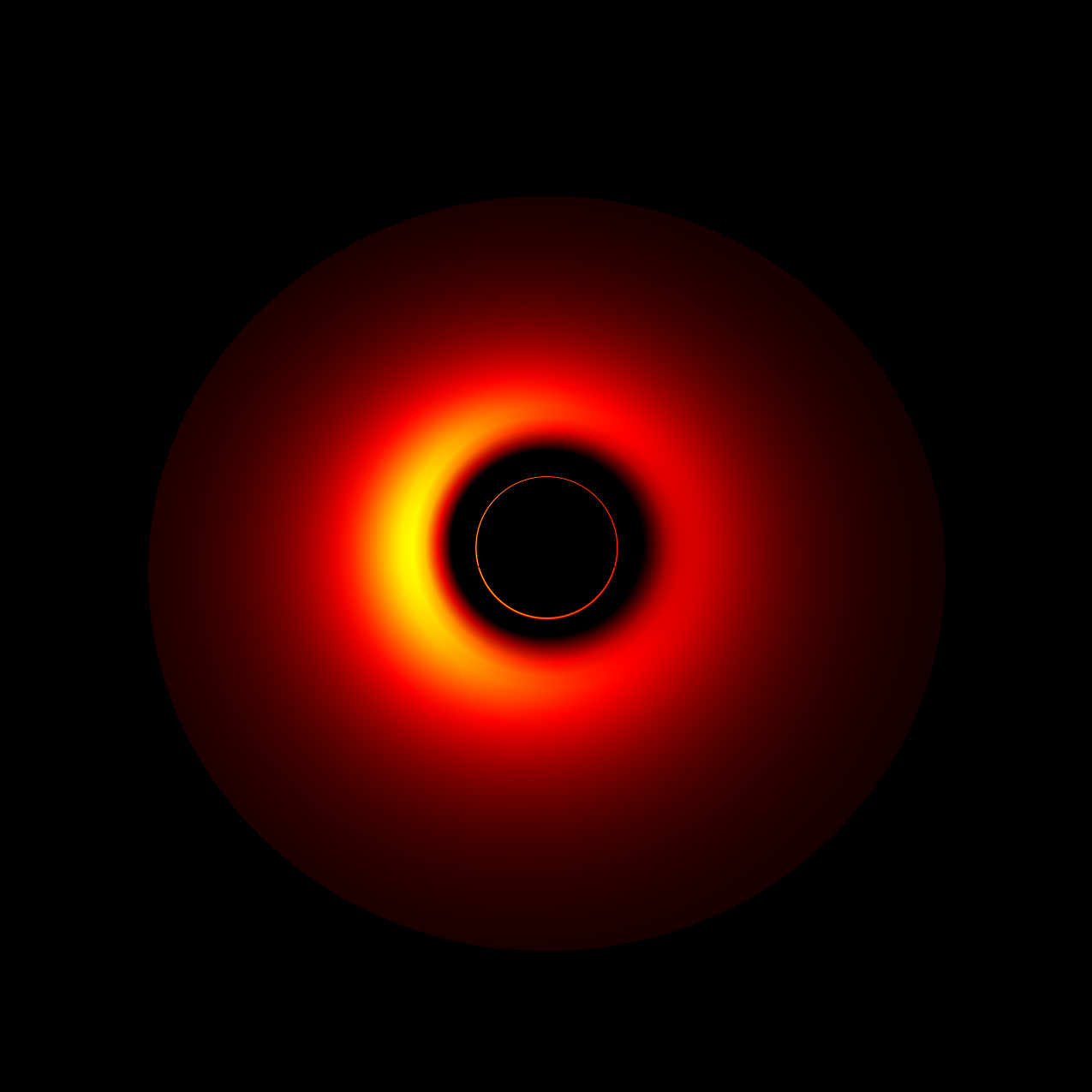} 
					\put(17,103){\color{black}\large $l=-0.3,\theta=17^{\circ}$} 
					\put(-10,48){\color{black} Y}
					\put(48,-10){\color{black} X}
				\end{overpic}
			\end{minipage}
			&
			\begin{minipage}[t]{0.28\textwidth}
				\centering
				\begin{overpic}[width=0.75\textwidth]{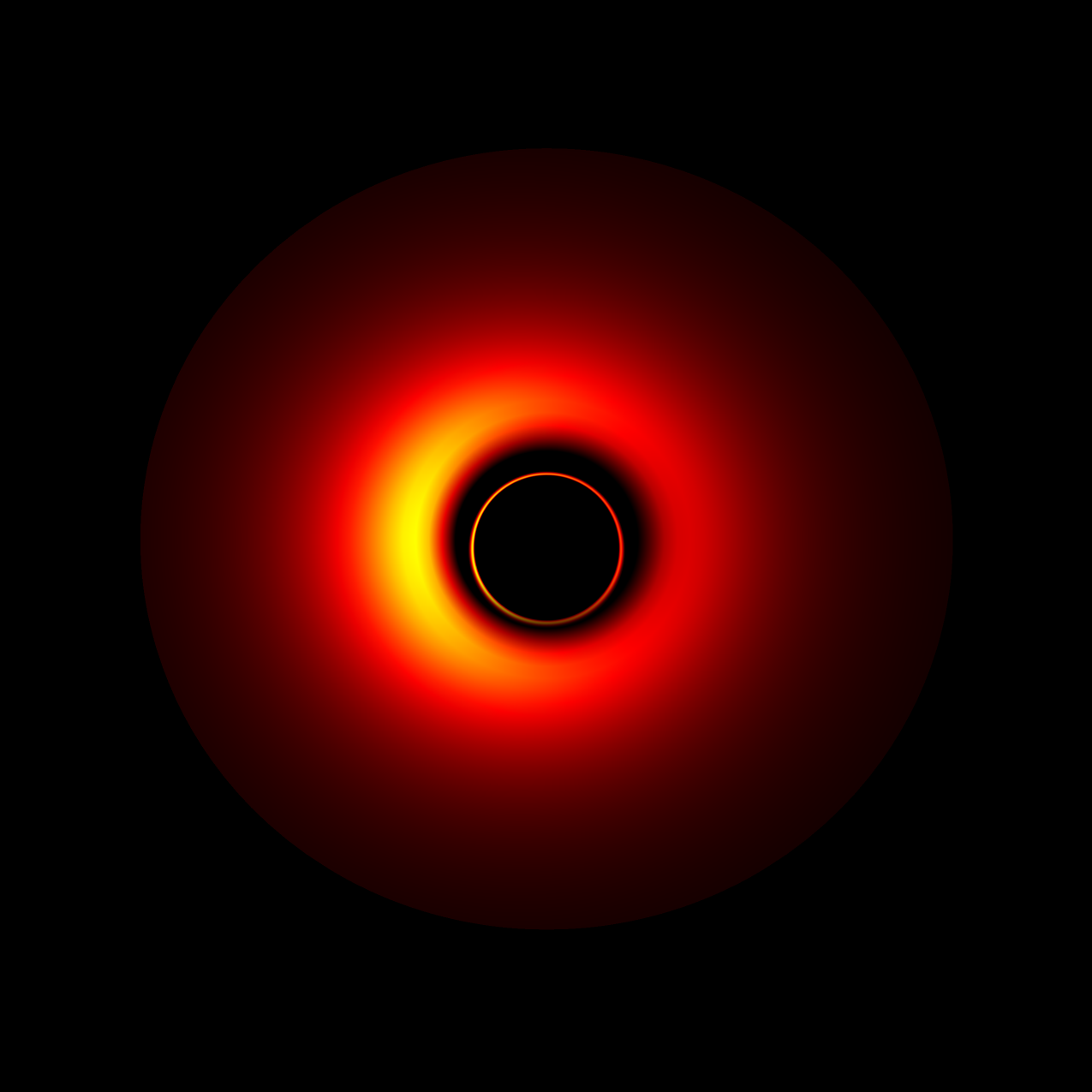}
					\put(24,103){\color{black}\large $l=0,\theta=17^{\circ}$} 
					\put(-10,48){\color{black} Y}
					\put(48,-10){\color{black} X}
				\end{overpic}
			\end{minipage}
			\vspace{30pt} 
			\\ 
			\begin{minipage}[t]{0.28\textwidth}
				\centering
				\begin{overpic}[width=0.75\textwidth]{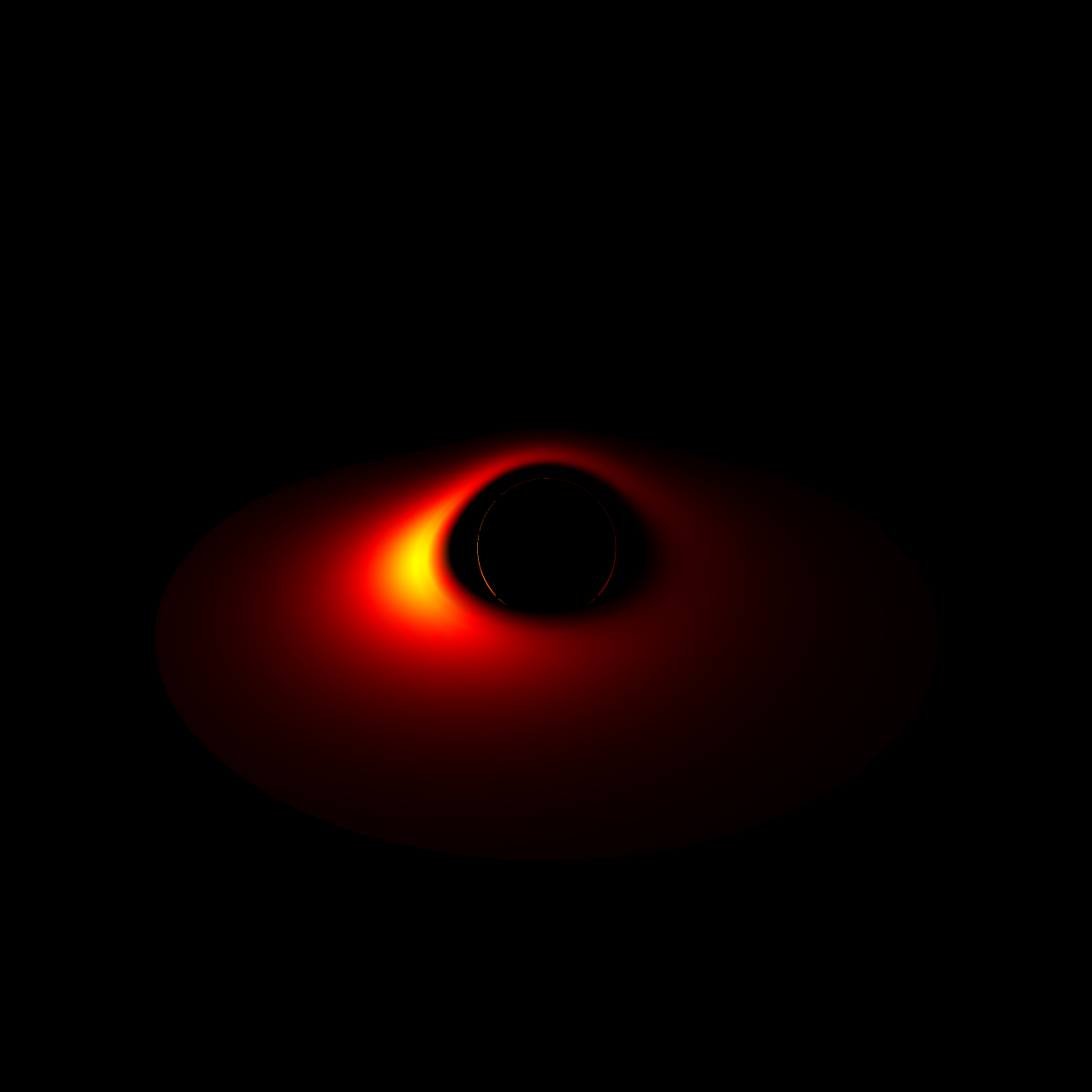} 
					\put(17,103){\color{black}\large $l=-0.5,\theta=53^{\circ}$} 
					\put(-10,48){\color{black} Y}
					\put(48,-10){\color{black} X}
				\end{overpic}
			\end{minipage}
			&
			\begin{minipage}[t]{0.28\textwidth}
				\centering
				\begin{overpic}[width=0.75\textwidth]{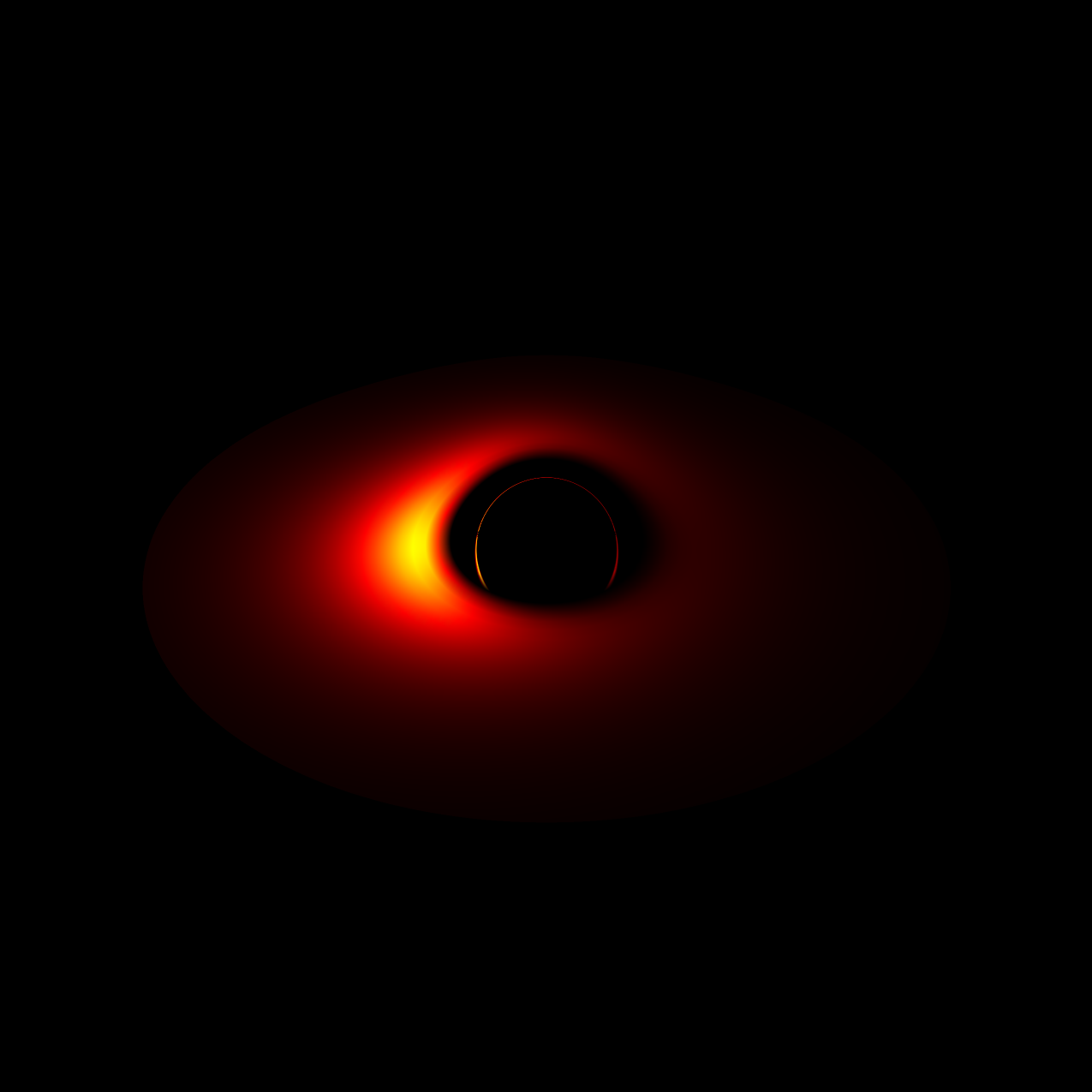} 
					\put(17,103){\color{black}\large $l=-0.3,\theta=53^{\circ}$} 
					\put(-10,48){\color{black} Y}
					\put(48,-10){\color{black} X}
				\end{overpic}
			\end{minipage}
			&
			\begin{minipage}[t]{0.28\textwidth}
				\centering
				\begin{overpic}[width=0.75\textwidth]{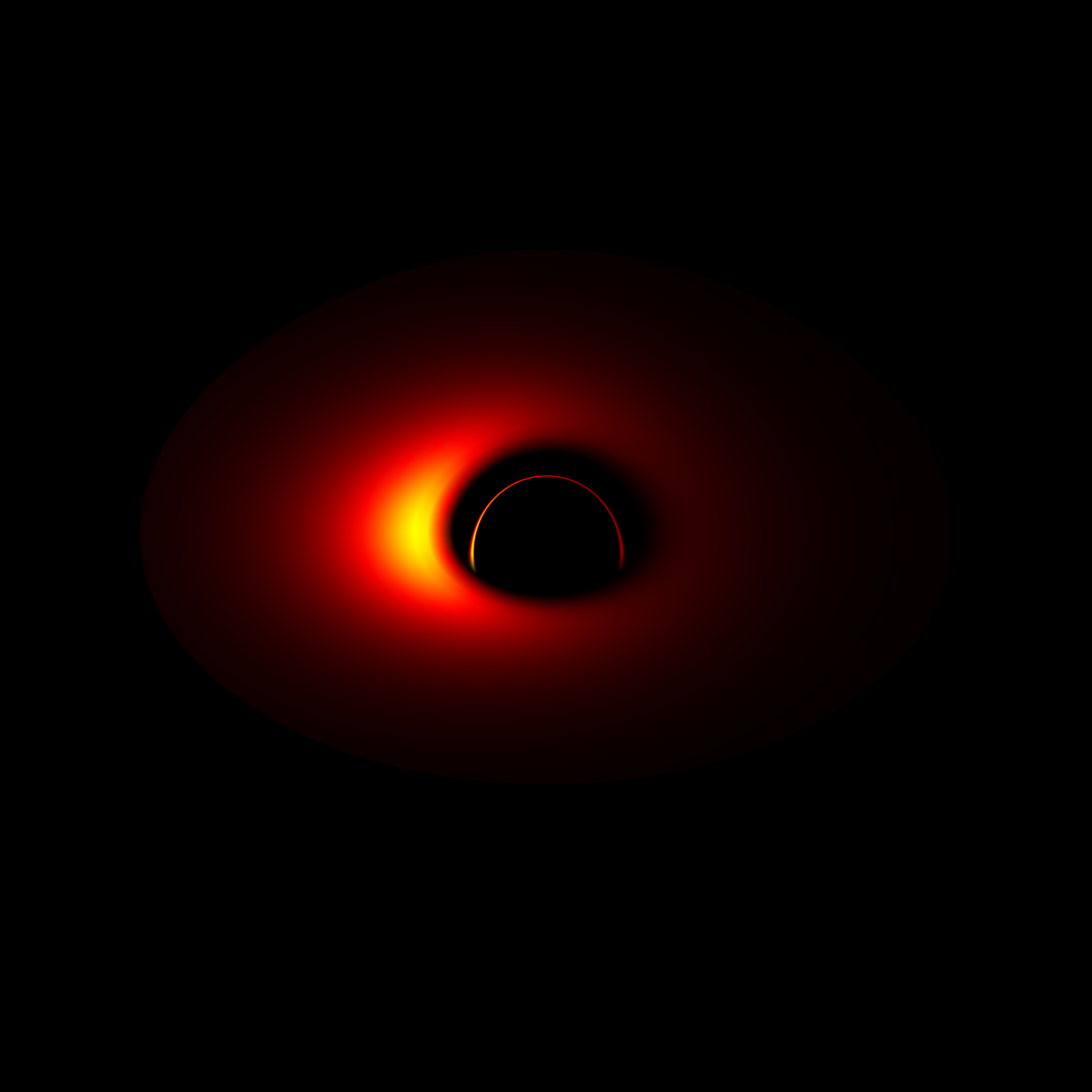} 
					\put(24,103){\color{black}\large $l=0,\theta=53^{\circ}$} 
					\put(-10,48){\color{black} Y}
					\put(48,-10){\color{black} X}
				\end{overpic}
			\end{minipage}
			\vspace{30pt} 
			\\
			\begin{minipage}[t]{0.28\textwidth}
				\centering
				\begin{overpic}[width=0.75\textwidth]{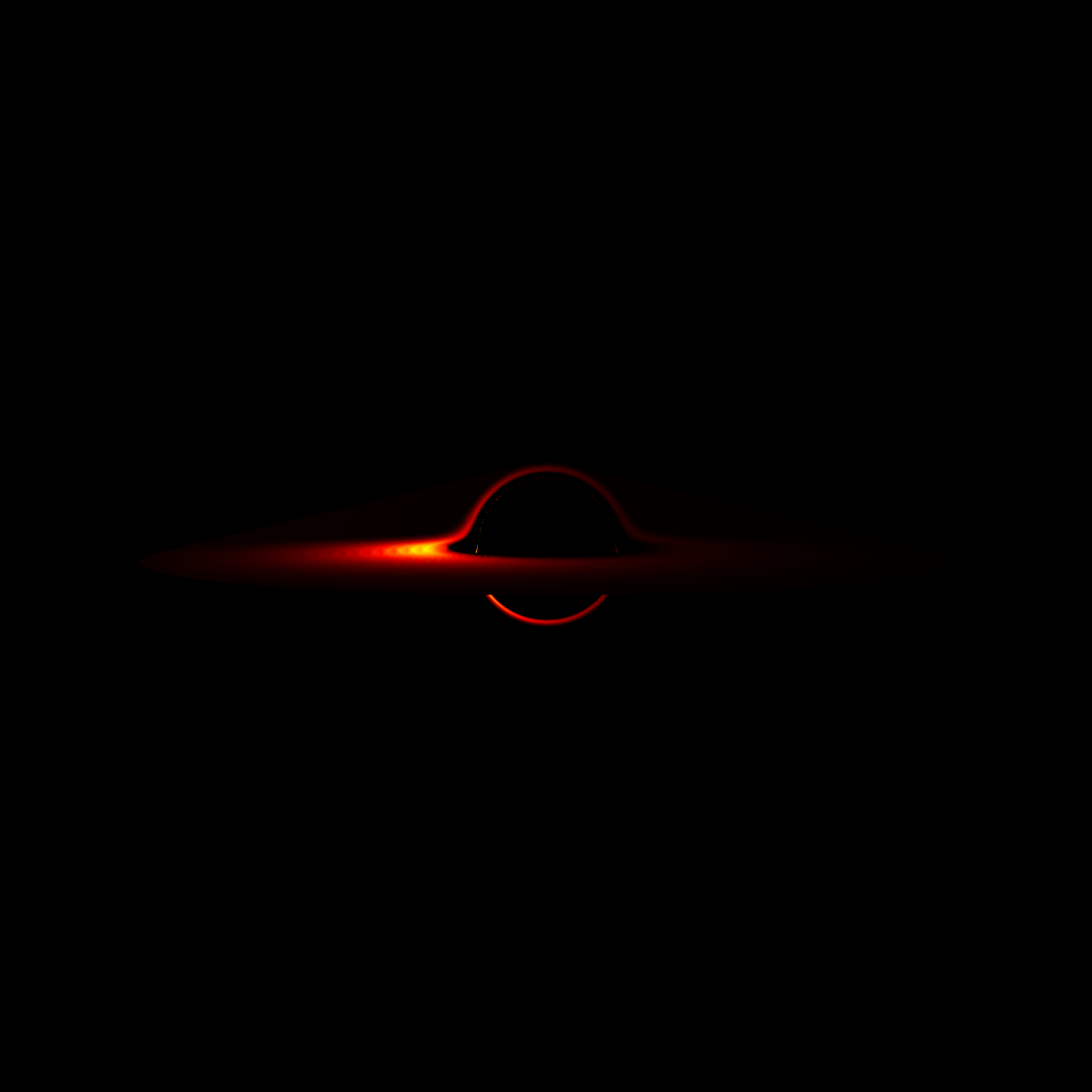}
					\put(17,103){\color{black}\large $l=-0.5,\theta=85^{\circ}$}
					\put(-10,48){\color{black} Y}
					\put(48,-10){\color{black} X}
				\end{overpic}
			\end{minipage}
			&
			\begin{minipage}[t]{0.28\textwidth}
				\centering
				\begin{overpic}[width=0.75\textwidth]{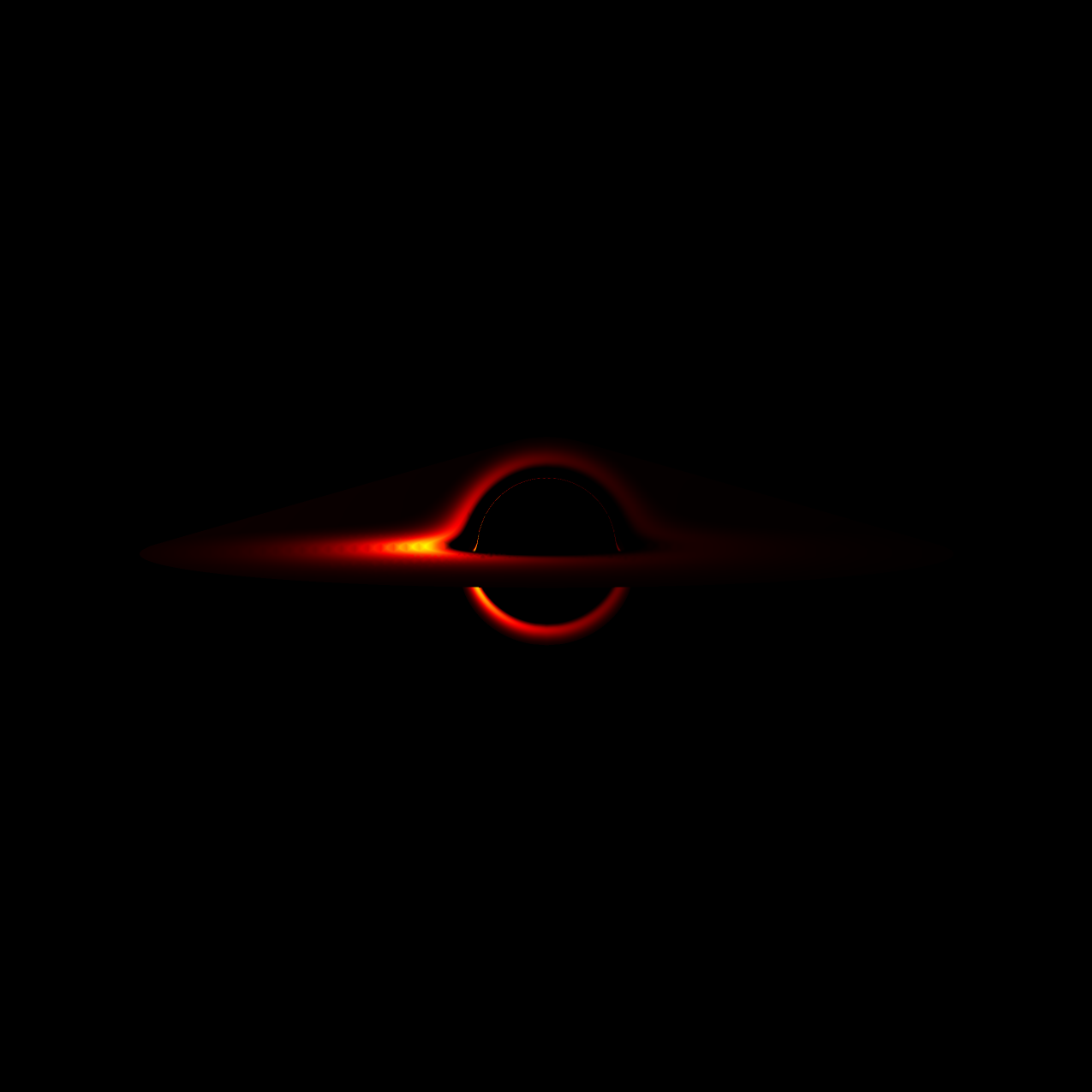}
					\put(17,103){\color{black}\large $l=-0.3,\theta=85^{\circ}$} 
					\put(-10,48){\color{black} Y}
					\put(48,-10){\color{black} X}
				\end{overpic}
			\end{minipage}
			&
			\begin{minipage}[t]{0.28\textwidth}
				\centering
				\begin{overpic}[width=0.75\textwidth]{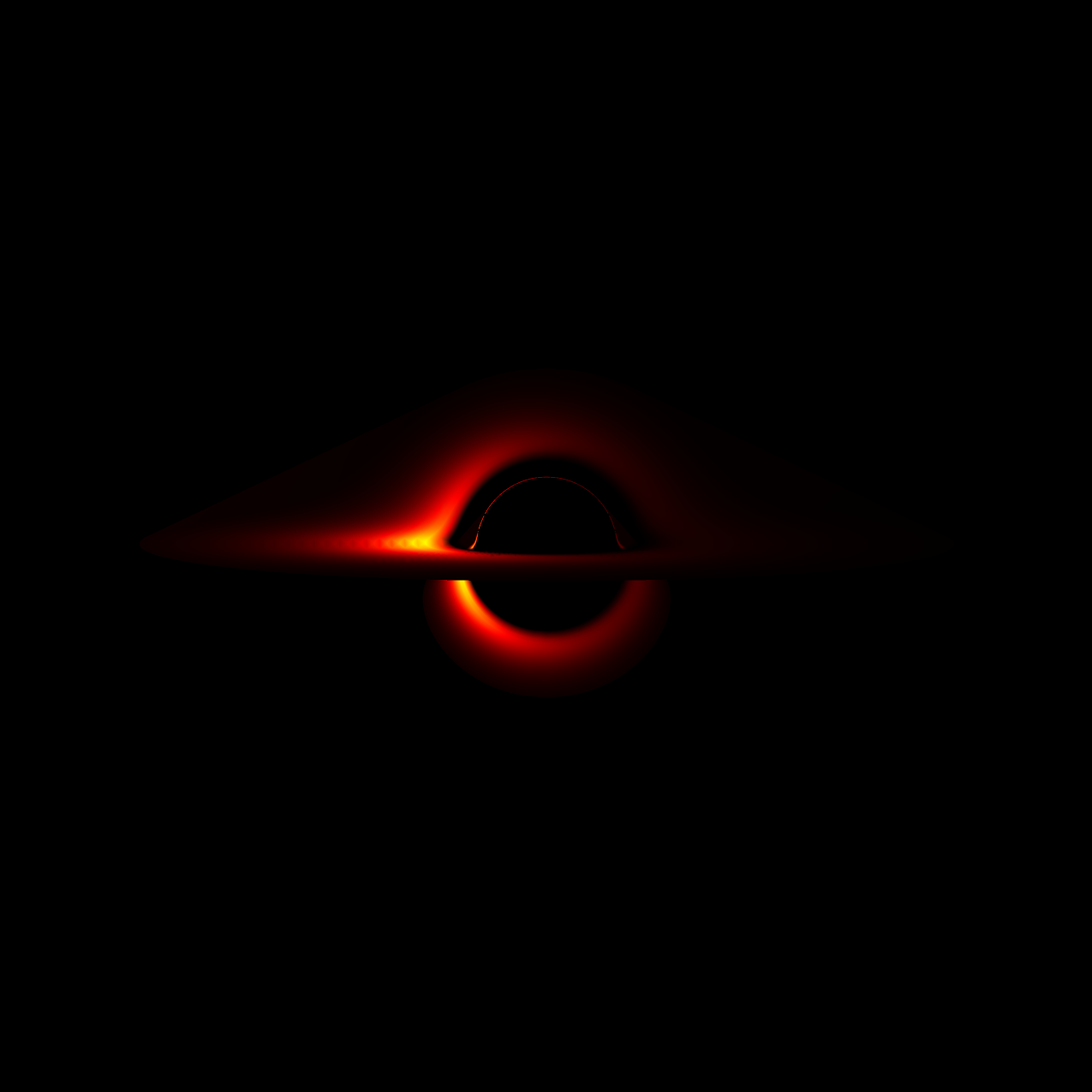}
					\put(24,103){\color{black}\large $l=0,\theta=85^{\circ}$}
					\put(-10,48){\color{black} Y}
					\put(48,-10){\color{black} X}
				\end{overpic}
			\end{minipage}
		\end{tabular}
		\caption{The complete apparent images of the thin accretion disk for different values of parameter $l$ and inclination angle.}
		\label{actualdiskimage}
\end{figure*}

To obtain the observable flux $F_{obs}$ at a specified point on the celestial sphere, the gravitational redshift $z$ must be taken into account. Therefore, we arrive at the relation:
\begin{equation}
		F_{obs}=\frac{F(r)}{(1+z)^{4}}.\label{Fobs}
\end{equation}
From Eqs. (\ref{fr}), (\ref{hongyi}), and (\ref{Fobs}), it follows that,
\begin{equation}
		F_{obs}=\frac{-\frac{\dot{M}\Omega_{,r}}{4\pi\sqrt{-g}(E-\Omega L)^{2}}\int^{r}_{r_{isco}}(E-\Omega L)L_{,r}dr}{(\frac{1+\Omega b\sin\theta\cos\alpha}{\sqrt{-g_{tt}-g_{\phi\phi}\Omega^{2}}})^{4}}.\label{Fobs2}
\end{equation}
According to the analysis above, we plotted the observed flux distribution of the accretion disk as shown in Fig.~\ref{actualdiskimage}. Each column, from top to bottom, corresponds to inclination angles of $17^{\circ}$, $53^{\circ}$, and $85^{\circ}$, while each row, from left to right, represents $l$ values of $-0.5$, $-0.3$, and $0$, respectively. The rightmost column corresponds to the Schwarzschild BH. As the observer's inclination angle increases, the flux distributions become strongly asymmetric, and the brightness distribution patterns are similar. Notably, the reduction in the thickness of the strongly lensed incoming flux with decreasing $l$ is an observationally relevant feature; however, in more complex accretion disk models (e.g., incorporating inhomogeneous mass accretion rate), this feature is degenerate with other astrophysical parameters—such as the mass accretion rate $\dot{M}$. This degeneracy implies that a direct interpretation of observed flux thickness variations as a sole indicator of $l$ may lead to ambiguities, and future observational analyses should account for the combined effects of $l$ and these astrophysical parameters.
	
\begin{figure*}[htbp]
		\centering
		\begin{tabular}{ccc}
			\begin{minipage}[t]{0.28\textwidth}
				\centering
				\begin{overpic}[width=0.75\textwidth]{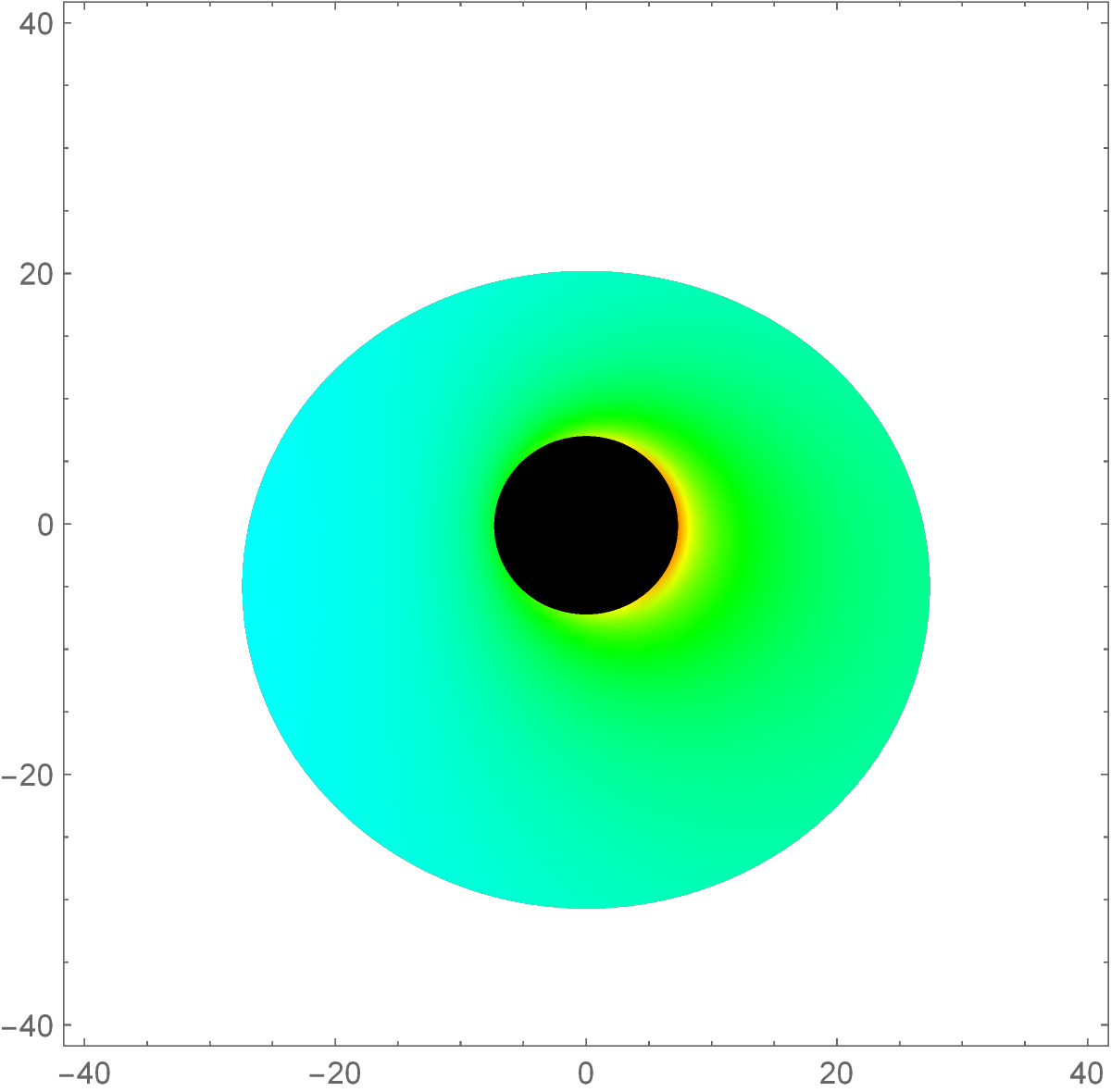}
					\put(20,100){\color{black}\large $l=-0.5,\theta=17^{\circ}$}
					\put(-8,48){\color{black} Y}
					\put(48,-10){\color{black} X}
				\end{overpic}
				\raisebox{0.05\height}{ 
					\begin{overpic}[width=0.07\textwidth]{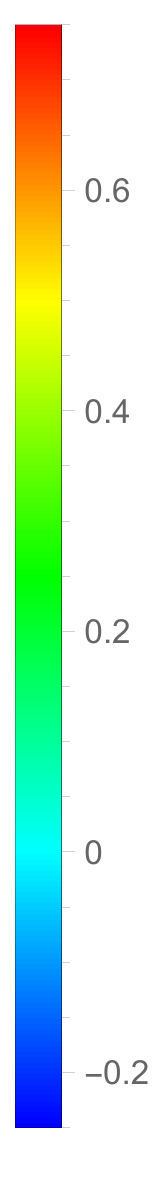}
						\put(0,103){\color{black}\large $z$} 
					\end{overpic}
				}
			\end{minipage}
			&
			\begin{minipage}[t]{0.28\textwidth}
				\centering
				\begin{overpic}[width=0.75\textwidth]{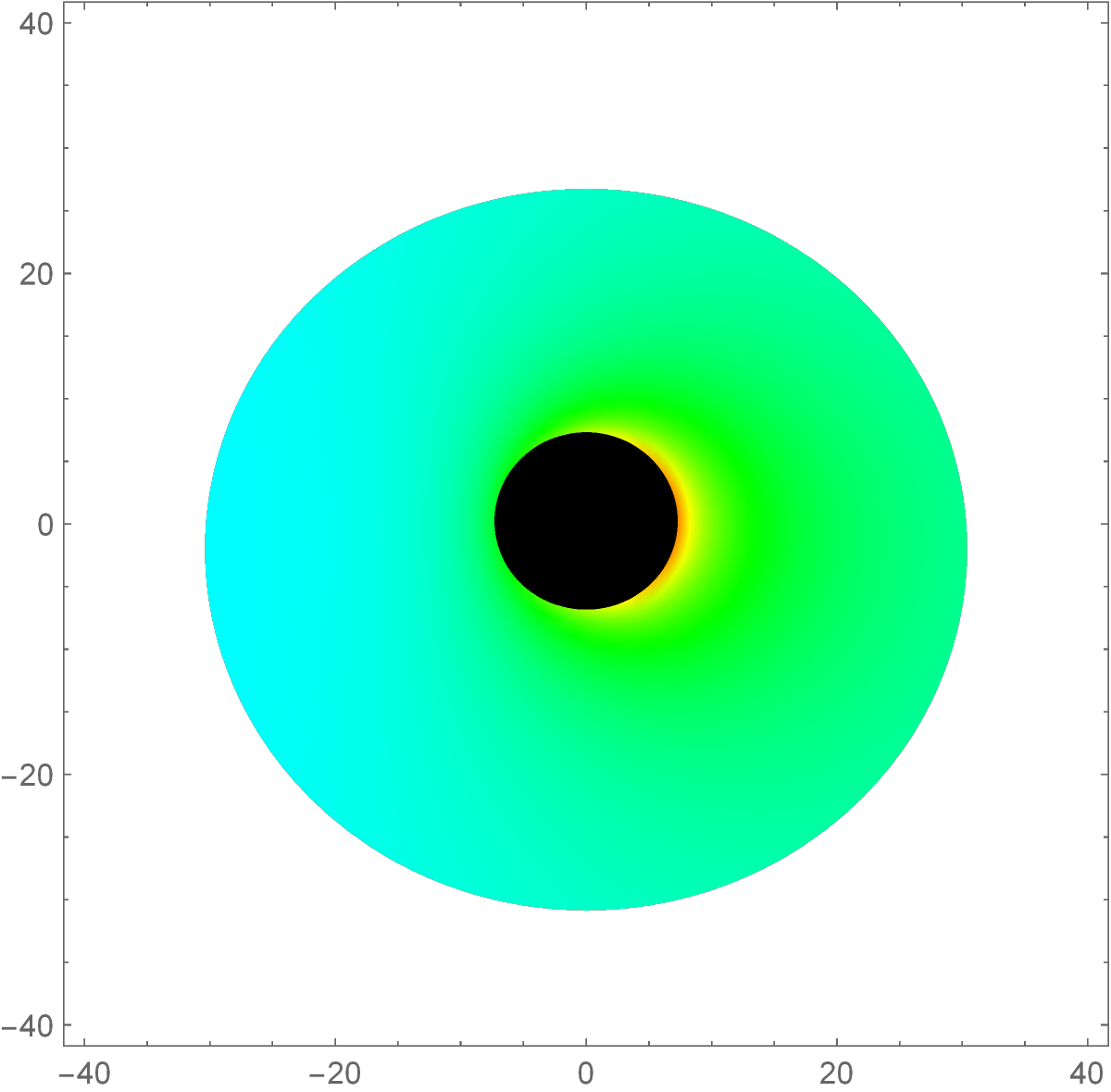}
					\put(17,100){\color{black}\large $l=-0.3,\theta=17^{\circ}$} 
					\put(-8,48){\color{black} Y}
					\put(48,-10){\color{black} X}
				\end{overpic}
				\raisebox{0.05\height}{ 
					\begin{overpic}[width=0.07\textwidth]{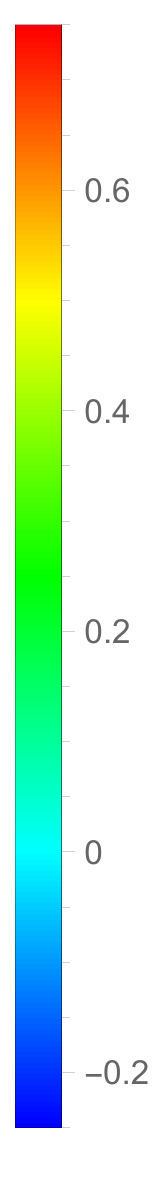} 
						\put(0,103){\color{black}\large $z$} 
					\end{overpic}
				}			
			\end{minipage}
			&
			\begin{minipage}[t]{0.28\textwidth}
				\centering
				\begin{overpic}[width=0.75\textwidth]{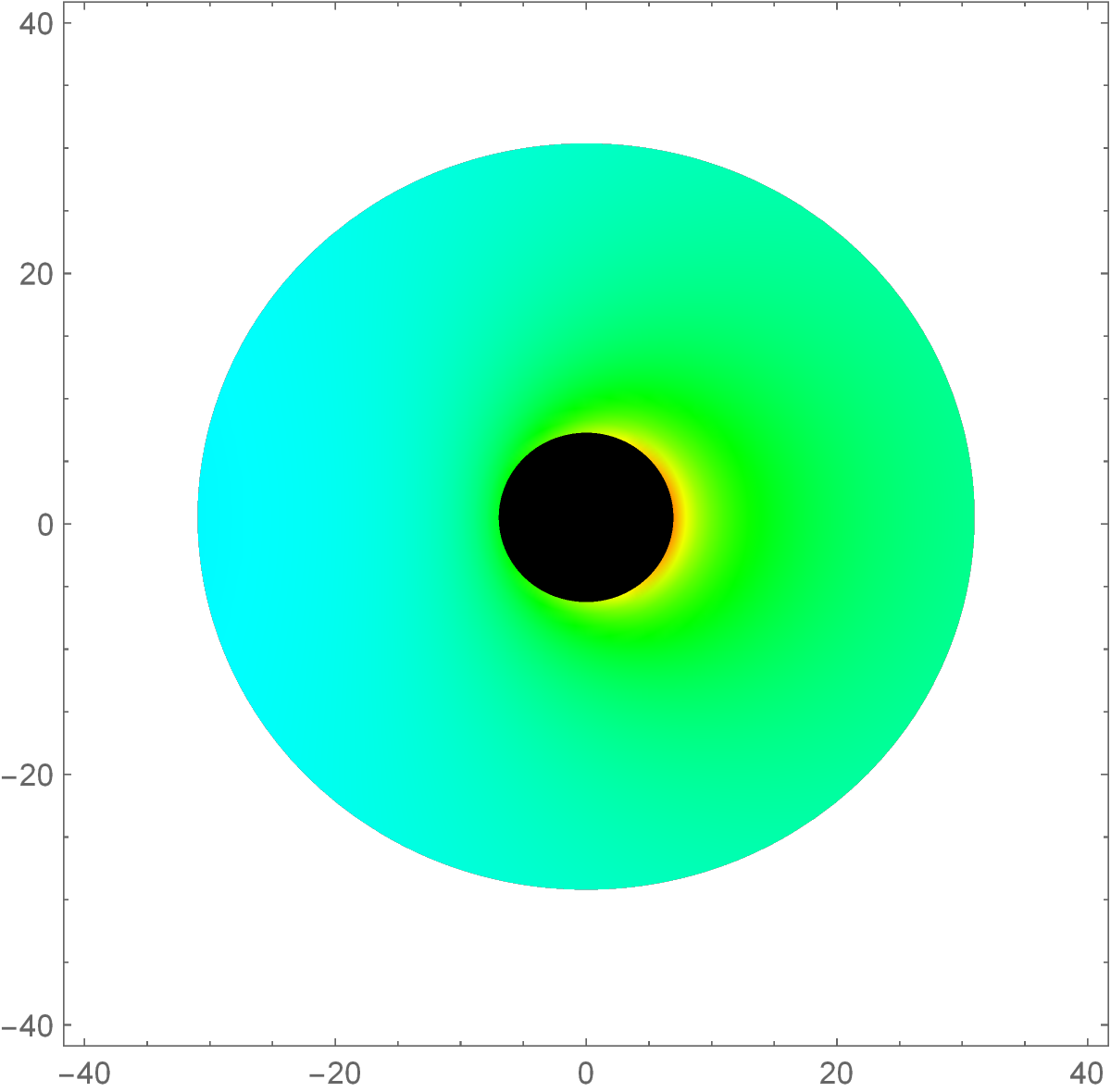}
					\put(25,100){\color{black}\large $l=0,\theta=17^{\circ}$} 
					\put(-8,48){\color{black} Y}
					\put(48,-10){\color{black} X}
				\end{overpic}
				\raisebox{0.05\height}{
					\begin{overpic}[width=0.07\textwidth]{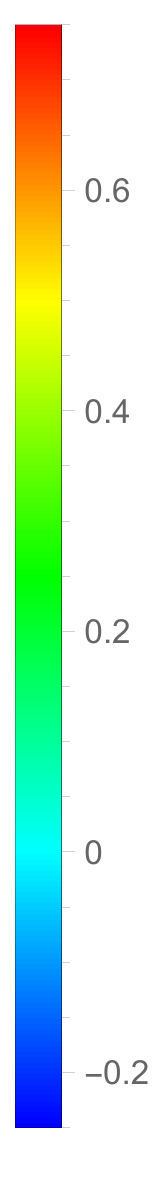}
						\put(0,103){\color{black}\large $z$}
					\end{overpic}
				}
			\end{minipage}
			\vspace{30pt} 
			\\ 
			\begin{minipage}[t]{0.28\textwidth}
				\centering
				\begin{overpic}[width=0.75\textwidth]{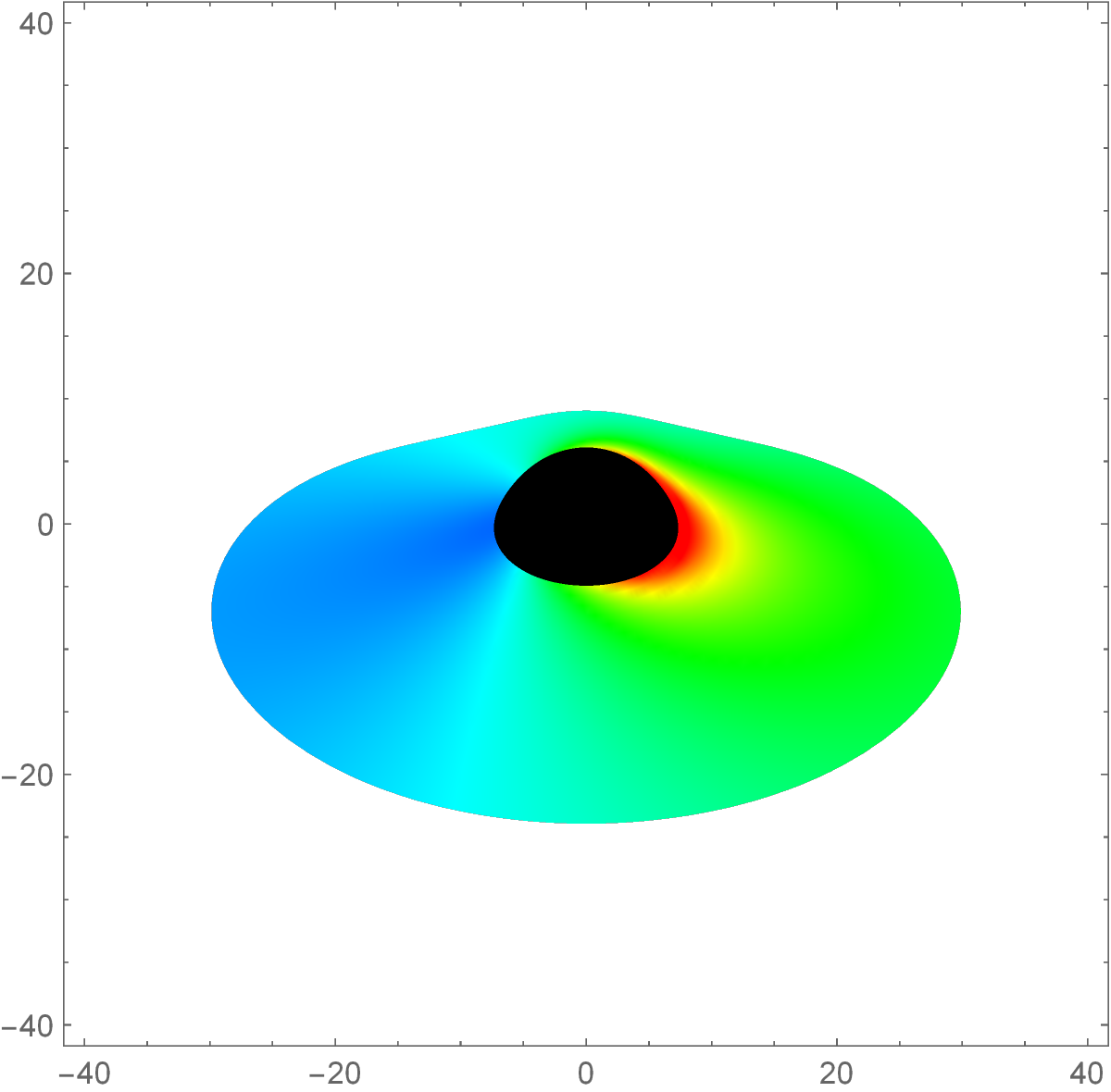}
					\put(20,100){\color{black}\large $l=-0.5,\theta=53^{\circ}$} 
					\put(-8,48){\color{black} Y}
					\put(48,-10){\color{black} X}
				\end{overpic}
				\raisebox{0.05\height}{
					\begin{overpic}[width=0.07\textwidth]{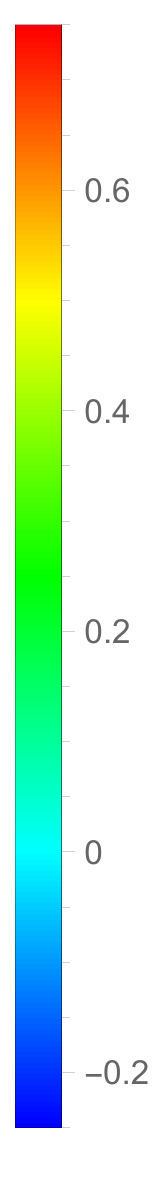} 
						\put(0,103){\color{black}\large $z$} 
					\end{overpic}
				}
			\end{minipage}
			&
			\begin{minipage}[t]{0.28\textwidth}
				\centering
				\begin{overpic}[width=0.75\textwidth]{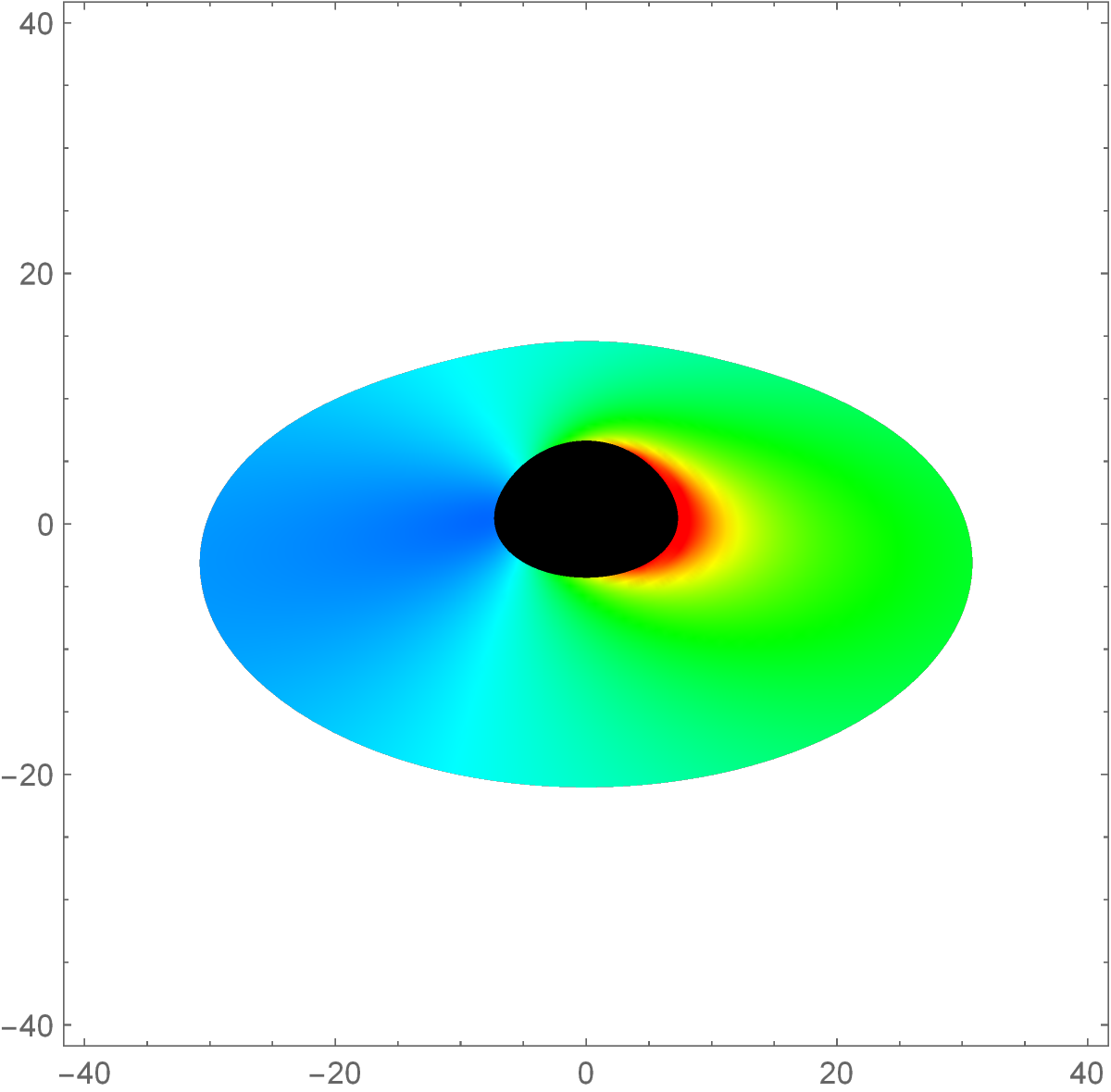} 
					\put(17,100){\color{black}\large $l=-0.3,\theta=53^{\circ}$} 
					\put(-8,48){\color{black} Y}
					\put(48,-10){\color{black} X}
				\end{overpic}
				\raisebox{0.05\height}{ 
					\begin{overpic}[width=0.07\textwidth]{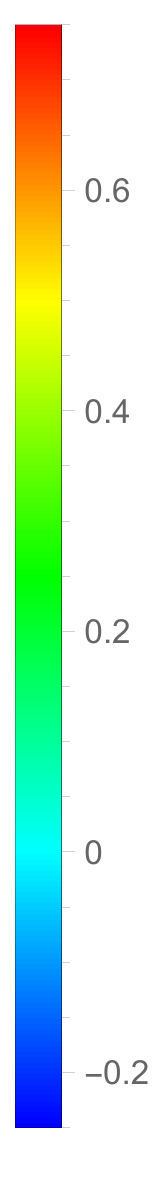} 
						\put(0,103){\color{black}\large $z$} 
					\end{overpic}
				}
			\end{minipage}
			&
			\begin{minipage}[t]{0.28\textwidth}
				\centering
				\begin{overpic}[width=0.75\textwidth]{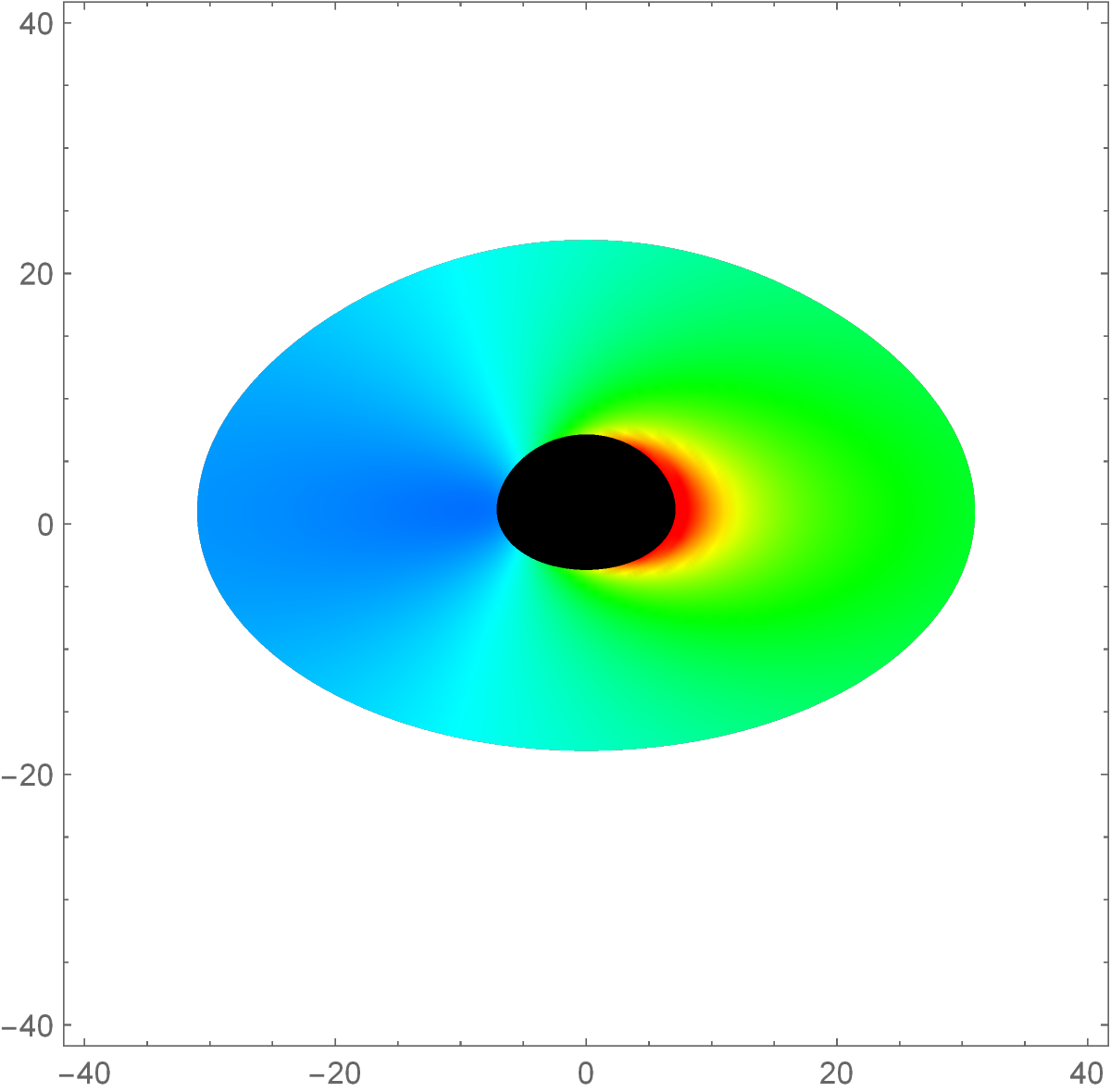} 
					\put(25,100){\color{black}\large $l=0,\theta=53^{\circ}$} 
					\put(-8,48){\color{black} Y}
					\put(48,-10){\color{black} X}
				\end{overpic}
				\raisebox{0.05\height}{
					\begin{overpic}[width=0.07\textwidth]{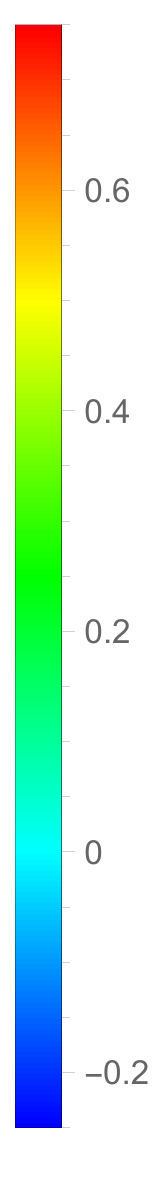}
						\put(0,103){\color{black}\large $z$} 
					\end{overpic}
				}
			\end{minipage}
			\vspace{30pt} 
			\\ 
			\begin{minipage}[t]{0.28\textwidth}
				\centering
				\begin{overpic}[width=0.75\textwidth]{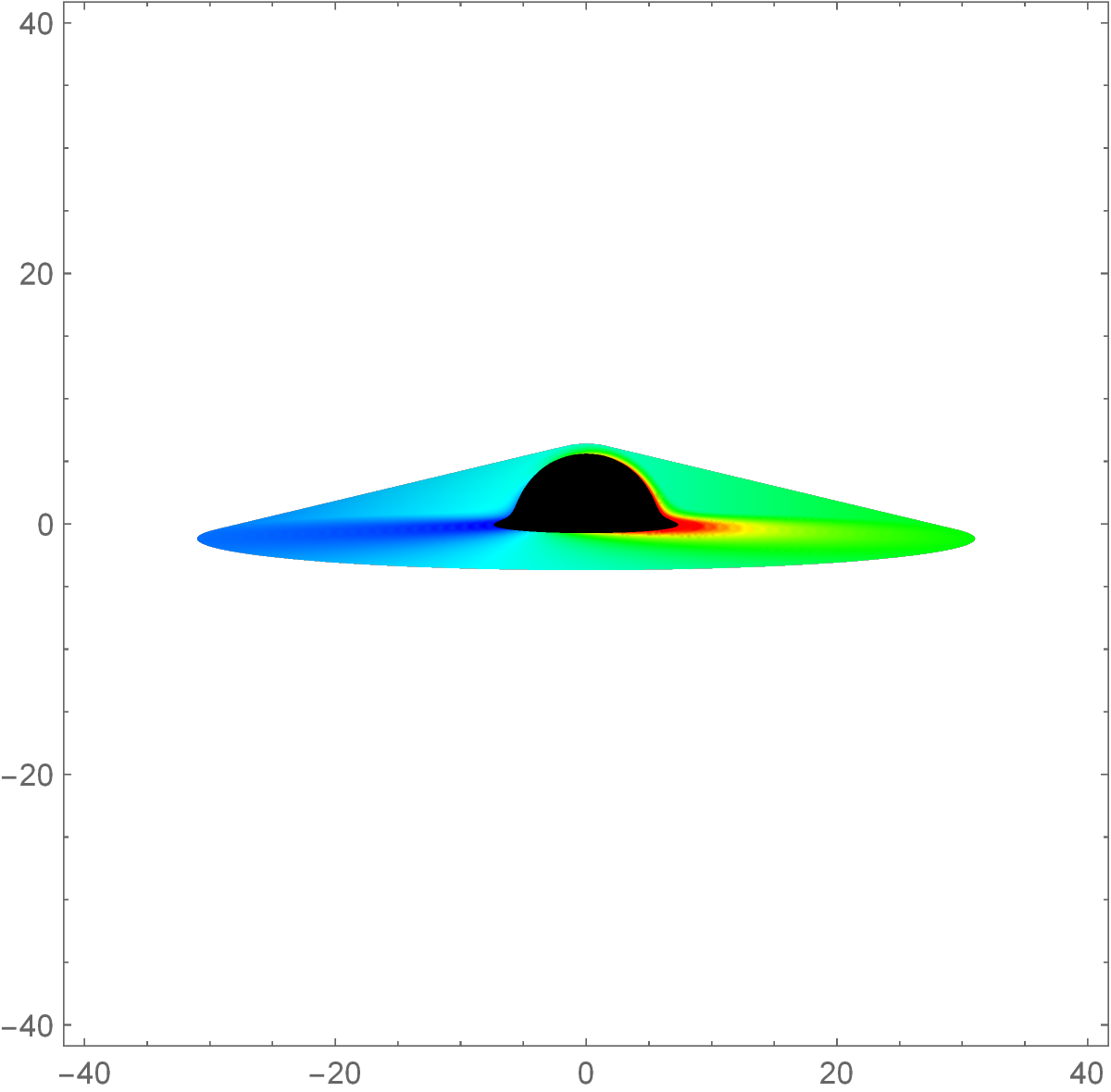} 
					\put(20,100){\color{black}\large $l=-0.5,\theta=85^{\circ}$}
					\put(-8,48){\color{black} Y}
					\put(48,-10){\color{black} X}
				\end{overpic}
				\raisebox{0.13\height}{ 
					\begin{overpic}[width=0.07\textwidth]{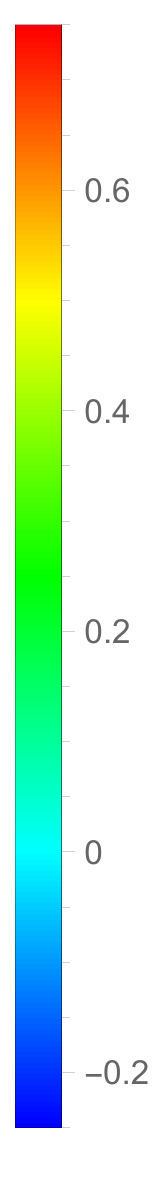}
						\put(0,103){\color{black}\large $z$} 
					\end{overpic}
				}
			\end{minipage}
			&
			\begin{minipage}[t]{0.28\textwidth}
				\centering
				\begin{overpic}[width=0.75\textwidth]{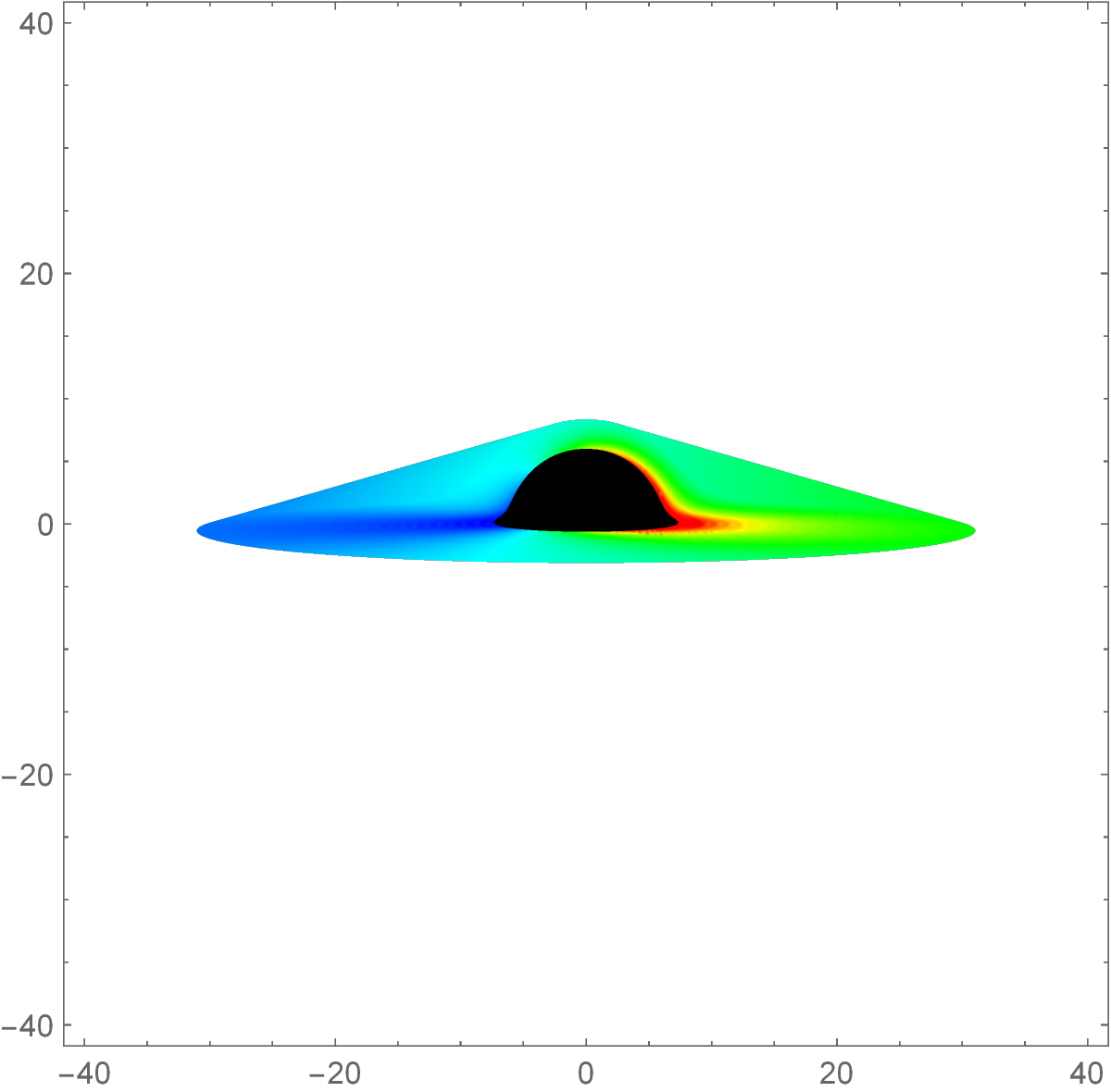} 
					\put(17,100){\color{black}\large $l=-0.3,\theta=85^{\circ}$} 
					\put(-8,48){\color{black} Y}
					\put(48,-10){\color{black} X}
				\end{overpic}
				\raisebox{0.13\height}{ 
					\begin{overpic}[width=0.07\textwidth]{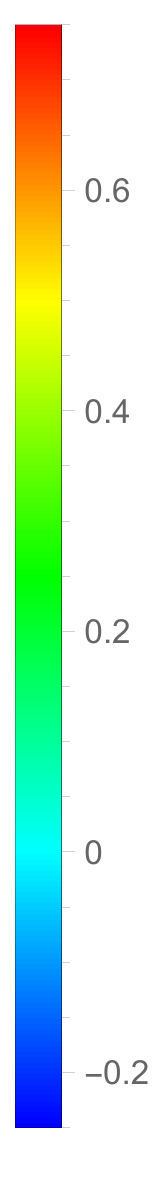} 
						\put(0,103){\color{black}\large $z$} 
					\end{overpic}
				}
			\end{minipage}
			&
			\begin{minipage}[t]{0.28\textwidth}
				\centering
				\begin{overpic}[width=0.75\textwidth]{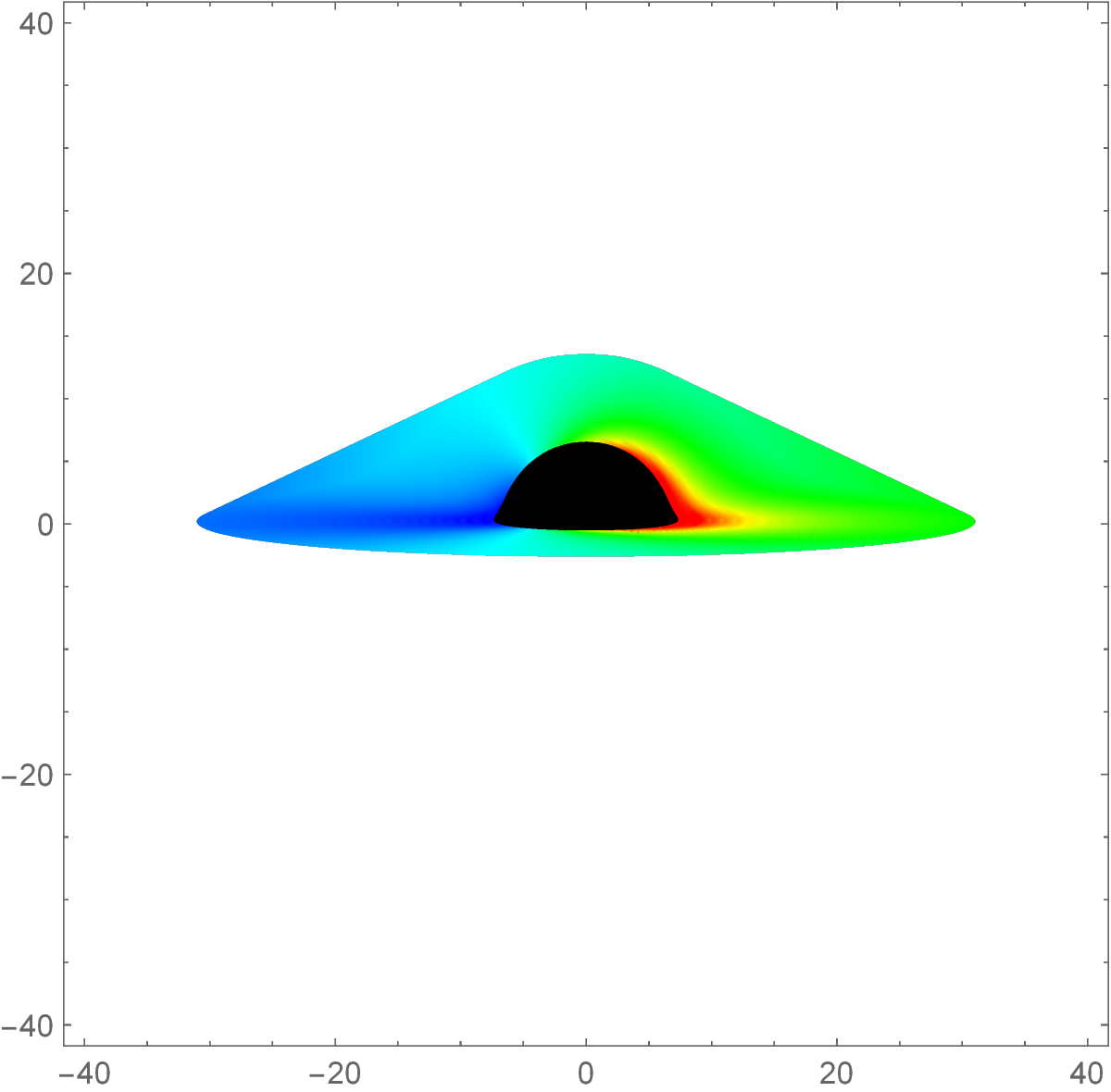}
					\put(25,100){\color{black}\large $l=0,\theta=85^{\circ}$}
					\put(-8,48){\color{black} Y}
					\put(48,-10){\color{black} X}
				\end{overpic}
				\raisebox{0.13\height}{ 
					\begin{overpic}[width=0.07\textwidth]{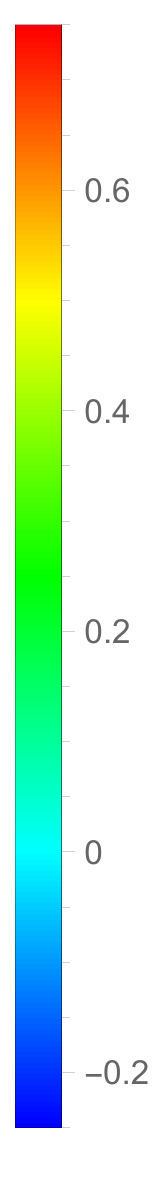} 
						\put(0,103){\color{black}\large $z$} 
					\end{overpic}
				}
			\end{minipage}
		\end{tabular}
		\caption{Continuous distributions of redshift $z$ in the direct images cast by the accretion disk in Schwarzschild-like spacetime under different values of parameter $l$ and inclination angles. Each column, from top to bottom, corresponds to inclination angles of $17^{\circ}$, $53^{\circ}$, and $85^{\circ}$, while each row, from left to right, represents $l$ values of $-0.5$, $-0.3$, and $0$, respectively. The rightmost column corresponds to the Schwarzschild BH.}
		\label{redshift}
\end{figure*}

\begin{figure*}[htbp]
		\centering
		\begin{tabular}{ccc}
			\begin{minipage}[t]{0.28\textwidth}
				\centering
				\begin{overpic}[width=0.75\textwidth]{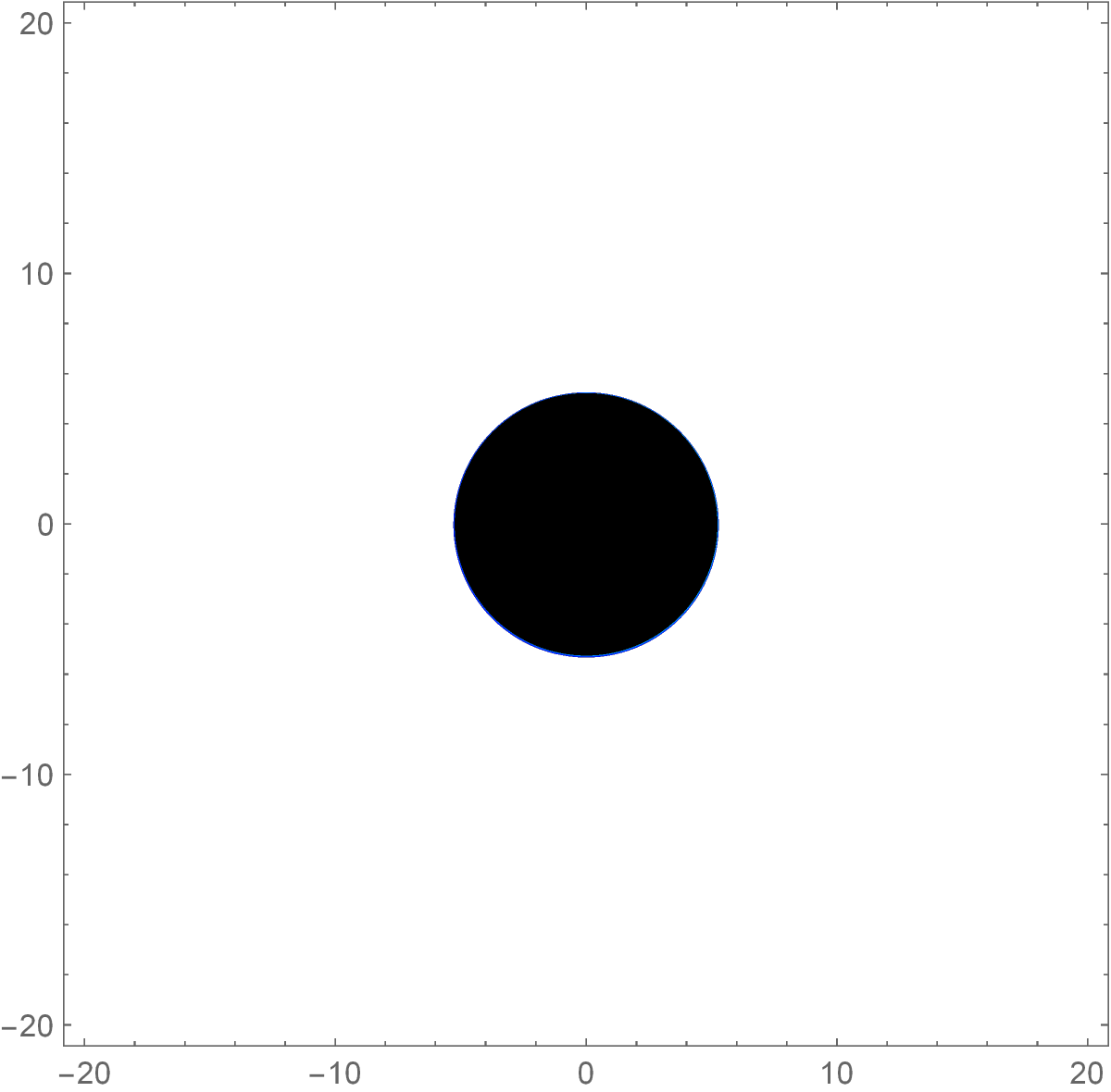}
					\put(20,100){\color{black}\large $l=-0.5,\theta=17^{\circ}$} 
					\put(-8,48){\color{black} Y}
					\put(48,-10){\color{black} X}
				\end{overpic}
				\raisebox{0.05\height}{ 
					\begin{overpic}[width=0.07\textwidth]{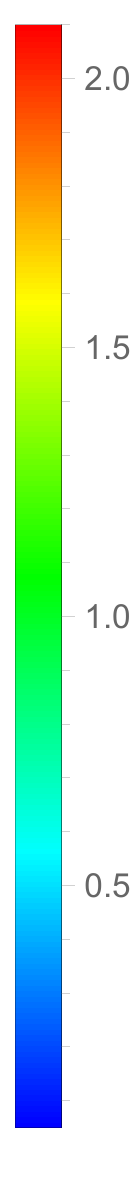}
						\put(0,103){\color{black}\large $z$}
					\end{overpic}
				}
			\end{minipage}
			&
			\begin{minipage}[t]{0.28\textwidth}
				\centering
				\begin{overpic}[width=0.75\textwidth]{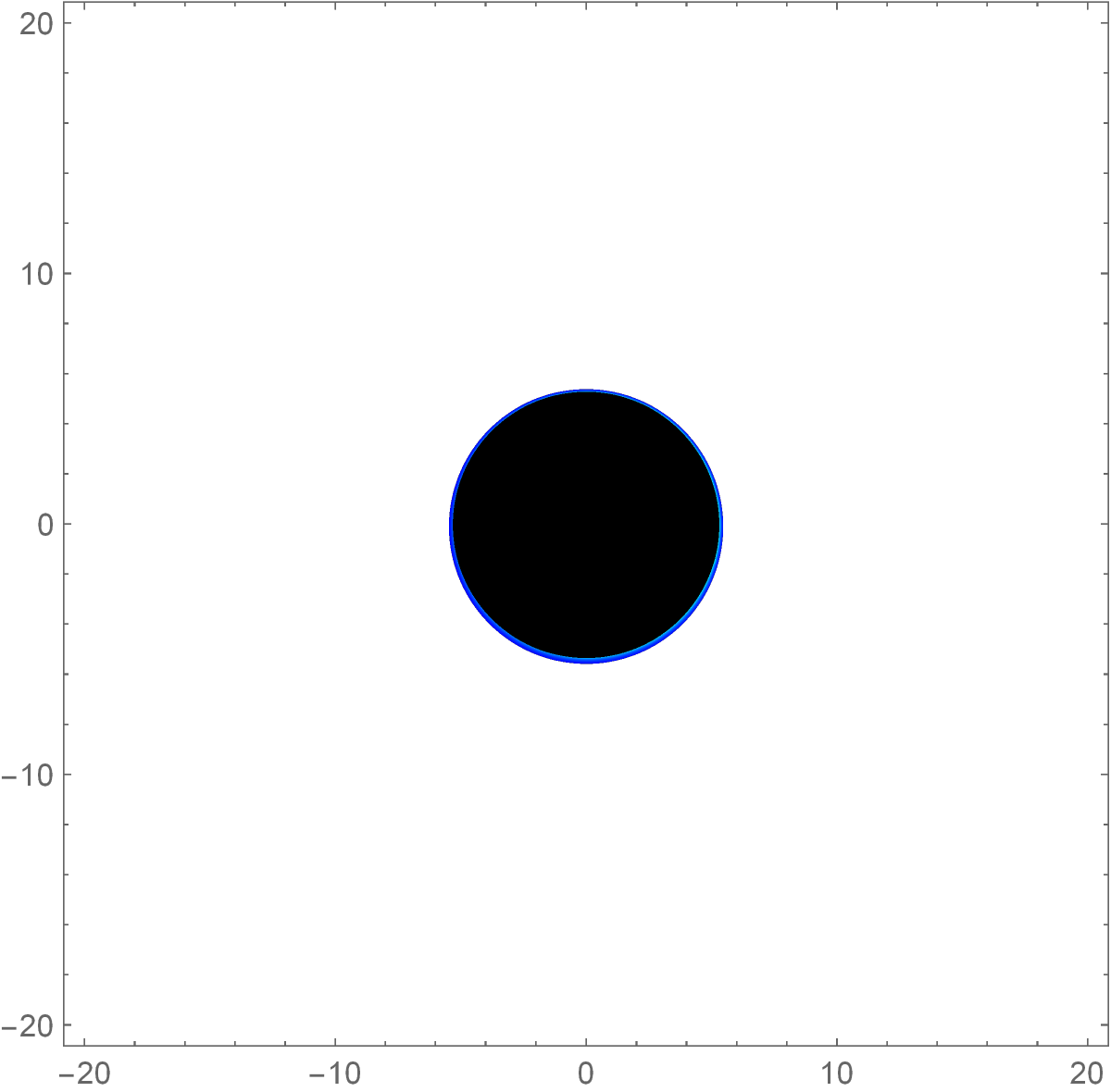} 
					\put(17,100){\color{black}\large $l=-0.3,\theta=17^{\circ}$} 
					\put(-8,48){\color{black} Y}
					\put(48,-10){\color{black} X}
				\end{overpic}
				\raisebox{0.05\height}{ 
					\begin{overpic}[width=0.07\textwidth]{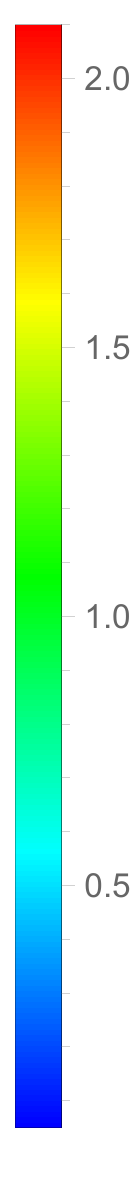}
						\put(0,103){\color{black}\large $z$}
					\end{overpic}
				}
			\end{minipage}
			&
			\begin{minipage}[t]{0.28\textwidth}
				\centering
				\begin{overpic}[width=0.75\textwidth]{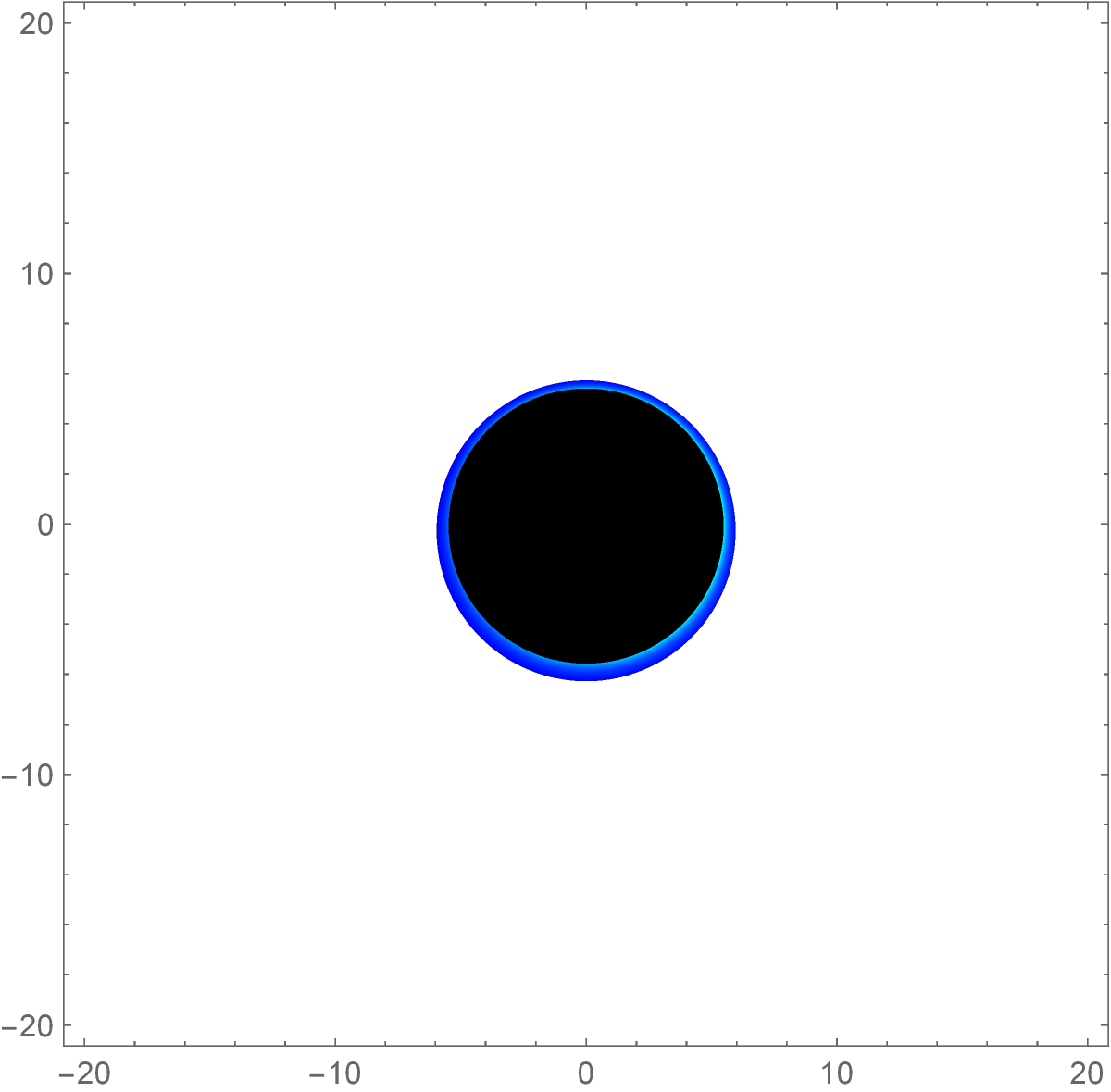} 
					\put(25,100){\color{black}\large $l=0,\theta=17^{\circ}$}
					\put(-8,48){\color{black} Y}
					\put(48,-10){\color{black} X}
				\end{overpic}
				\raisebox{0.05\height}{ 
					\begin{overpic}[width=0.07\textwidth]{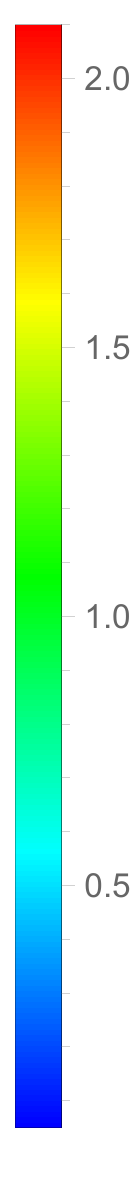}
						\put(0,103){\color{black}\large $z$} 
					\end{overpic}
				}
			\end{minipage}
			\vspace{30pt} 
			
			\\ 
			\begin{minipage}[t]{0.28\textwidth}
				\centering
				\begin{overpic}[width=0.75\textwidth]{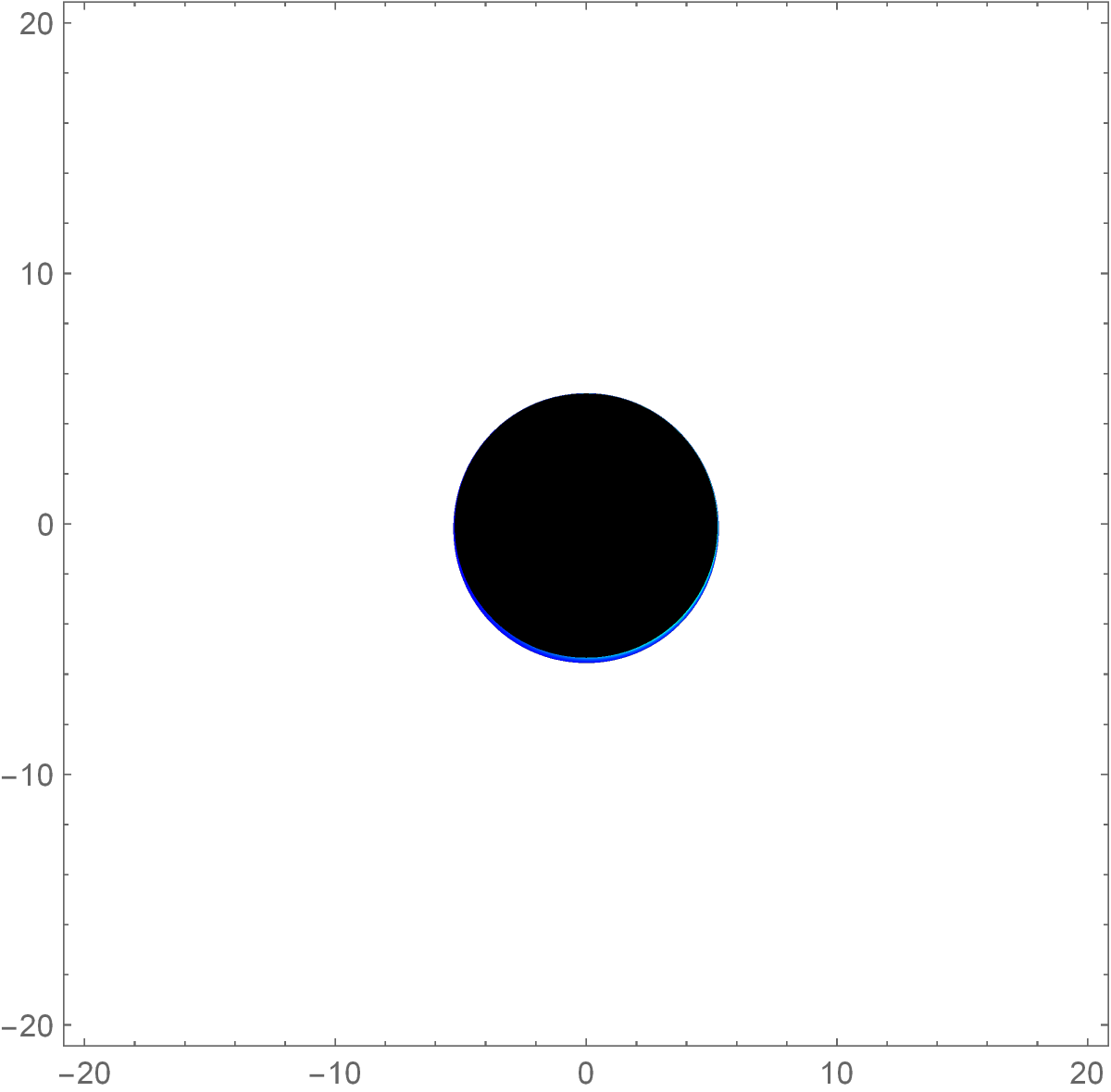}
					\put(20,100){\color{black}\large $l=-0.5,\theta=53^{\circ}$}
					\put(-8,48){\color{black} Y}
					\put(48,-10){\color{black} X}
				\end{overpic}
				\raisebox{0.05\height}{
					\begin{overpic}[width=0.07\textwidth]{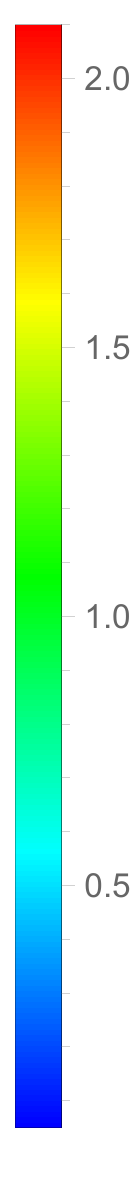}
						\put(0,103){\color{black}\large $z$} 
					\end{overpic}
				}
			\end{minipage}
			&
			\begin{minipage}[t]{0.28\textwidth}
				\centering
				\begin{overpic}[width=0.75\textwidth]{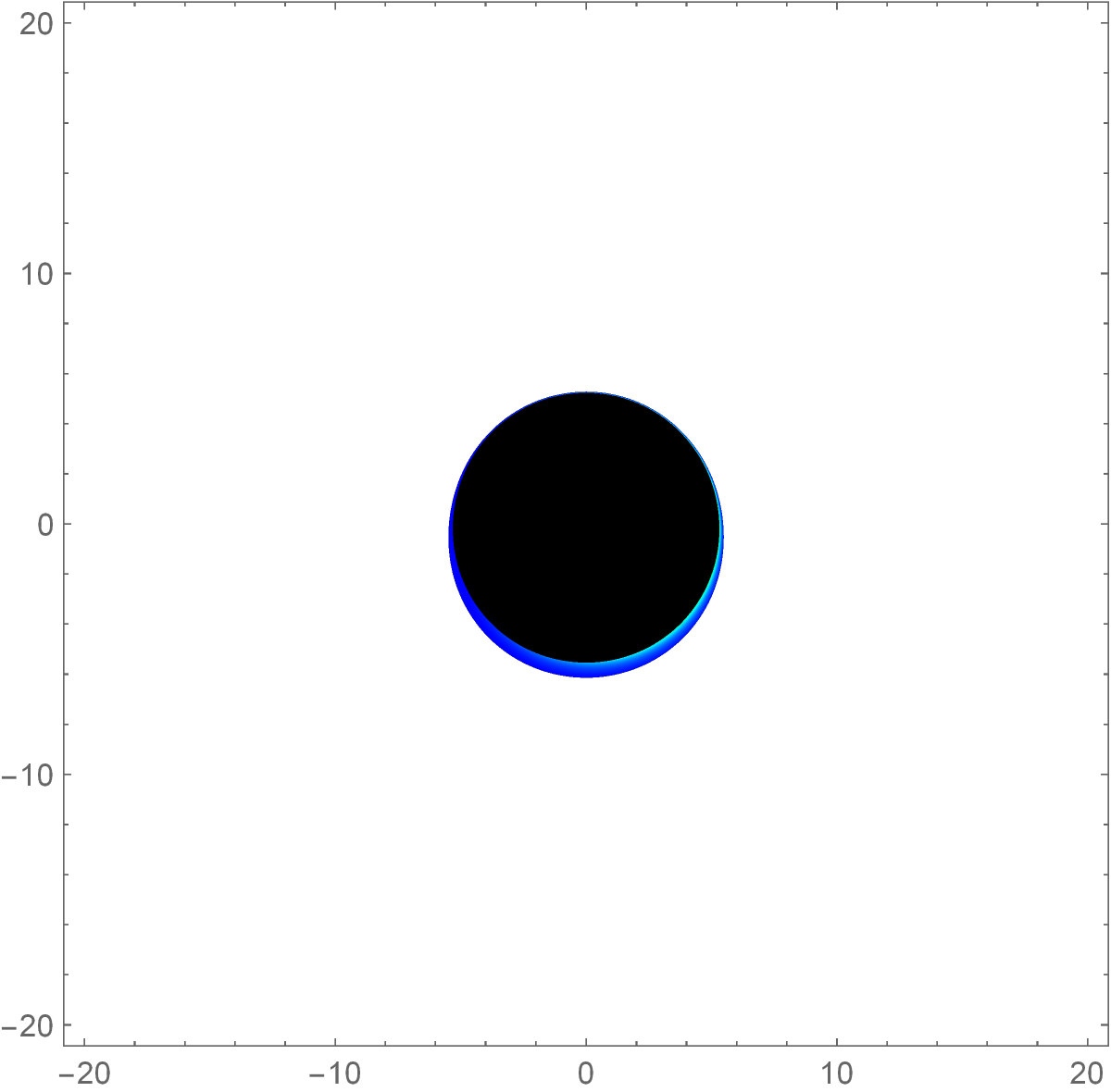}
					\put(17,100){\color{black}\large $l=-0.3,\theta=53^{\circ}$} 
					\put(-8,48){\color{black} Y}
					\put(48,-10){\color{black} X}
				\end{overpic}
				\raisebox{0.05\height}{ 
					\begin{overpic}[width=0.07\textwidth]{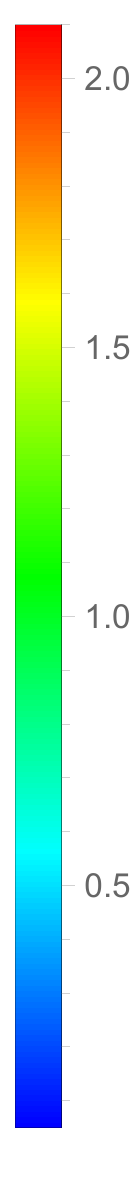} 
						\put(0,103){\color{black}\large $z$}
					\end{overpic}
				}
			\end{minipage}
			&
			\begin{minipage}[t]{0.28\textwidth}
				\centering
				\begin{overpic}[width=0.75\textwidth]{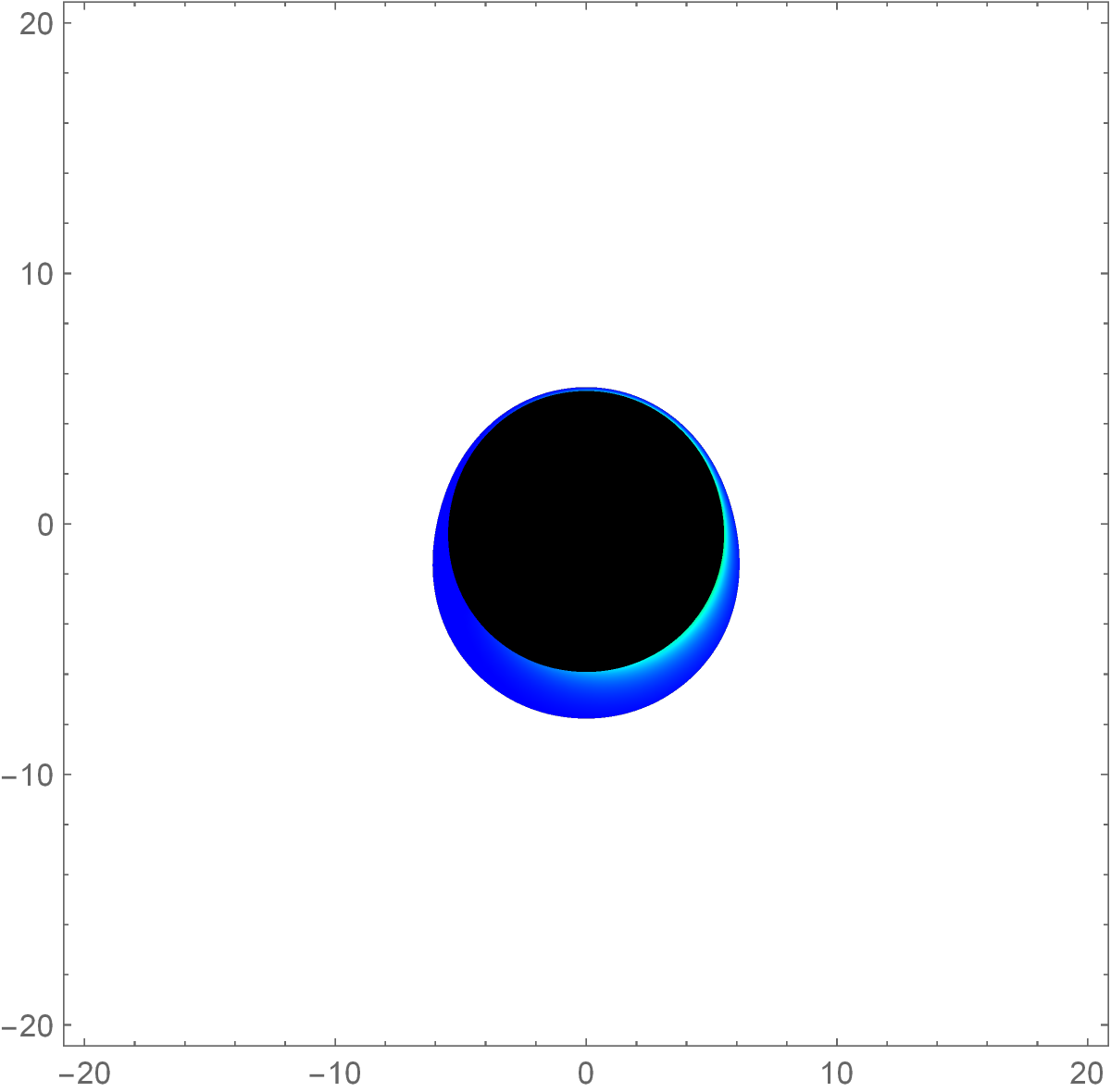} 
					\put(25,100){\color{black}\large $l=0,\theta=53^{\circ}$} 
					\put(-8,48){\color{black} Y}
					\put(48,-10){\color{black} X}
				\end{overpic}
				\raisebox{0.05\height}{ 
					\begin{overpic}[width=0.07\textwidth]{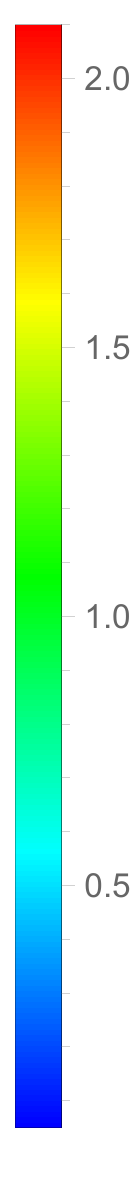} 
						\put(0,103){\color{black}\large $z$} 
					\end{overpic}
				}
			\end{minipage}
			\vspace{30pt} 
			
			\\ 
			\begin{minipage}[t]{0.28\textwidth}
				\centering
				\begin{overpic}[width=0.75\textwidth]{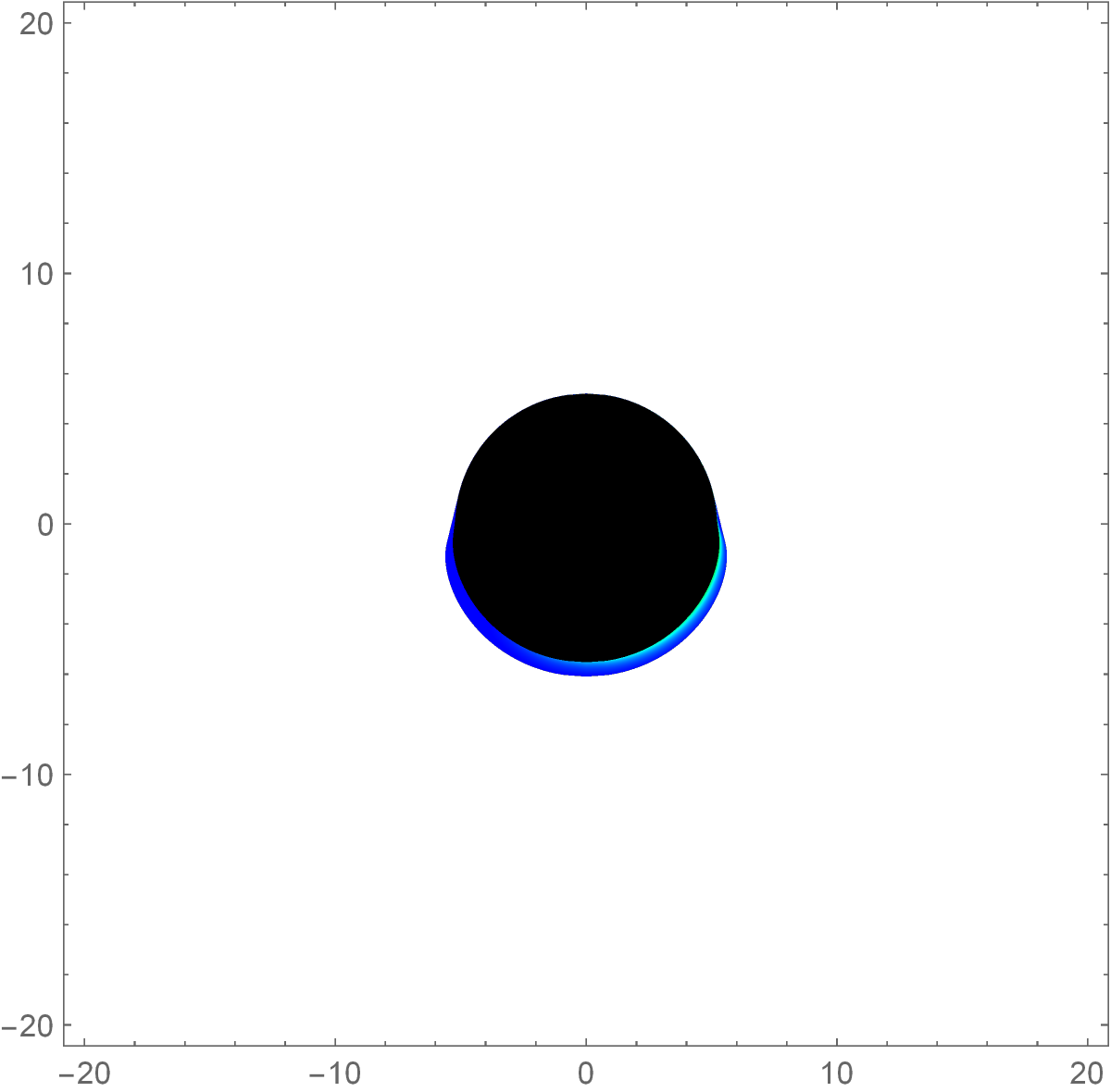}
					\put(20,100){\color{black}\large $l=-0.5,\theta=85^{\circ}$}
					\put(-8,48){\color{black} Y}
					\put(48,-10){\color{black} X}
				\end{overpic}
				\raisebox{0.13\height}{ 
					\begin{overpic}[width=0.07\textwidth]{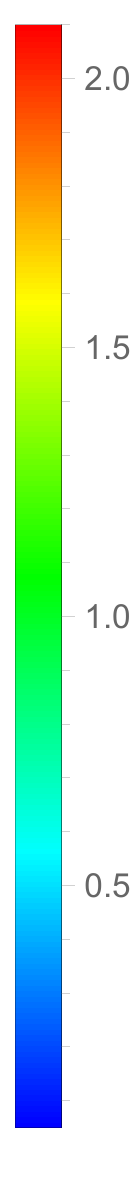}
						\put(0,103){\color{black}\large $z$} 
					\end{overpic}
				}
			\end{minipage}
			&
			\begin{minipage}[t]{0.28\textwidth}
				\centering
				\begin{overpic}[width=0.75\textwidth]{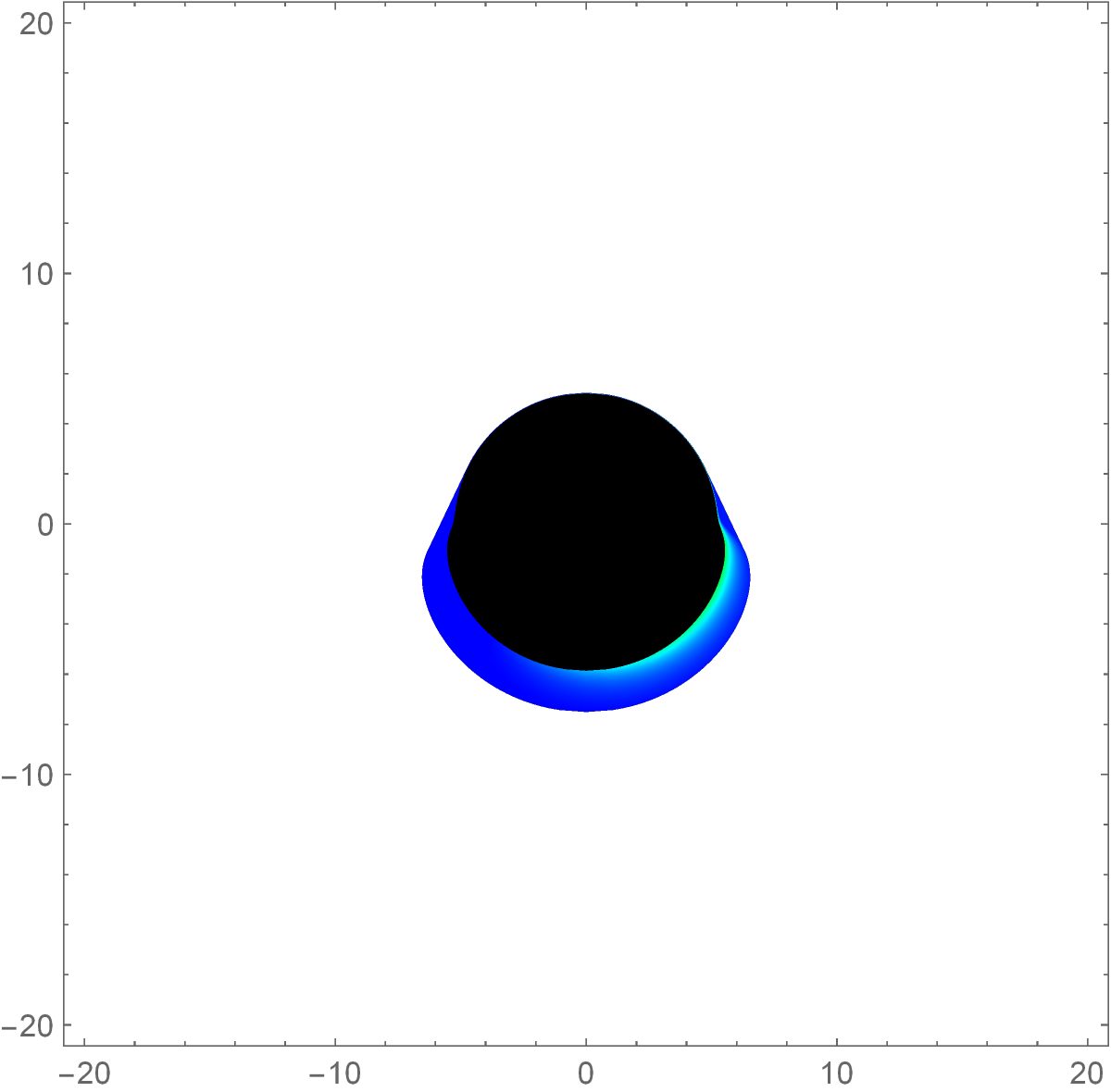}
					\put(17,100){\color{black}\large $l=-0.3,\theta=85^{\circ}$}
					\put(-8,48){\color{black} Y}
					\put(48,-10){\color{black} X}
				\end{overpic}
				\raisebox{0.06\height}{ 
					\begin{overpic}[width=0.07\textwidth]{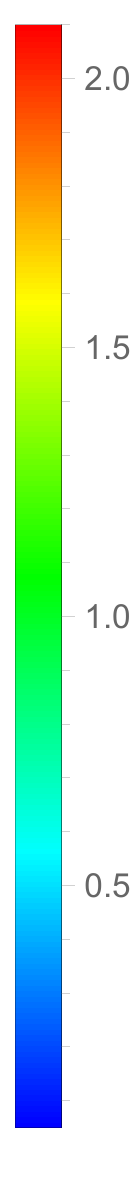} 
						\put(0,103){\color{black}\large $z$} 
					\end{overpic}
				}
			\end{minipage}
			&
			\begin{minipage}[t]{0.28\textwidth}
				\centering
				\begin{overpic}[width=0.75\textwidth]{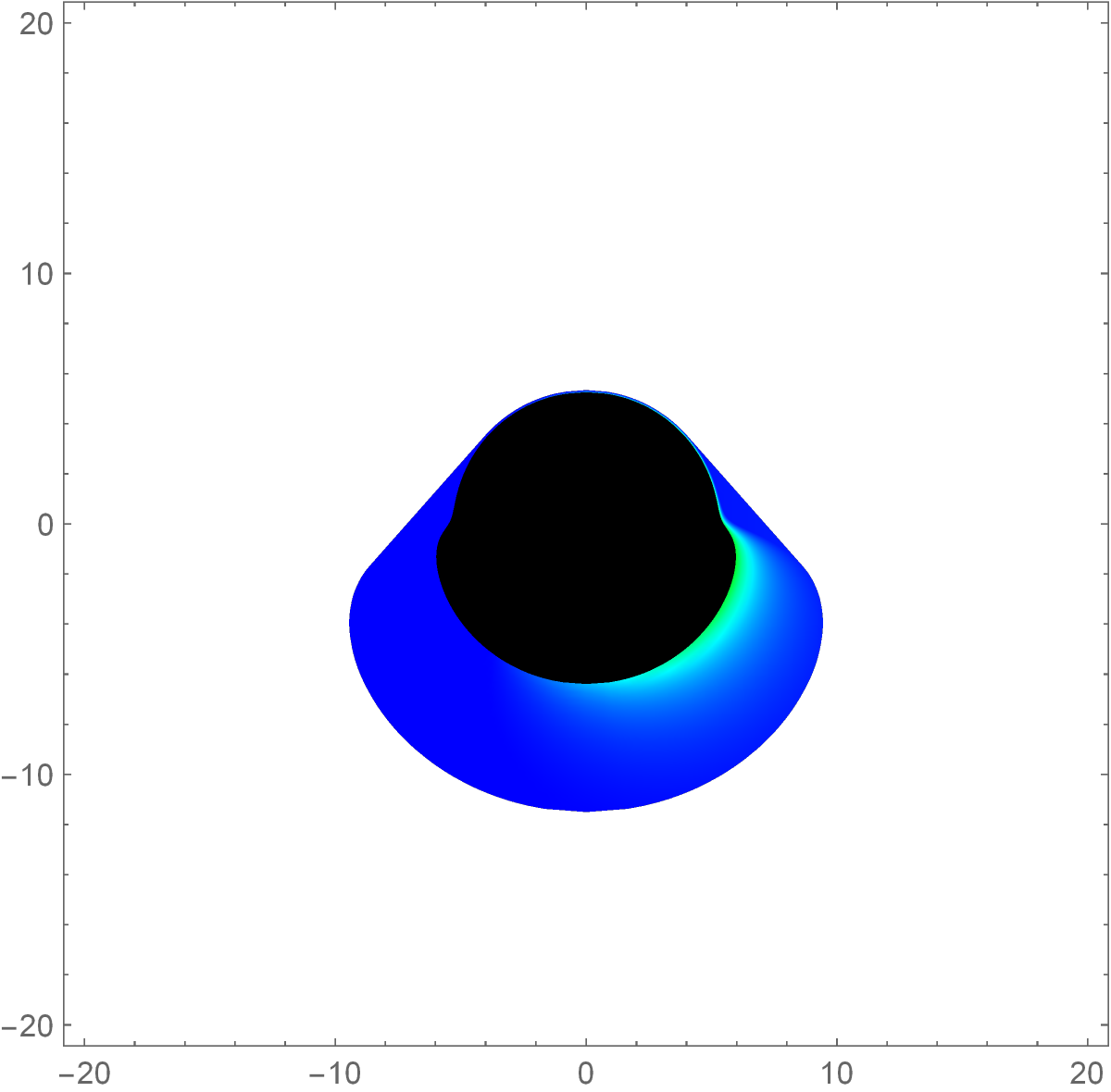}
					\put(25,100){\color{black}\large $l=0,\theta=85^{\circ}$}
					\put(-8,48){\color{black} Y}
					\put(48,-10){\color{black} X}
				\end{overpic}
				\raisebox{0.06\height}{ 
					\begin{overpic}[width=0.07\textwidth]{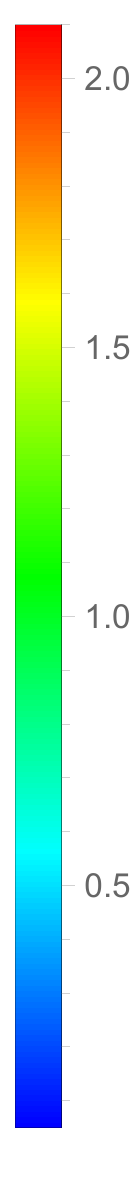} 
						\put(0,103){\color{black}\large $z$} 
					\end{overpic}
				}
			\end{minipage}
		\end{tabular}
		\caption{Similar to Fig.~\ref{redshift}, but for the secondary images.}
		\label{redshift2}
\end{figure*}

We investigate the redshift distributions (contour map of redshift z) in the direct images for various inclination angles and parameter $l$, as demonstrated in Fig.~\ref{redshift}. The dependency of the redshift on parameter $l$ is also shown in the secondary images, which are plotted in Fig.~\ref{redshift2}. For high inclination angles, the blueshift $(z<0)$ in the left half of the plate exceeds the gravitational redshift $(z>0)$ due to the presence of the BH. Conversely, for low inclination angles, no blueshift distribution is observed. Additionally, we find that the region with high redshift values $z$ increases as the parameter $l$ decreases.
\section{Conclusion} 
\label{section5}
In this study, within the framework of bumblebee gravity, we have studied the physical properties and the optical appearance of a thin accretion disk around a Schwarzschild-like BH. As the LSB parameter $l$ decreases, the energy flux, temperature, and emission spectrum of the accretion disk exhibit a systematic enhancement, accompanied by an increase in overall luminosity. Furthermore, with decreasing $l$, the direct image stretches outward horizontally and compresses inward vertically, while the secondary image contracts inward in both directions, resulting in an overall reduction in size. Additionally, as the observational inclination angle increases, the distortion and asymmetry of the images become more pronounced. Our investigation into the redshift distribution reveals that the redshift factor grows as $l$ decreases. Moreover, our analysis reveals that the relationship between the photon's impact parameter $b$ and its deflection angle $\varphi$ when it reaches the circular orbit of timelike particles plays a crucial role in elucidating the imaging mechanism of the thin accretion disk.
	
By uncovering these effects, our research not only deepens our understanding of how modified gravity theories influence BH physics but also offers potential avenues for distinguishing between different gravitational models through observational data. The implications of our findings suggest that bumblebee gravity could provide a richer framework for investigating the complex interplay between LSB and the observable properties of BHs.
	
\begin{acknowledgments}
	This research was supported by the National Natural Science Foundation of China (Grant No. 12265007).
\end{acknowledgments}

\end{document}